\documentclass[aip, jmp, amsmath, amssymb, showkeys, reprint]{revtex4-1}

\usepackage{graphicx} 
\usepackage{dcolumn} 
\usepackage{bm} 
\usepackage{mathtools}
\usepackage{hyperref}
\usepackage{caption}
\usepackage{subcaption}
\usepackage{tabularx}
\usepackage{amsmath}
\usepackage{xcolor}
\usepackage{comment}
\usepackage{soul}
\usepackage{setspace}
\usepackage[capitalize]{cleveref}
\usepackage{caption}

\onecolumngrid

\begin{document}

\preprint{AIP/123-QED}

\title{Effect of channel dimensions and Reynolds numbers on the turbulence modulation for particle-laden turbulent channel flows}

\author{Naveen Rohilla}
\author{Siddhi Arya}
\author{Partha Sarathi Goswami}
\email{psg@iitb.ac.in}
\affiliation{Department of Chemical Engineering, Indian Institute of Technology Bombay, Mumbai - 400076 (India)}

\date{\today}

\begin{abstract}
The addition of particles to turbulent flows changes the underlying mechanism of turbulence and leads to turbulence modulation. Different temporal and spatial scales for both phases make it challenging to understand turbulence modulation via one parameter. The important parameters are particle Stokes number, mass loading, particle Reynolds number, fluid bulk Reynolds number, etc., that act together and affect the fluid phase turbulence intensities. In the present study, we have carried out the large eddy simulations for different system sizes ($2\delta/d_p = 54, 81,$ and 117) and fluid bulk Reynolds numbers ($Re_b = 5600$ and 13750) to quantify the extent of turbulence attenuation. Here, $\delta$ is the half-channel width, $d_p$ is the particle diameter, and $Re_b$ is the fluid Reynolds number based on the fluid bulk velocity and channel width. The point particles are tracked with the Lagrangian approach. The scaling analysis of the feedback force shows that system size and fluid bulk Reynolds number are the two crucial parameters that affect the turbulence modulation more significantly than the other. It is found that the extent of turbulence attenuation increases with an increase in system size for the same volume fraction while keeping the Reynolds number fixed. But, for the same volume fraction and fixed channel dimension, the extent of attenuation is low at a higher Reynolds number. The streamwise turbulent structures are observed to become lengthier and fewer with an increase in system size for the same volume fraction and fixed bulk Reynolds number. However, the streamwise high-speed streaks are smaller, thinner, and closely spaced for higher Reynolds numbers than the lower ones for the same volume fraction. In particle statistics, it is observed that the scaled particle fluctuations increase with the increase in system size while keeping the Reynolds number fixed. However, the scaled particle fluctuations decrease with the increase in fluid bulk Reynolds number for the same volume fraction and fixed system size. The present study highlights the scaling issue for designing industrial equipment for particle-laden turbulent flows.

\end{abstract}

\keywords{particle-laden flows,  LES, turbulence modulation, } 

\maketitle

\section{\label{sec:intro}Introduction}

The Particle laden turbulent flows are ubiquitous in different industrial processes such as combustion, coating, fluidization, pneumatic transport of solids, electrostatic precipitators, and natural processes like raindrop formation, dust storms, etc. The complex interactions between the gas and solid phases lead to the preferential accumulation of high inertia particles at low vorticity zones, crossing trajectories, turbophoresis induced inhomogeneous distribution of particles, and fluid phase turbulence modulation. Turbulence modulation is an important area of investigation as the extent of turbulence modulation directly impacts the heat and mass transfer properties. Turbulence modulation depends on many parameters such as Stokes number, particle Reynolds number, fluid Reynolds number, the ratio of integral length scale to the particle diameter, solid mass loading, etc. In an early work, \citet{Gore1989} divided the augmentation and attenuation region based on the integral length scale ratio to the particle diameter ($L/d_p$). The authors found that if the ratio is less than 0.1, the attenuation will be observed; otherwise, augmentation will be observed. \citet{Tanaka2008} presented two dimensionless numbers based on the Stokes number and particle Reynolds number ($Re_p$), and differentiated the regions of augmentation and attenuation based on these two dimensionless numbers. They found that region of attenuation exist between two augmentation regions. \citet{yu2021modulation} have proposed the criteria to predict the attenuation or augmentation for turbulent kinetic energy. The authors have performed fully resolved simulations for an upward channel for two bulk Reynolds numbers ($Re_b$) of 5746 and 12000, $\rho_p/\rho_f = 2-100$, and $\phi = 0.003 - 0.0236$. Here, $\phi$ is the solid volume fraction, $\rho_p$ is the particle density, and $\rho_f$ is the fluid density. 
For  $Re_p < 50$, $\rho_p/\rho_f = 2$, and $ 2\delta/d_p  = 20$, an increase in bulk Reynolds number from 5746 to 12000 leads to less attenuation for $\phi = 0.0236$ (Fig.~11 of their paper). The authors observed that at low $Re_p$, the turbulence attenuation is observed across the channel; at intermediate $Re_p$, the turbulence intensity is enhanced in the channel center and diminished in the near wall. However, at sufficiently high $Re_p$, the turbulence intensity is increased across the channel width. \citet{yu2017effects} have simulated the fully resolved DNS at friction Reynolds number of 180, $\phi = 0.84\%$, and $\rho_p/\rho_f = 1 - 104.2$. The effect of density ratio on the turbulence modulation in a turbulent channel flow with constant pressure gradient and without considering the gravity effect is studied. 
The drag modification with varying density ratio is not monotonic, and authors found large drag at a density ratio of 10.42 compared to unity and 104.2. It is mentioned that the reduction in Reynolds stress results in drag reduction, but it is counteracted by the increase in particle-induced stress and leads to an increase in overall drag\cite{yu2017effects, costa2021near}. 

In their experimental work, \citet{KulickJD1994} found that turbulence attenuation increases with increased Stokes number, mass loading, and distance from the wall. The authors studied the turbulence modulation at $Re_\tau = 644$ based on half-channel width and friction velocity. They didn't observe the attenuation in fluid fluctuations for low Stokes numbers (fluid time scale based on $k/\epsilon$ at the centerline) of 0.57 (glass particles). However, a significant decrease in the streamwise fluid fluctuations was observed for Stokes number of 3 (copper particles) at a mass loading of 0.8. \citet{li2001numerical} carried out DNS studies at low $Re_\tau = 125$ for the vertical channel flows and studied the effect of mass loading, Stokes number, and density ratio. A complete turbulence collapse was observed for the high mass loading, $\phi_m = 2$. \citet{YAMAMOTO2001} performed large eddy simulations (LES) for a vertical channel at Reynolds number ($Re_\tau$) of 644 and emphasized the importance of particle-particle collisions even for low volume fraction cases. The authors observed very small attenuation even at mass loading of one for the copper particle of $St=\tau_pU_{cl}/\delta$ = 70 where $\tau_p$ is the particle relaxation time, $\delta$ is the half-channel width, and $U_{cl}$ is the channel centerline velocity. \citet{duque2021influence} performed the LES of turbulent channel flows for three different Reynolds numbers of $Re_\tau = 180$, 365, and 950. For mass loading of 2.96, a complete turbulence collapse was observed at $Re_\tau = 180$,  while a negligible turbulence modification was observed at $Re_\tau = 950$. 
The Stokes number can be expressed as $St = \tau_p/\tau_k = (1/18) (d_p/\eta)^2 (\rho_p/\rho_f)$. Here, $\tau_k$ is the Kolmogorov velocity scale, $\eta$ is the Kolmogorov length scale, and $\tau_p$ is the particle relaxation time. Thus, the Stokes number can be modified by varying the density ratio or changing particle diameter. The effect of both the parameters was investigated by  \citet{shen2022turbulence} for homogenous isotropic turbulence with fully resolved simulations. In their study, it was observed that particles always lead to turbulence attenuation; attenuation was found to be larger for high-density particles when the particle diameters were the same. However, the attenuation is smaller for the larger particles for the fixed density ratio due to vortex shedding behind the particles when the particle Reynolds number is significantly larger. 
\citet{lee2015modification} performed DNS to examine the effect of Stokes number on the turbulence modulation for the channel flow. The authors observed that the smaller particles ( $ St^+ = 0.5 $) act as the energy source for the high-low speed streaks that may increase the instability and gives rise to the new quasi-streamwise vortices. However, the turbulence was decreased for the high $St^+$ particles. An increase in the number of quasi-streamwise vortices was observed at $St^+ = 0.5$, while the number is less for the high Stokes numbers.

The effect of particle loading on the turbulent structure, viz. streak spacing in turbulent channel flows, has been described by different authors \cite{zhao2010turbulence, zhou2020non, duque2021influence}.
\citet{zhou2020non} observed a non-monotonic behavior of streamwise fluid fluctuations with an increase in mass loading. However, the wall-normal and spanwise fluid fluctuations decreased with an increase in particle mass loading. It was observed that the streamwise fluctuations increased up to a mass loading of 0.75, and a further increase in mass loading to 0.96 caused a decrease in streamwise fluid fluctuations. The authors presented two phenomena related to the number and spacing of the near-wall streaks and modifications of the strength of vortices. First, the near-wall vortices become fewer and wider, and the spacing between the streaks increases with an increase in mass loading. This results in a decrease in the streamwise fluid fluctuations. Second, the vortices and streaks aligned in the streamwise direction, and enhanced the strength of the streaks, increasing the streamwise fluctuations beyond the mass loading of 0.75. \citet{zhao2010turbulence} and \citet{white2008mechanics} mentioned that the particles affect the regeneration cycle of near-wall turbulence by reducing the strength and the number of quasi-streamwise vortices. This leads to wider and more stable near-wall streaks. \citet{dritselis2008numerical} concluded that the particles provide torque in the opposite direction to that of the mean vortex, which leads to the weaker and wider mean fluid structure. \citet{ghosh2022dynamics} discussed the role of inter-particle collisions in breaking up the high-low speed streaks due to particles in the context of particle-laden shear flows. The turbulence modulation and particle clustering phenomena have been reviewed by \citet{balachandar2010turbulent, brandt2022particle}.  
A number of studies on particle-laden turbulent channel flows have also been summarized by \citet{muramulla2020disruption}, \citet{wang2020effect}, and \citet{rohilla2022applicability}.





\citet{Tanaka2008} have proposed two dimensionless numbers (particle momentum numbers) to classify the regimes of turbulence modulation, which are based on the particle Reynolds number and Stokes number. 
One of the particle momentum numbers based on the Stokes number is given as, $Pa_{St} = \frac{1}{54 \sqrt{2}} \frac{Re_b^2}{St_K^{1/2}} \left( \frac{\rho_p}{\rho_f}\right)^{3/2} \left( \frac{d_p}{L}\right)^{3}$, when $v_k \sim v_{rel}$ in the limit of $Re_p \sim d_p/\eta$. Here, $St_K$ is the Stokes number based on the Kolmogorov time scale, $v_k$ and $\eta$ are the Kolmogorov velocity and length scales, and $v_{rel}$ represents the relative velocity between the fluid and particle phases. 
Another dimensionless number based on the particle Reynolds number is defined as $Pa_{Re} = \frac{1}{18} \frac{\rho_p}{\rho_f} \left(\frac{d_p}{L}\right)^3 \frac{Re_b^2}{Re_p}$. The particle Reynolds number is not an independent variable. However, other parameters such as density ratio, bulk Reynolds number, and the ratio of particle size to integral length scale can be input parameters. The effect of mass loading or the density ratio on the turbulence modulation for the channel flows \cite{KulickJD1994, li2001numerical, wu2006experimental, lee2015modification, capecelatro2018transition, fong2019velocity, muramulla2020disruption, zhou2020non}, homogenous isotropic turbulence \cite{ten2004fully, yeo2010modulation, shen2022turbulence}, homogenous shear \cite{ahmed2000mechanisms, yousefi2020modulation}, decaying turbulence \cite{lucci2010modulation, gao2013lattice, schneiders2017direct, luo2017fully}, Couette flow  \cite{richter2013momentum, richter2014modification, richter2015turbulence, ghosh2022dynamics} have been reported. The different particle-laden experimental works are summarized by \citet{wu2006experimental} and \citet{fong2019velocity}. Effect of channel dimensions (ratio of the integral length scale to the particle diameter) and Reynolds number on the turbulence modulation in the limit of high Stokes number and low solid volume loading is scarce in the literature \cite{yu2017effects, yu_2021}. In the case of LES, there are limited studies where turbulence modulation has been explored with the variation of volume fraction/mass loading and Reynolds number \cite{YAMAMOTO2001, duque2021influence, rohilla2022applicability}. In this work, we have studied the effect of $L/d_p$ and Reynolds numbers on turbulence modulation and found that these are the two important parameters that quantify the attenuation level relative to volume fraction. The present study highlights the important scaling issue which should be dealt while comparing the simulation and experimental results.

Paper outline: In the second section, the governing equations for fluid and particle phases are discussed briefly, along with the simulation parameters used in the present study. The third section presents the turbulence modulation of fluid fluctuations and particle properties due to the variation in volume fractions, system sizes, and Reynolds numbers. The conclusions are discussed in the last section.

\section{Governing equations}\label{sec:Governing_equations}
\subsection{Fluid Phase Equations}

The incompressible fluid phase is described in the Eulerian framework using Navier-Stokes equations. In the LES approach, the instantaneous mass and momentum equations are defined using the filtered quantities written as,
\begin{equation}
\frac{\partial \widetilde{u}_i}{\partial x_i}=0
\label{continuty}
\end{equation}
and,
\begin{multline}
\frac{\partial \widetilde{u}_i}{\partial t} + \frac{\partial \widetilde{u}_i\widetilde{u}_j}{\partial x_j} = - \frac{1}{\rho_f} \frac{\partial \widetilde{p}}{\partial x_i} + {\nu} \frac{\partial^2 \widetilde{u}_i}{\partial x_j \partial x_j} \\ +  \frac{\partial (\widetilde{u}_i\widetilde{u}_j - \widetilde{u_i u_j})}{\partial x_j}+ \frac{\widetilde{f_i}}{\rho_f}.
\label{LES eqn}
\end{multline}
Where $\widetilde{u}_i$ and  $\widetilde{p} $ are the filtered velocity and pressure, respectively, $\nu$ is the kinematic viscosity, and $ \rho_f$ is the fluid density. The ($\tilde{.}$) denotes the filtered quantity throughout the manuscript. $\widetilde{f_i}$ is the feedback force per unit volume caused by the dispersed phase. The feedback force includes the drag and the lift on the particles, and is written as,
\begin{equation}
\widetilde{f_i}= - \sum (F_{i,I}^d + F_{i,I}^l)
\end{equation}
Here, $F_{i,I}^d$ and $F_{i,I}^l$ are the drag and lift forces on the $I^{th}$ particle. The third term on the right-hand side of Eqn.~(\ref{LES eqn}) is referred to as the subgrid-scale (SGS) stress term, which requires closure. The dynamic one-equation model has been used as a fluid SGS model.  A correction on the lift force term is applied following \citet{mei1992approximate} and \citet{loth2009equation}.
The importance of the inclusion of lift force is discussed by \citet{costa2020interface}. In the present methodology, a finite volume based opensource software CFDEM \cite{kloss2012models, el2021theories} has been used. A second-order backward scheme for time derivative and a second-order Gauss linear differencing scheme for pressure gradient, viscous diffusion, and advection terms have been adopted.

 \subsubsection{Dynamic one-equation model}
In the dynamic one-eqaution model, the subgrid scale stress ($\tau_{ij}$) is expressed in terms of the eddy viscosity ($\nu_t$) and filtered strain rate ($\widetilde{S}_{ij}$) as $\tau_{ij} = -2\nu_t \widetilde{S}_{ij}$. The eddy viscosity is calculated as,
\begin{equation}
\nu_t = C \widetilde{\Delta} \sqrt{k_{SGS}}.
\label{eddy_vis}
\end{equation}
Here, $  k_{SGS} = \frac{1}{2}[\widetilde{u_i u_i} - \widetilde{u}_i \widetilde{u}_i]$ is the subgrid scale kinetic energy, $\widetilde{\Delta}$ is the cube root of grid volume, and $C$ is a parameter calculated dynamically during the simulation. The $k_{SGS}$ is calculated using the following equation (Eqn.~\ref{K_sgs}).
\begin{equation}
\frac{\partial k_{SGS}}{\partial t} +  \frac{ \partial (k_{SGS} \widetilde{u}_i)}{\partial x_j} = \frac{\partial }{\partial x_j} \left[(\nu + \nu_t) \frac{\partial k_{SGS}}{\partial x_j}\right] - \epsilon - \tau_{ij} \frac{\partial \widetilde{u}_i }{\partial x_j}.
\label{K_sgs}
\end{equation}
Where $\epsilon = C_\epsilon k_{SGS}^{3/2}/\widetilde{\Delta}$, and $C_\epsilon = 1.048$ is the model coefficient.
In this model, the model constant appearing in eddy viscosity ($C$ in Eqn.~\ref{eddy_vis}) is calculated dynamically \cite{pomraning2002dynamic} which is explained as, 
\begin{equation}
\tau_{ij} - \frac{2}{3} \delta_{ij}  k_{SGS} = -2 C_1 \widetilde{\Delta} k_{SGS}^{1/2} \widetilde{S}_{ij}.
\label{tau_is}
\end{equation}
Using the second or test filter, the subgrid stress at the test filter level can be written as,
\begin{equation}
T_{ij} - \frac{2}{3} \delta_{ij}  K_{SGS} = -2 C_2 \widetilde{\widetilde{\Delta}} K_{SGS}^{1/2} \widetilde{\widetilde{S}}_{ij}.
\end{equation}
Where $ K_{SGS} = \frac{1}{2}[\widetilde{\widetilde{u_i u_i}} - \widetilde{\widetilde{u}}_i \widetilde{\widetilde{u}}_i]$. The calculation of $ C $ involves Germano's identity \cite{germano1991dynamic}, 
\begin{equation}
 L_{ij} =  T_{ij} - \widetilde{\tau_{ij}} = \widetilde{\widetilde{u_i } \widetilde{u_j }} - \widetilde{\widetilde{u_i}} \widetilde{\widetilde{u_i}},
 \label{Lij_2}
\end{equation}
and,
\begin{equation}
K_{SGS} = \widetilde{k_{SGS}} + \frac{1}{2} L_{ii}.
\end{equation}
The formulation involves two assumptions. First, the subgrid stress scales with filter size, then $C_1$ and $C_2$ are replaced by $C$. And second, $C$ is taken outside the integral in Eqn.~\ref{Lij} as the variation of $C$ is low in space.  Using Eqns.~\ref{tau_is} - \ref{Lij_2}, the deviatoric part of the $L_{ij}$ is written as,
\begin{equation}
L_{ij}^d  = L_{ij} - \frac{1}{3} L_{kk} \delta_{ij}
 = \alpha_{ij} C - \widetilde{\beta_{ij }C},
 \label{Lij}
 \end{equation}
 and,
 \begin{eqnarray}
M_{ij} =  \alpha_{ij} - \widetilde{\beta_{ij}}.
\label{Mij}
\end{eqnarray}
Where $ \alpha_{ij} =  -2 \widetilde{\widetilde{\Delta}} K_{SGS}^{1/2} \widetilde{\widetilde{S_{ij}}}$ and $  \beta_{ij} = -2 \widetilde{\Delta} k_{SGS}^{1/2} \widetilde{S_{ij}}$. Then, 
\begin{equation}
C = \frac{1}{2}\frac{\langle M_{ij}  L_{ij} \rangle }{\langle M_{kl} M_{kl} \rangle }.
\label{C_dyna}
\end{equation}
Here, $C_s ( = \sqrt{C})$ is the Smagorinsky coefficient. The angular brackets in Eqn.~\ref{C_dyna} denote the plane averaging. The present work uses a box filter as a test filter. This model does not require the Van-driest damping as in the Smagorinsky model to satisfy the wall scaling.

\subsection{Particle phase description}
The solid individual particle is tracked in the Lagrangian framework. The particle-particle and particle-wall collisions are considered to be elastic in nature. The particle motion of the dispersed phase is described using Newton's second law of motion, which is defined as,
\begin{equation}
m_p \frac{dv_{i,I}}{dt} = F_{i,I}^D + F_{i,I}^L + \sum_{I\neq J}^{} F_{i,IJ} + F_{i,Iw} + m_pg.
\label{pEqn}
\end{equation}
Where $m_p$ is the mass of the particle, $v_{i,I}$  is the velocity of $I^{th}$ particle, $F_{i,I}^D$ and $ F_{i,I}^L$ are the drag and lift force acting on the particle, respectively. The g is the gravitational acceleration, $F_{i,IJ}$ is the interaction force between $ I^{th} $ and $ J^{th} $ particle, and $F_{i,Iw}$ is the interaction force between $I^{th }$ particle and the wall. The drag force is calculated using the Schiller-Naumann correlation \cite{naumann1935drag} given by Eqn.~\ref{drag_law}. 
\begin{equation}
F_{i,I}^D = 3 \pi \mu d_p (\widetilde{u}_{i,I}(x,t) - v_{i,I}) (1+ 0.15Re_p^{0.687})
\label{drag_law}
\end{equation} 
A third-order accurate scheme has been used to interpolate the fluid velocity at the particle location to calculate the drag and the lift forces. The particle size is smaller than the grid size. A point particle approximation has been used for the present study \cite{boivin2000prediction, balachandar2009scaling, balachandar2010turbulent, zhao2010turbulence, dritselis2011large, Mallouppas2013, marchioli2017large}.

The soft-sphere (spring dashpot) \cite{cundall1979discrete, crowe1998multiphase} interaction model has been used to capture the particle-particle and particle-wall collisions. The particle collisions are important to capture even at a low solid volume fraction($\phi \sim 10^{-4}$)~\cite{YAMAMOTO2001} . The collisions lead to momentum transfer to all the directions and affect the particle phase statistics. During the collision, the repulsive force on the particle depends on the extent of overlapping of the colliding particles. This is modeled using a spring-dashpot system where the dashpot includes the energy loss associated with the collision. The motion of particles is divided into normal and tangential components, along with overlapping distances and forces. Considering two colliding particles as $i$ and $j$, the normal ($\mathbf{F}_{nij}$) and tangential ($\mathbf{F}_{tij}$) components of the forces are modeled as,
 \begin{equation}
 	\mathbf{F}_{nij} = (-k_n\delta^{3/2}_{n} - \eta_{nj}\mathbf{G.n})\mathbf{n},
 \end{equation}
  \begin{equation}
  	\mathbf{F}_{tij} = -k_t \delta_{t} - \eta_{tj} \mathbf{G}_{ct}.
  \end{equation}   
Here, the subscripts `n' and `t' refer to the normal and tangential directions. $\delta$ is the overlap distance, $k$, and $\eta$ are the stiffness and damping coefficient, $\mathbf{G}$ is the relative velocity vector of $i^{th}$ particle relative to $j^{th}$ particle ($\mathbf{G = v_i - v_j}$), and $\mathbf{n}$ is the unit vector in the direction of the line connecting the centers of both the particles. $\mathbf{G}_{ct}$ is the slip velocity at the contact point, and defined as $\mathbf{G}_{ct} = \mathbf{G- (G.n)n}$.

\subsection{Simulation parameters}\label{sec:Simulation_parameters}
Simulations are performed at two bulk Reynolds numbers ($Re_b = \bar{u}\times 2\delta/\nu$) and three different channel dimensions ($2\delta/d_p$). Where $\delta$ is the half-channel width, $\bar{u}$ is the average fluid velocity, and $d_p$ is the particle diameter. The bulk Reynolds numbers are 5600 and 13750, and three different channel dimensions used are 54, 81, and 117. Lengths of the vertical channel considered are  $8\pi\delta$, $2\delta$, and  $(4/3)\pi\delta$ in streamwise (x), wall-normal (y), and spanwise (z) directions, respectively. The finite volume method is applied to solve the filtered Navier-Stokes Eqns.~\ref{continuty} and \ref{LES eqn}. The fluid phase is resolved with $ 128*65*64 $ and $ 192*99*96 $ grid points ($x*y*z$) for bulk Reynolds numbers of 5600 and 13750, respectively. The grid resolution is $\triangle x^+ = 35$, $\triangle z^+ = 12$ for $Re_b = 5600$, and $\triangle x^+ = 51$,  $\triangle z^+ = 17$  for $Re_b = 13750$ for the LES models. Earlier studies\cite{Wang1996, Salmanzadeh2010} have also used similar grid resolution . Here, ($^+$) symbol indicates the quantities normalized with viscous scales. In the wall-normal direction, the first grid point is such that $y^+$  is less than one. No-slip boundary condition is applied at the walls. The time step is taken such that the CFL number is nearly 0.2 for all the constant bulk flow rate simulations. The Dynamic one-equation (Eqns.~\ref{eddy_vis} - \ref{C_dyna}) LES model is used to close the SGS stress term in the filtered Navier-stokes equation.

\begin{table}[]
\centering
\begin{tabular}{|c|c|c|c|c|c|}
\hline
Case & $Re_b$            & $2\delta/d_p$ & $St$ & $St_v$    \\ \hline
$A_1$& 5600              & 54            & 179.53 & 4108.8    \\ \hline
$B_1$ & 5600& 81            & 74.45  & 1703.21   \\ \cline{1-5} 
$C_1$ &       5600       & 117           & 38.71  & 885.27    \\ \hline
$B_2$ & 13750              & 81            & 73.12 & 3304.4    \\ \hline
           $C_2$ &   13750 & 117            & 38.02  & 1720.8   \\ \cline{2-5}  \hline
\end{tabular}
\caption{$Re_b$ is the fluid bulk Reynolds number, $\delta$ is half channel width, $d_p$ is the particle diameter, $St$ is the Stokes number based on the integral time scale of fluid ($2\delta/\bar{u}$), and $St_v$ is the Stokes number based on the viscous time scale. }
\label{Stokes_number}
\end{table}

\begin{figure*}
\centering	
	\begin{subfigure}[b]{1\textwidth}
	\minipage{0.45\textwidth}
	\includegraphics[width=\textwidth]{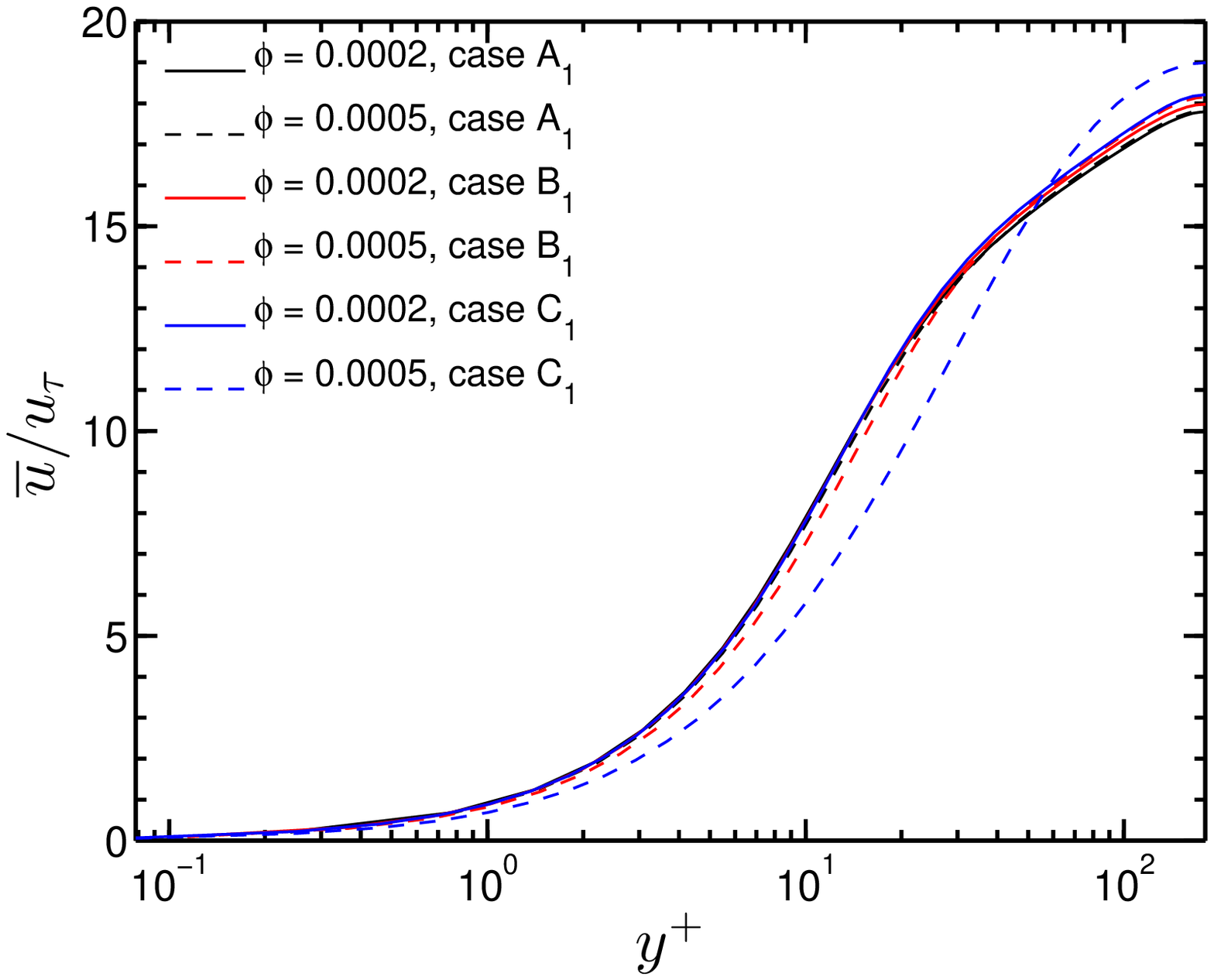}
	\caption{}
	\endminipage 
	\minipage{0.45\textwidth}
	\includegraphics[width=\textwidth]{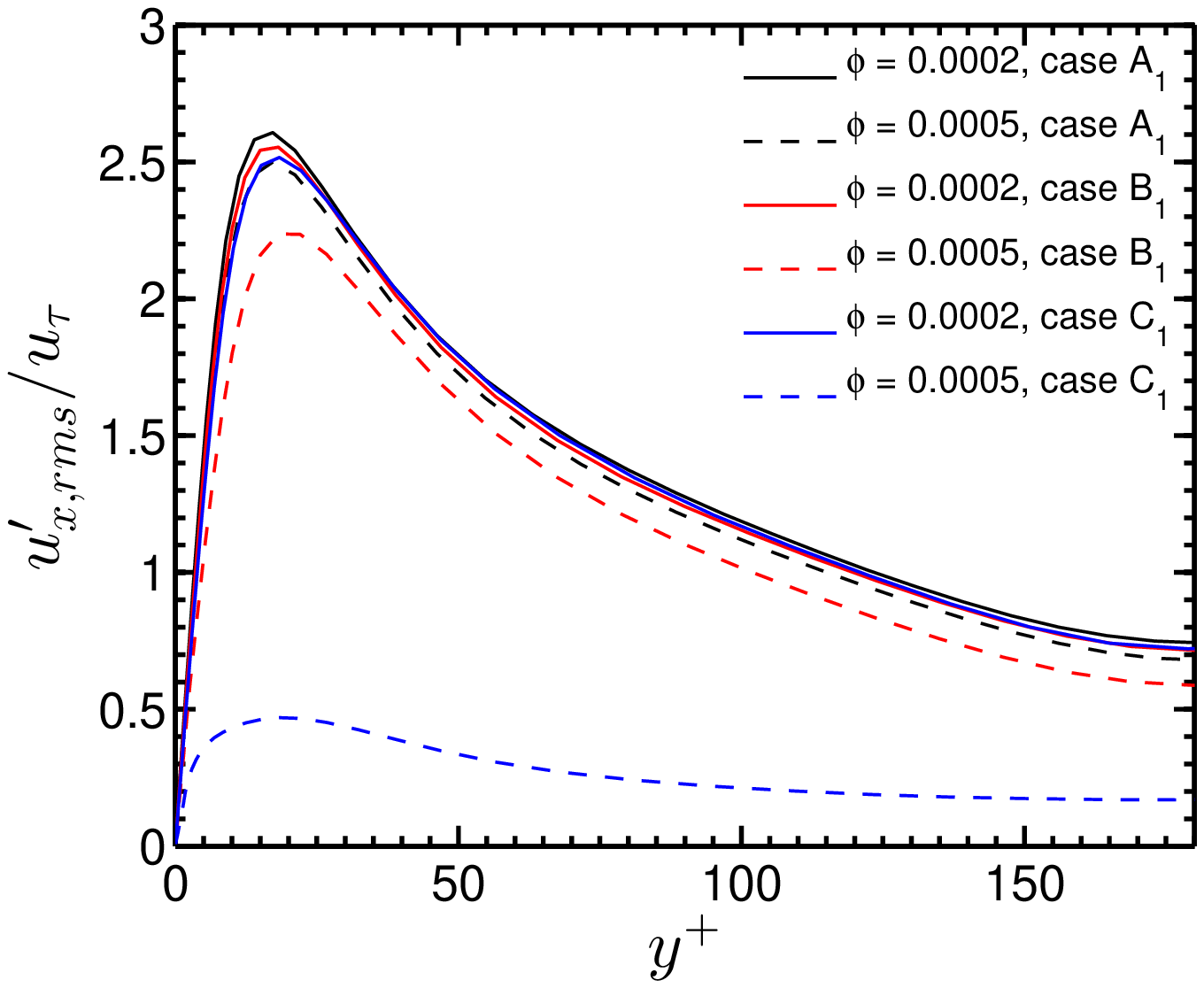}
	\caption{}
	\endminipage \\
	\minipage{0.45\textwidth}
	\includegraphics[width=\textwidth]{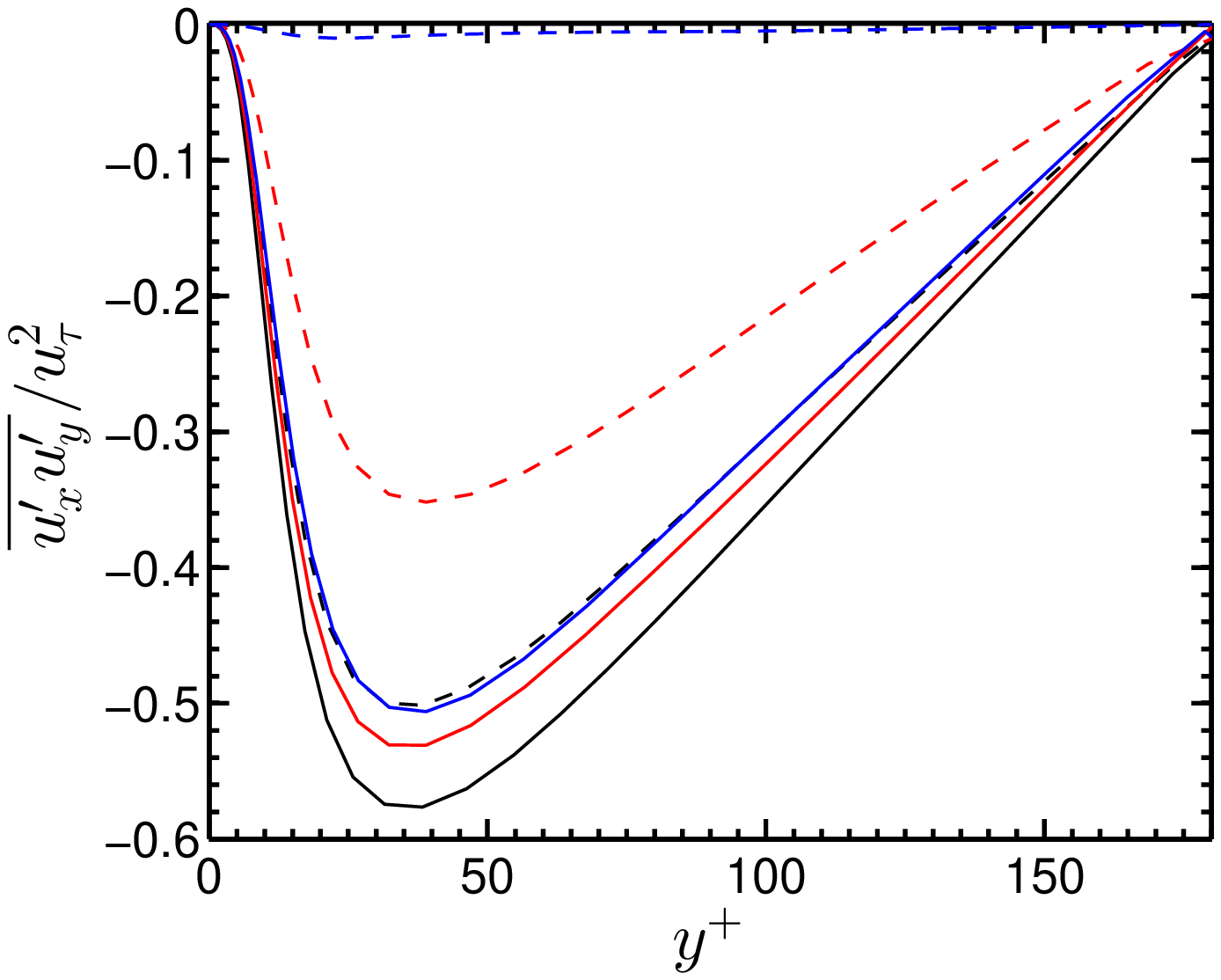}
	\caption{}
	\endminipage	
	\minipage{0.45\textwidth}
	\includegraphics[width=\textwidth]{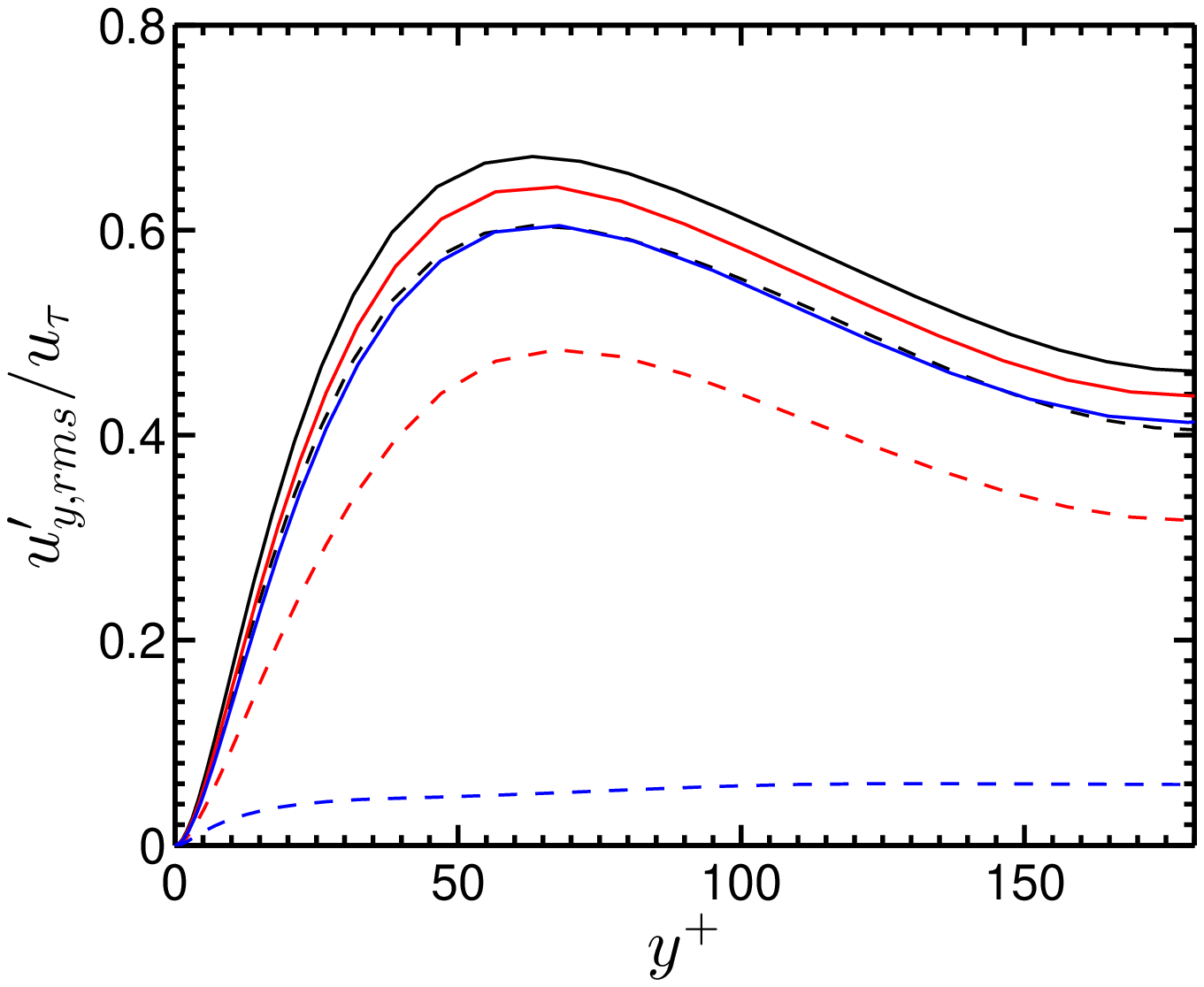}
	\caption{}
	\endminipage 
	\end{subfigure} 
	\caption{The fluid velocity profiles along the wall-normal direction for different volume fractions at $Re_b = 5600$. The fluid velocities and wall-normal distance are normalized with viscous units. Fig.~(a) Streamwise mean velocity, (b) Streamwise fluctuations, (c) Cross-stream stress, and (d) Wall-normal fluid fluctuations. The legends for the Figs.~(c and d) are same as in Fig.~(a).}
	\label{profile_Ldp}
\end{figure*}

The particle diameter is considered to be $39 \mu m$. The particle densities are considered as $2000 kg/m^3$ and $800 kg/m^3$. As the particle to fluid density ratio is higher, the Basset history and buoyancy effects are neglected in the particle's equation of motion. Simulation parameters like Reynolds numbers, channel dimensions, and the particle Stokes numbers are provided in Table~\ref{Stokes_number}. The Stokes number based on the fluid integral length scale and viscous scales are denoted by $St = \tau_p/\tau_f$ and $St_v = \tau_p/\tau_{fv}$. Here, the particle relaxation time ($\tau_p$) is defined as, $\tau_p = \rho_p d_p^2/18\mu_f$, the fluid integral time scale is defined as, $\tau_f = 2\delta/\overline{u}$, and viscous time scale is defined as $\tau_{fv} = \nu/u_\tau^2$ where $u_\tau$ is the friction velocity.  Here, unladen friction velocity is taken to calculate $St_v$ in Table~\ref{Stokes_number}. Various authors \citep{YAMAMOTO2001, Kuerten2005, Kuerten2006, marchioli2008some, dritselis2011large, zhou2020non} have used time scale based on the friction velocity to define the Stokes number. However, it is to be noted that the turbulence modulation  with the increase in particle mass loading leads to a modification in friction velocity, shown in Fig.~\ref{profile_Ldp} (a). For that reason, we have used $2\delta/\bar{u}$ as the fluid time scale.

\section{Results }
\label{section:results}

\subsection{Effect of system size}

In this section, the simulations are performed to analyze the dynamics of fluid and solid phases with different channel widths for a range of solid volume fractions at $Re_b = 5600$. The channel dimensions are considered such that the ratios of channel width ($2\delta$) and particle diameter ($d_p$) are 54, 81, and 117. Here, channel width ($2\delta$) varies for different channel configurations. However, the computational domain is kept as $8\pi\delta \times 2\delta \times (4/3)\pi\delta$ for all the cases. The comparison of current unladen LES with available literature \cite{Moser1999, Salmanzadeh2010, li2012experimental, muramulla2020disruption} is shown in the appendix for $Re_b = 5600$ and 13750. In Fig.~\ref{profile_Ldp} (a), the fluid streamwise mean velocity is plotted for all three channels and two volume fractions, $\phi = 2\times10^{-4}$ and $5\times10^{-4}$. It is observed that there is no variation in the mean profiles for the case $A_1$ when the volume fraction is increased from $\phi = 2\times10^{-4}$ to $5\times10^{-4}$. For case $B_1$, there is a marginal decrease of about 5\% in mean velocity in the buffer region as the volume fraction is increased from $\phi = 2\times10^{-4}$ to $5\times10^{-4}$. However, the reduction in the mean velocity is more significant for the case $C_1$ at $\phi = 5\times10^{-4}$. The normalized fluid fluctuations as a function of the wall-normal distance are plotted in Fig.~\ref{profile_Ldp} (b-d). In the case of streamwise fluid fluctuations, there is only 4\% and 9\% decrease in the peak value as the volume fraction is increased from $\phi = 2\times10^{-4}$ to $5\times10^{-4}$ for the case $A_1$ and $B_1$, respectively. However, the decrease is almost one order of magnitude for the case $C_1$ for the same increase in volume fraction (Fig.~\ref{profile_Ldp} (b)). In the case of cross-stream fluid velocity fluctuations, there is 13\% and 34\% decrease as the volume fraction is changed from $\phi = 2\times10^{-4}$ to $5\times10^{-4}$ for the case $A_1$ and $B_1$, respectively. However, the decrease is of almost one order of magnitude for similar volume fraction change in case of the larger channel of $2\delta/d_p = 117$ (case $C_1$), Fig.~\ref{profile_Ldp} (c). A similar behavior is observed for wall-normal fluid fluctuations, shown in Fig.~\ref{profile_Ldp} (d).

\begin{figure*}[htb]
	\begin{subfigure}[b]{1\textwidth}
	\minipage{0.45\textwidth}
	\includegraphics[width=\textwidth]{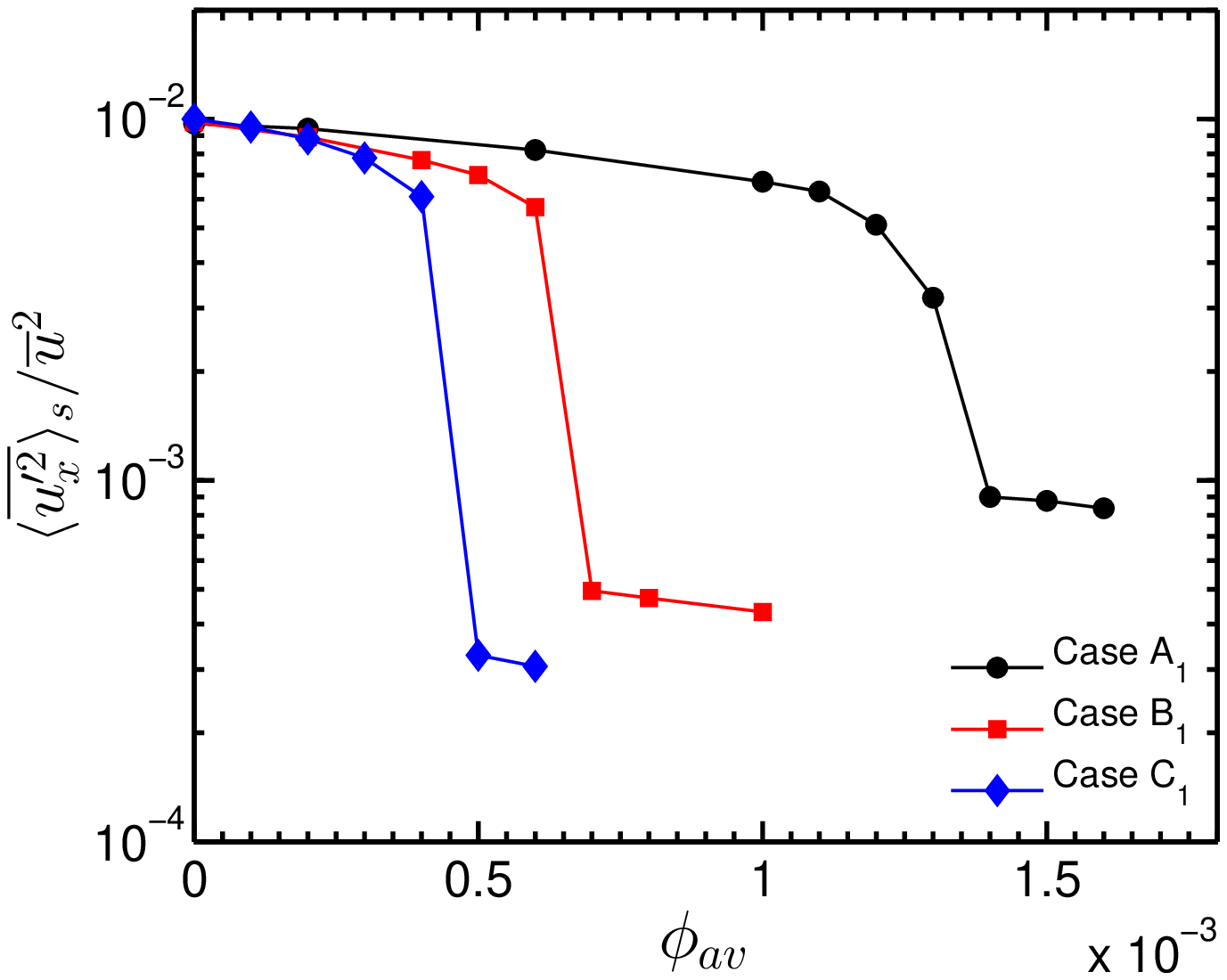}
	\caption{}
	\endminipage 
	\minipage{0.45\textwidth}
	\includegraphics[width=\textwidth]{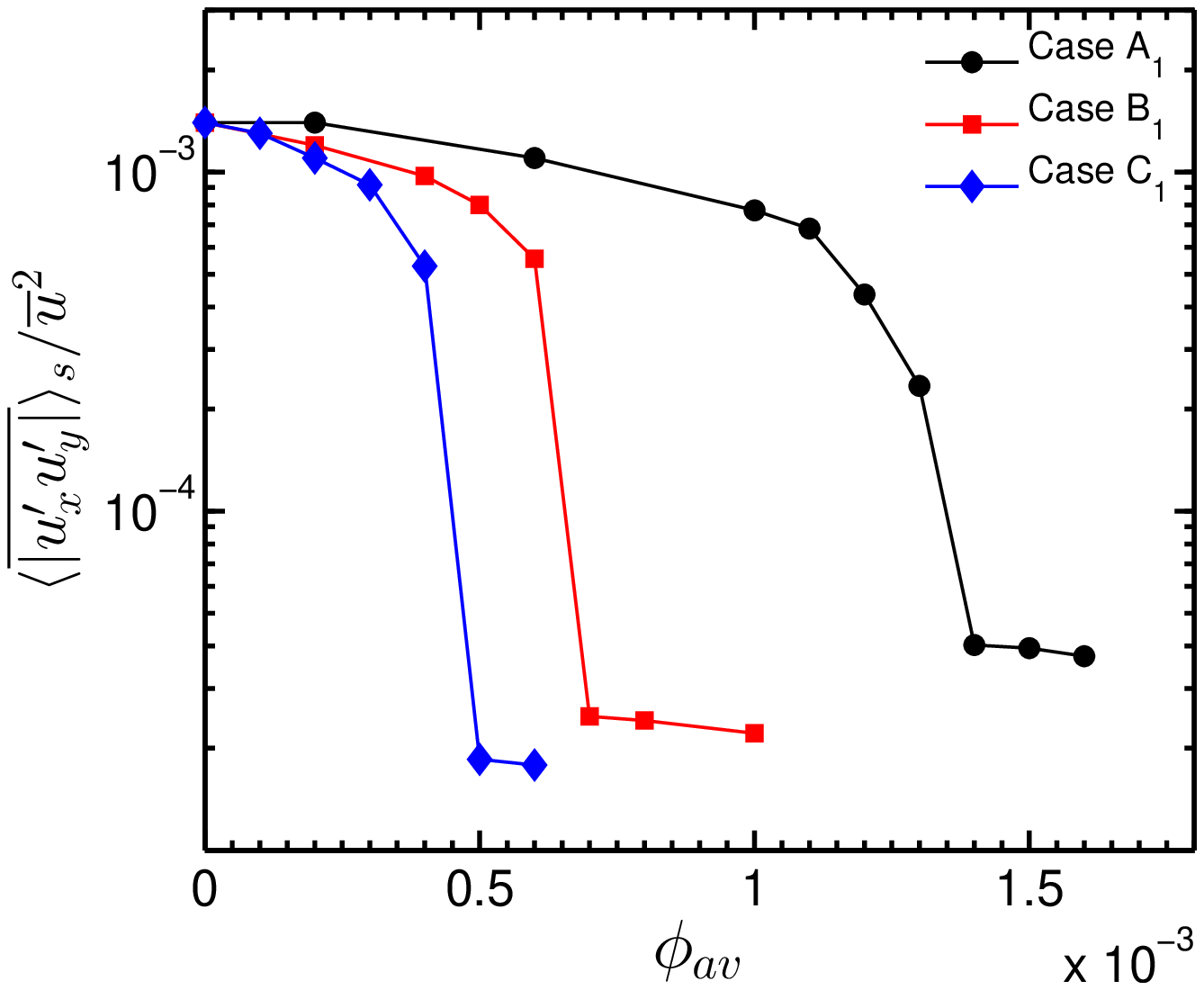}
	\caption{}
	\endminipage \\
	\minipage{0.45\textwidth}
	\includegraphics[width=\textwidth]{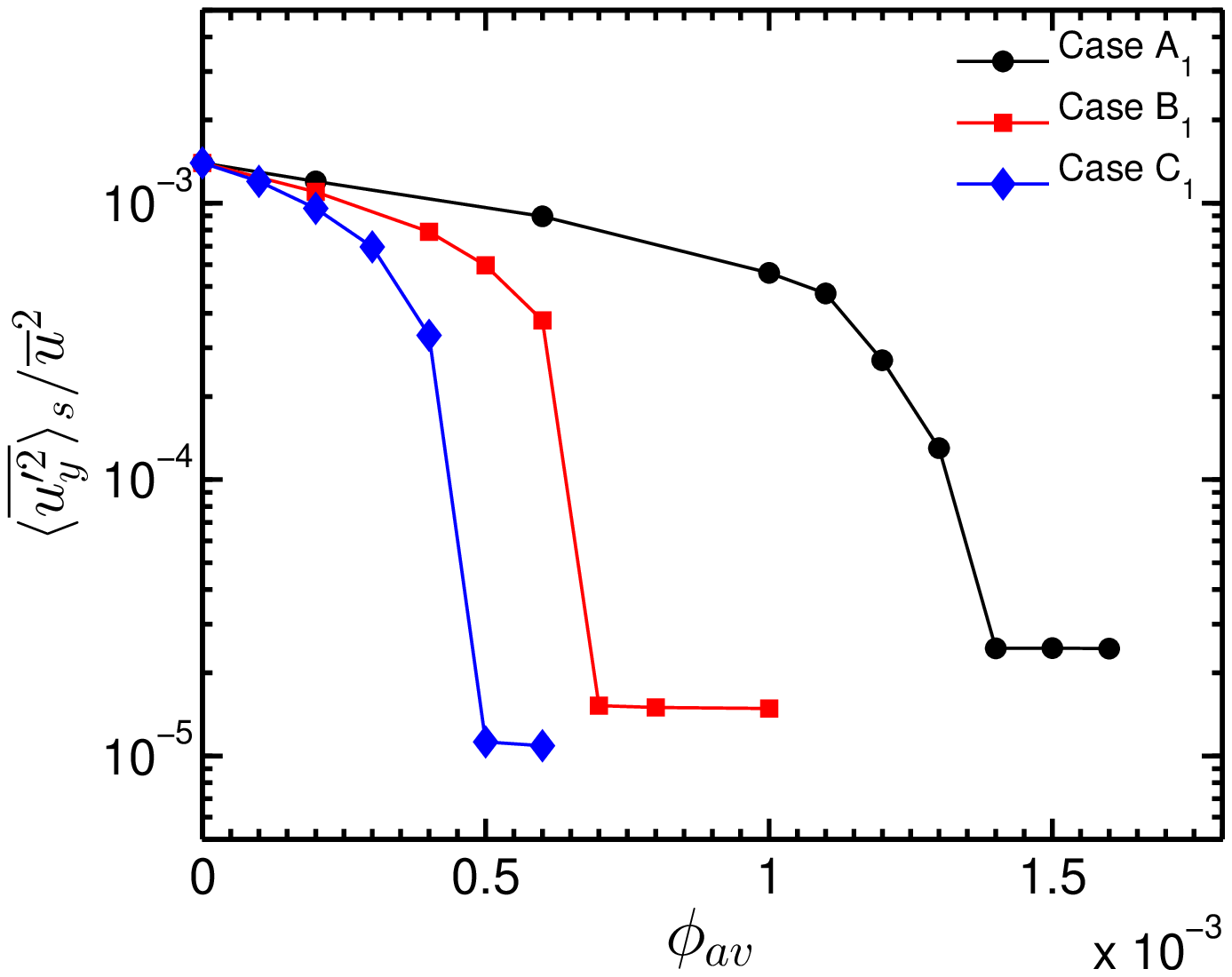}
	\caption{}
	\endminipage	
	\minipage{0.45\textwidth}
	\includegraphics[width=\textwidth]{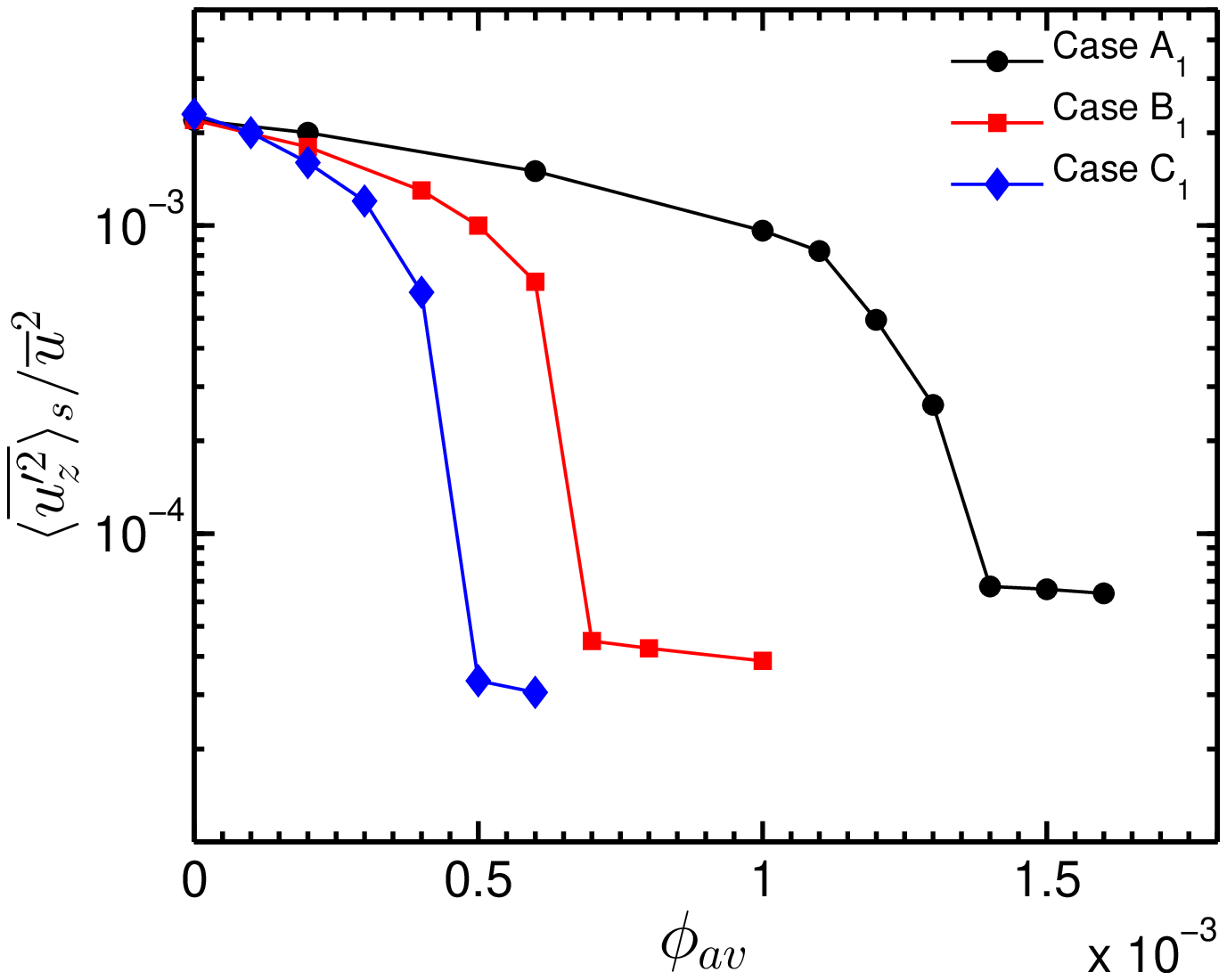}
	\caption{}
	\endminipage 
	\end{subfigure} 
	\caption{The average fluid fluctuations across the channel width for a volume fraction range at $Re_b = 5600$. The fluid fluctuations are normalized with the square of bulk fluid velocity. Fig.~(a) Streamwise, (b) Cross-stream, (c) Wall-normal, and (d) Spanwise fluid fluctuations.}
	\label{u_mod_Re180}
\end{figure*}

To quantify the variation of turbulence intensity, we compute channel averaged fluid fluctuations as defined by 
\begin{equation}
\langle \star \rangle_s = \frac{1}{\delta} \int_{0}^{\delta} dy \langle \star \rangle.
\label{avg_fluid_fluc}
\end{equation}
In Fig.~\ref{u_mod_Re180} (a-d), the average fluid fluctuations across the channel width are plotted for a range of volume fractions. The average fluctuations predicted by different channels match at lower volume fractions. The fluid fluctuations decrease with an increase in volume fraction, and a complete turbulence suppression is observed at a higher particle volume loading for all three cases. The particle loading at which complete turbulence collapse occurs is called critical particle volume loading (CPVL). The attenuation observed in the fluid fluctuations is higher for the case $C_1$. The CPVL is $14\times 10^{-4}$, $7\times 10^{-4}$, and $5\times 10^{-4}$ for the cases $A_1$, $B_1$ and $C_1$, respectively. It is interesting to note that the fluid fluctuations decrease with an increase in system size for the same volume fraction and fixed Reynolds number, although, Stokes number is decreased (Table~\ref{Stokes_number}). Such an observation is in contrary to the earlier findings\cite{li2001numerical, dritselis2011large, lee2015modification, muramulla2020disruption, kumaran2020turbulence} that the fluid fluctuations decrease with an increase in Stokes number for $St_K > 1$. Here, $St_K$ is the Stokes number based on the Kolmogorov time scale. It is worth noting that, in the earlier studies Stokes number was changed by changing particle relaxation time for a fixed fluid time scale. However, in the present study, particle relaxation time is fixed and the fluid time scale changes with the system size. The results show that there is a reduction in the fluid fluctuations with the decrease in considered Stokes numbers. This raises the question whether the Stokes number is a correct parameter to represent the effect of particle inertia on the turbulence modulation for a range of system size and in the limit  $St_K > 1$. Addressing such a question is important if one is interested to quantify turbulence modulation in an unified manner as a function of fluid and particle inertia. Another interesting implication is that the understanding gained will help to compare the simulation results with the experiments.


\subsection{Effect of Reynolds number of fluid phase}

This section examines the turbulence modulation for two channel dimensions of $2\delta/d_p = 81$ and 117. The unladen fluid statistics are validated against the DNS data of \citet{Moser1999}, experimental data of \citet{li2012experimental}, and LES data of \citet{Salmanzadeh2010} in the appendix, shown in Fig.~\ref{unladen_Re395}. In Fig.~\ref{laden_Re395}, the fluid statistics for two volume fractions of $\phi = 5\times 10^{-4}$ and $10^{-3}$ are presented for cases $B_1$ and $B_2$. In Fig.~\ref{laden_Re395} (a), it is observed that there is no change in the streamwise mean velocity profiles as the volume fraction is increased from $\phi = 5\times 10^{-4}$ to $10^{-3}$ for the case $B_2$. However, in case of $B_1$, there is a decrease of fluid streamwise mean velocity in the buffer region and an increase in the mean velocity near the channel center for a similar change in volume fraction (shown in Fig.~\ref{laden_Re395} a). Fig.~\ref{laden_Re395} shows that there is about a 5\% decrease in the peak value of streamwise fluid fluctuations as the volume fraction is increased from $\phi = 5\times 10^{-4}$ to $10^{-3}$ for the case $B_2$. However, a decrease of almost one order of magnitude in the peak value is observed for the case $B_1$ for the same variation of volume fraction. There is a decrease of nearly 14\% and 11\% in the peak value of the Reynolds stress and wall-normal fluctuations, respectively, as the volume fraction is increased from $\phi = 5\times 10^{-4}$ to $10^{-3}$ for the $Re_b = 13750$. However, in the case of $Re_b = 5600$, there is a decrease of almost one order of magnitude in the peak value of the Reynolds stress and the wall-normal fluctuations.

\begin{figure*}[htb]
	\begin{subfigure}[b]{1\textwidth}
	\minipage{0.45\textwidth}
	\includegraphics[width=\textwidth]{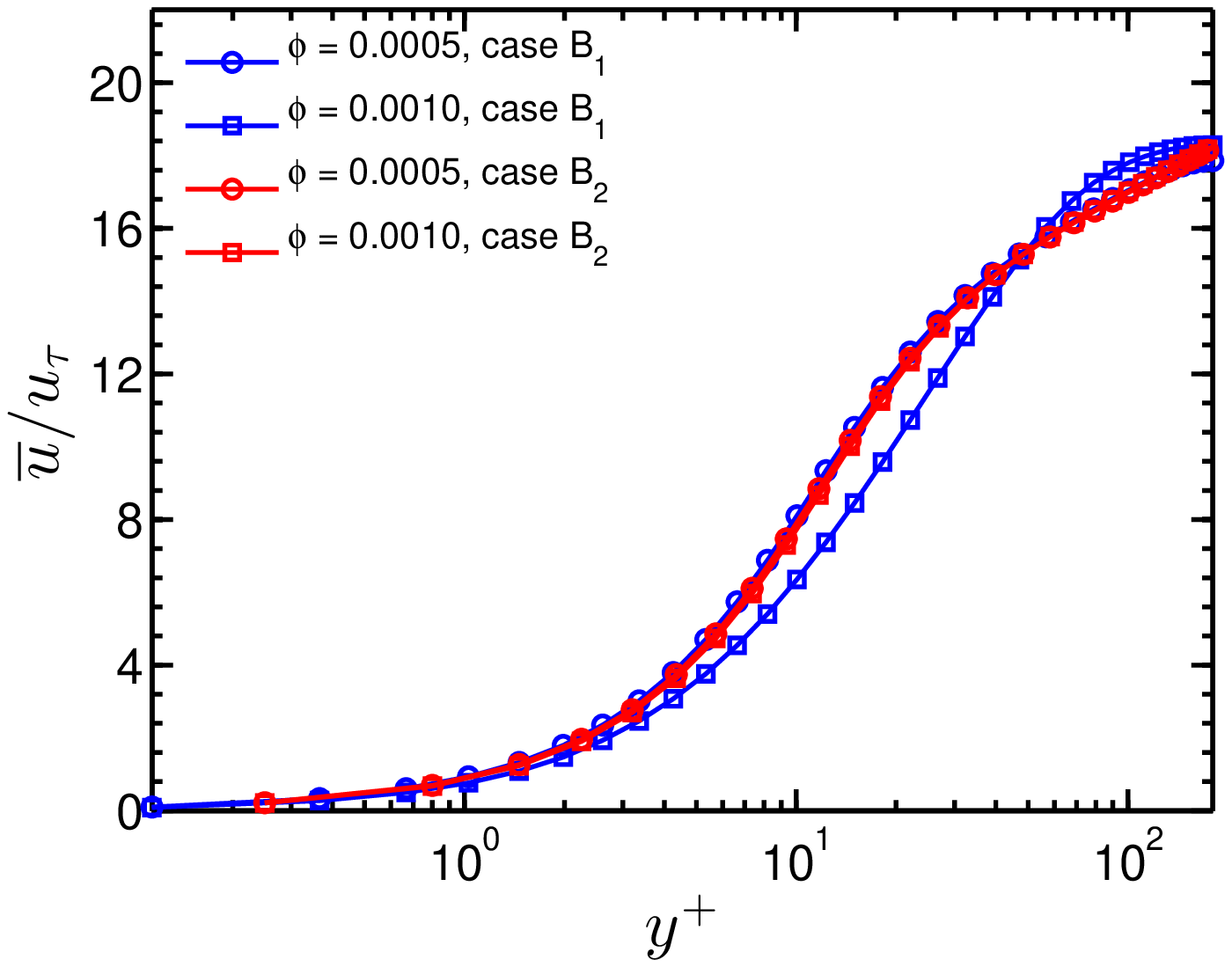}
	\caption{}
	\endminipage 
	\minipage{0.45\textwidth}
	\includegraphics[width=\textwidth]{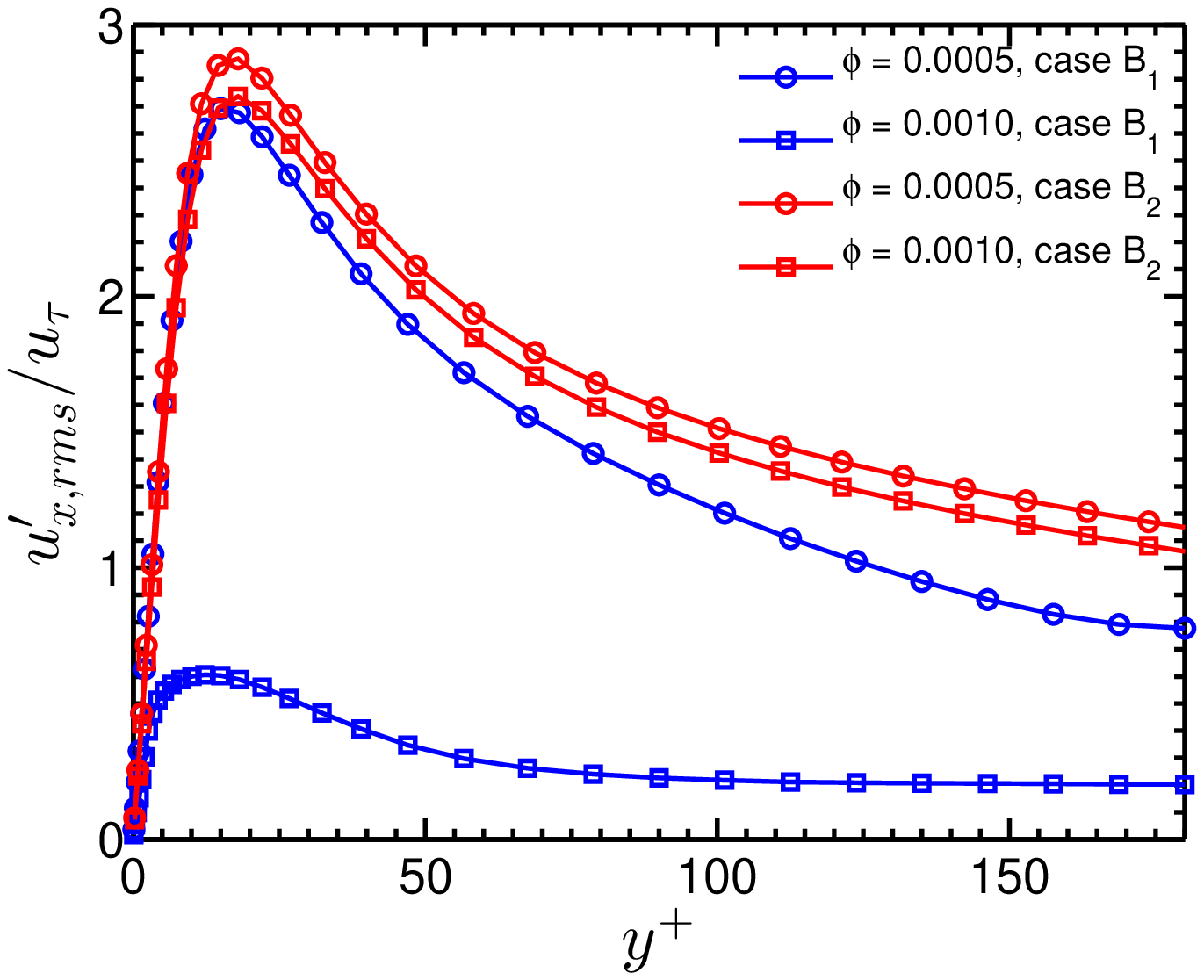}
	\caption{}
	\endminipage \\
	\minipage{0.45\textwidth}
	\includegraphics[width=\textwidth]{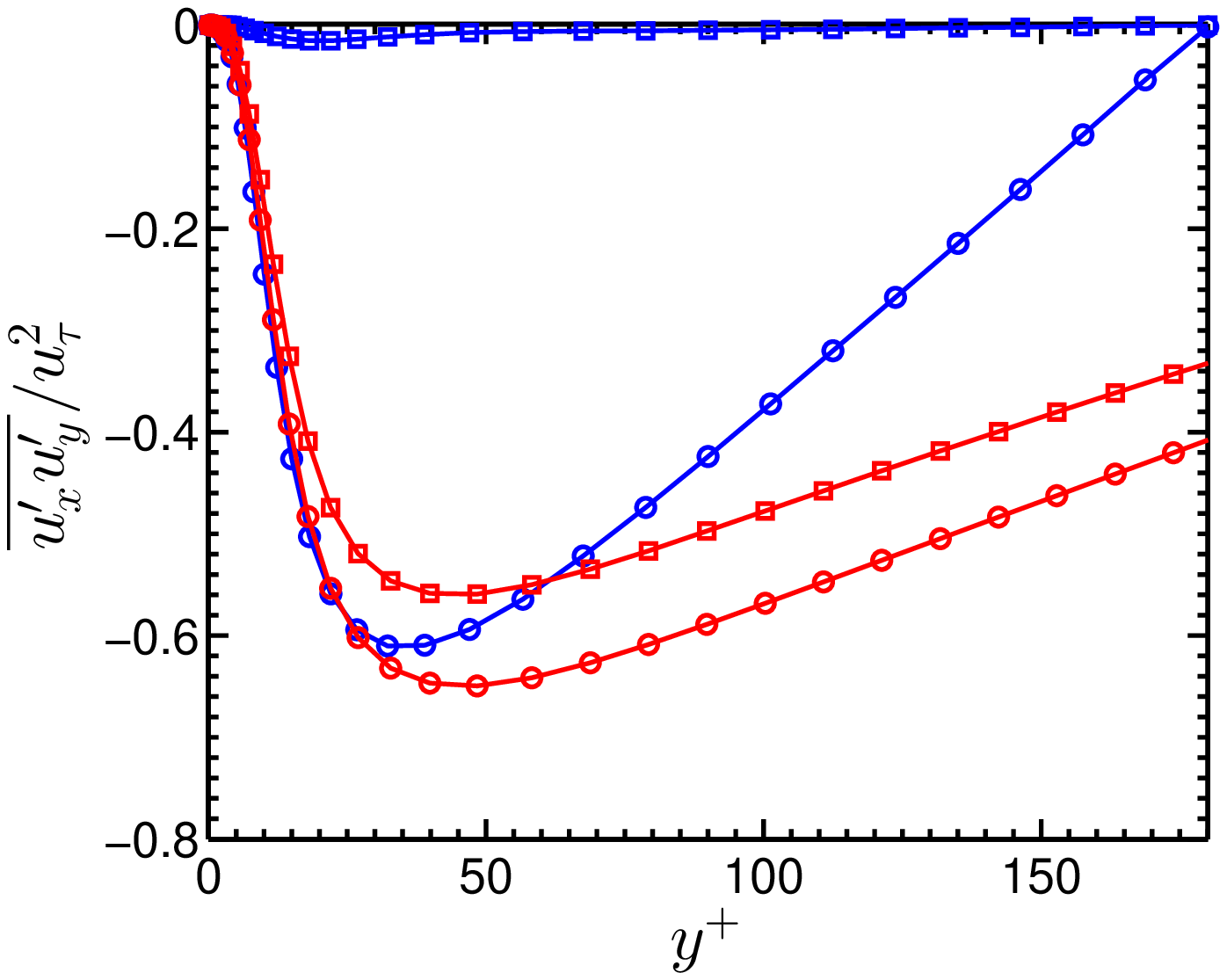}
	\caption{}
	\endminipage	
	\minipage{0.45\textwidth}
	\includegraphics[width=\textwidth]{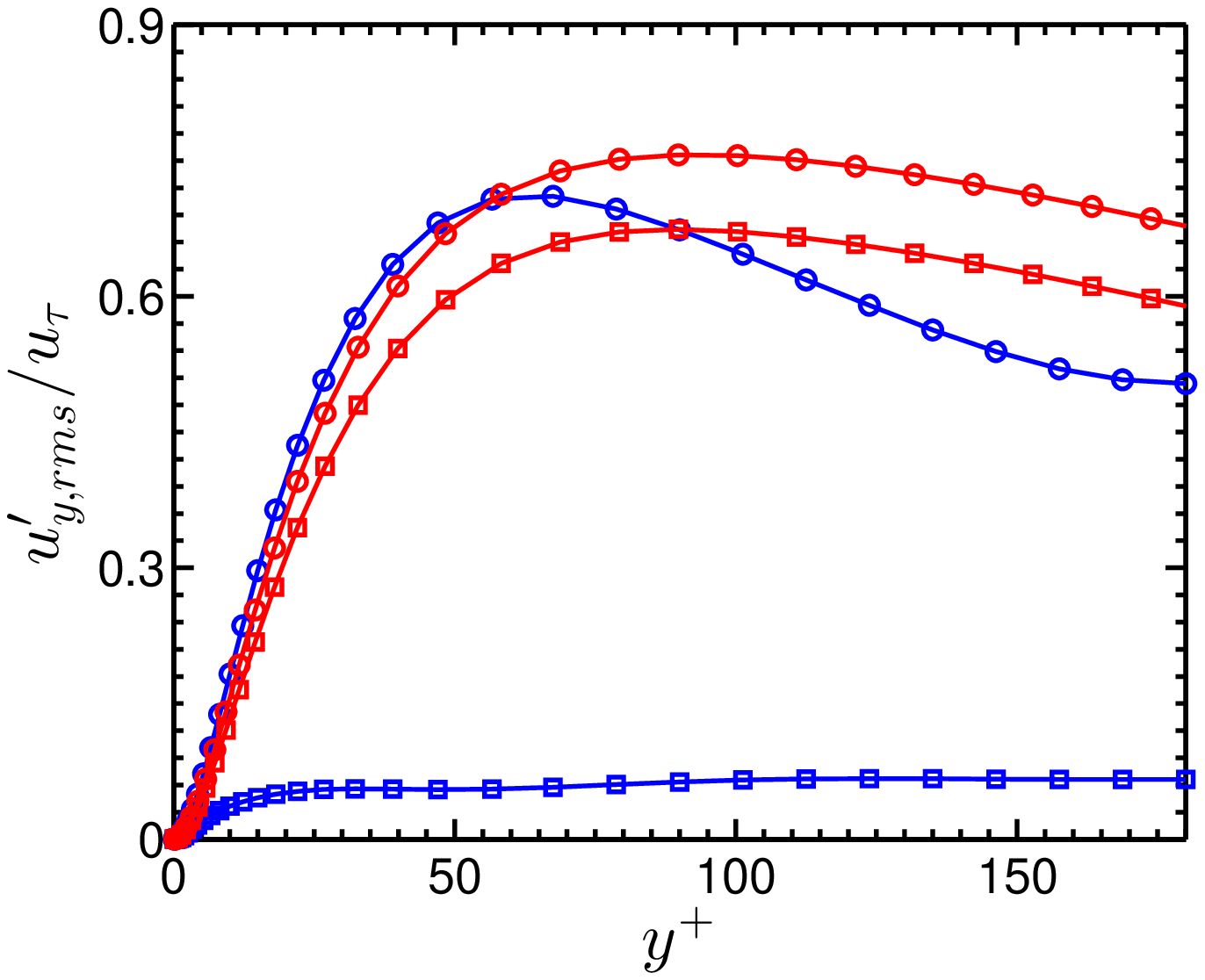}
	\caption{}
	\endminipage 
	\end{subfigure} 
	\caption{The fluid velocity profiles are compared for cases $B_1$ and $B_2$ for different volume fractions. The fluid velocities and wall-normal distance are normalized with viscous units. Fig.~(a) Streamwise mean velocity, (b) Streamwise fluctuations, (c) Cross-stream stress, and (d) Wall-normal fluid fluctuations. The legends for the Figs.~(c and d) are the same as in Fig.~(a).}
	\label{laden_Re395}
\end{figure*}

\begin{figure*}[htb]
	\begin{subfigure}[b]{1\textwidth}
	\minipage{0.45\textwidth}
	\includegraphics[width=\textwidth]{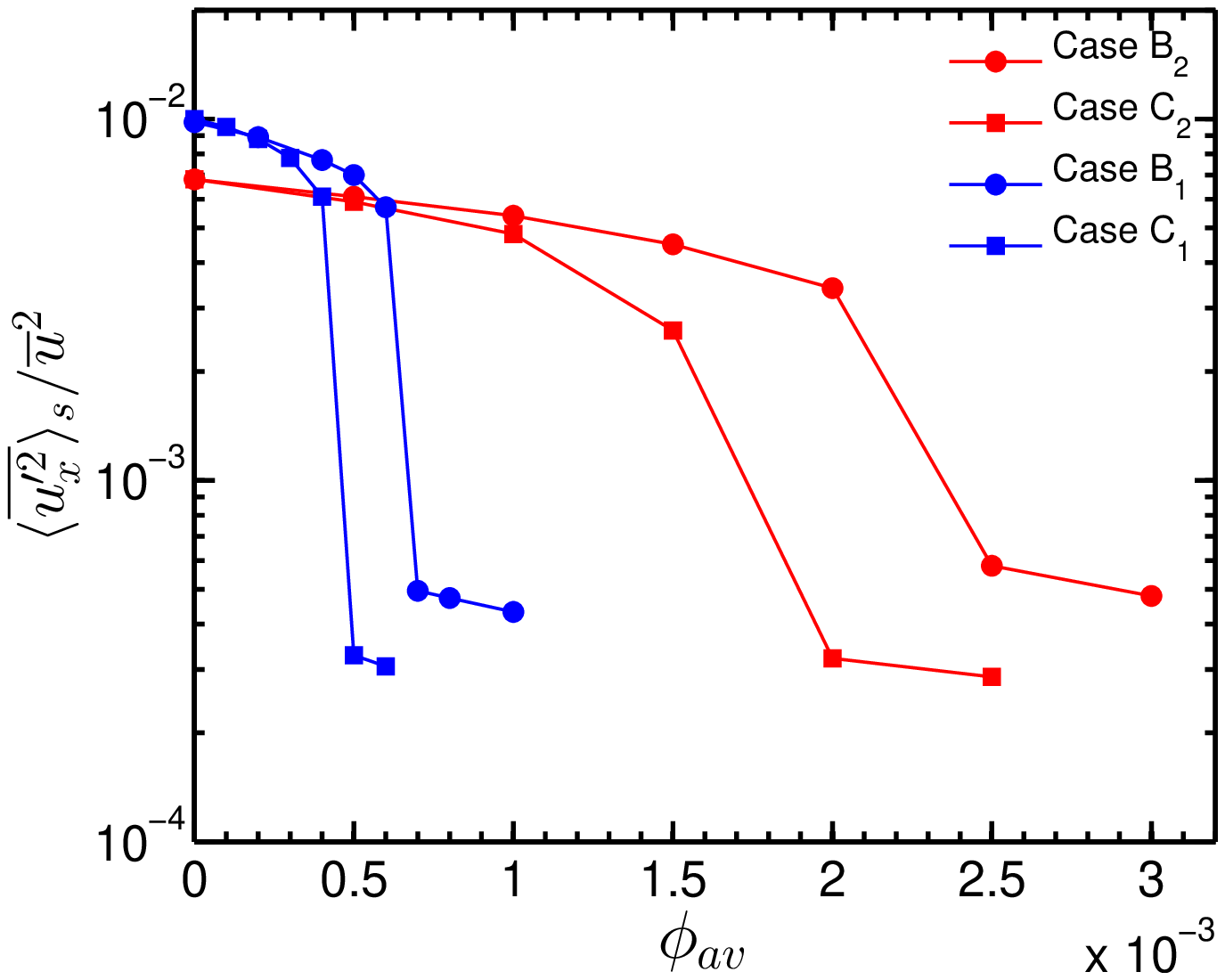}
	\caption{}
	\endminipage 
	\minipage{0.45\textwidth}
	\includegraphics[width=\textwidth]{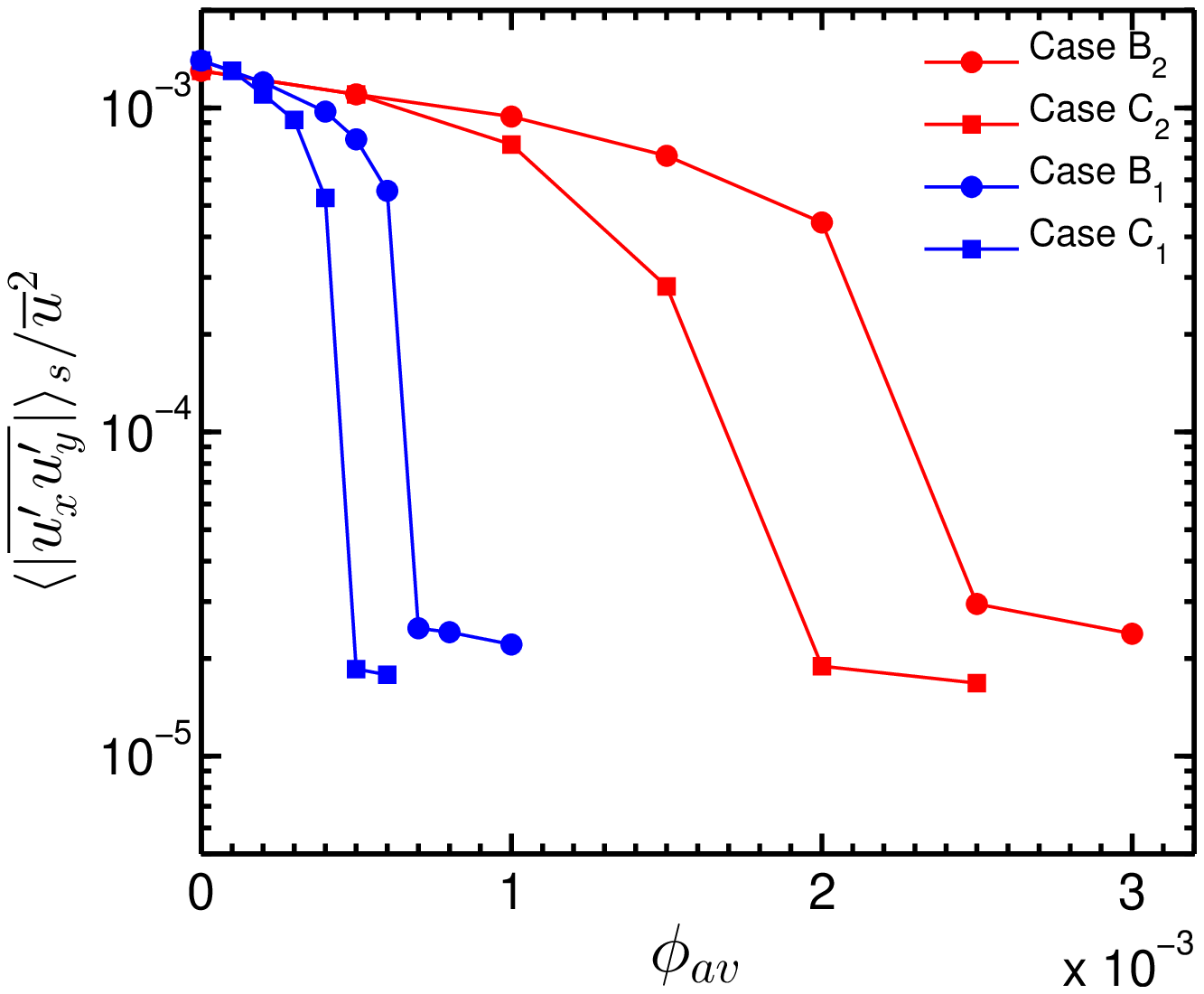}
	\caption{}
	\endminipage \\
	\minipage{0.45\textwidth}
	\includegraphics[width=\textwidth]{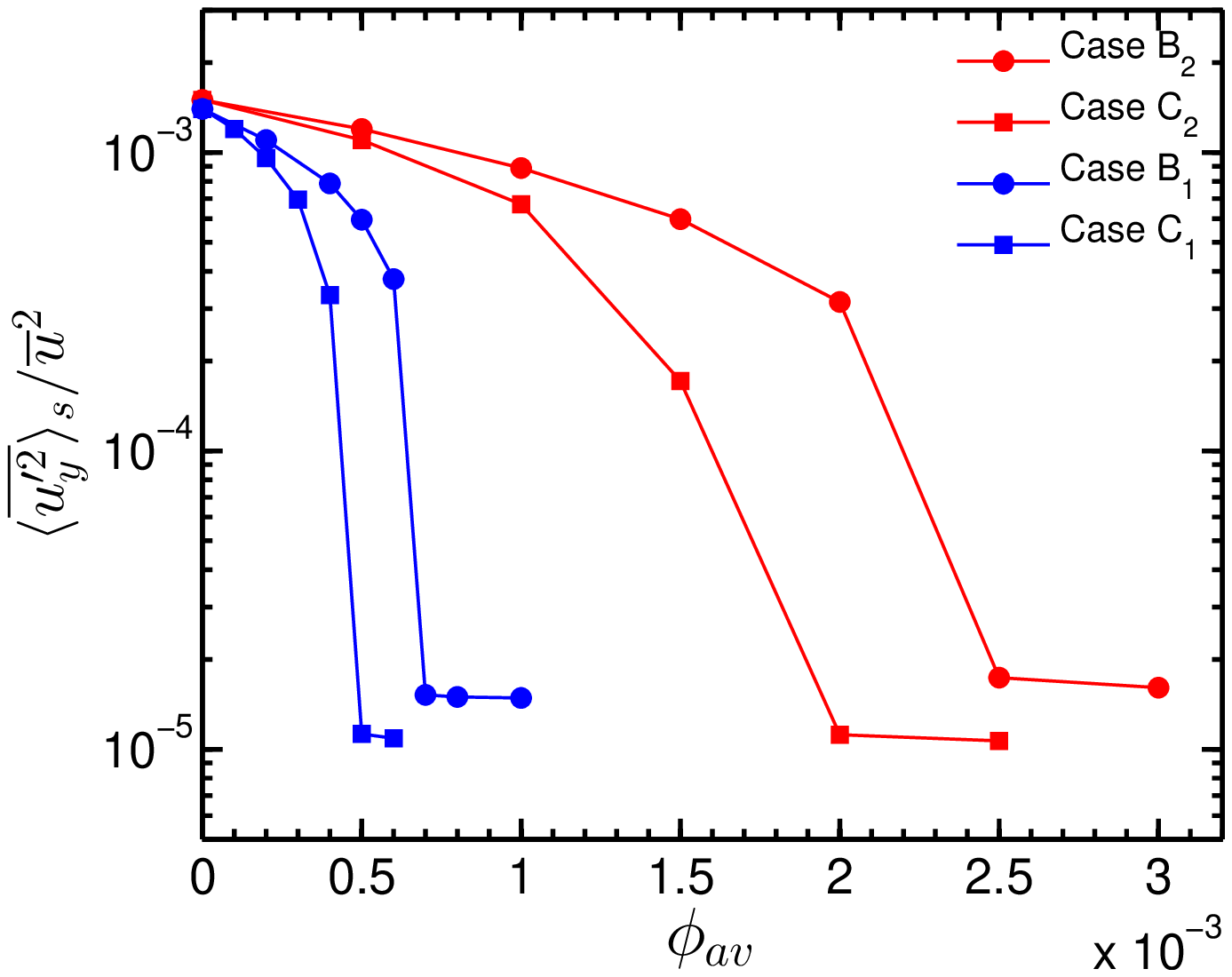}
	\caption{}
	\endminipage	
	\minipage{0.45\textwidth}
	\includegraphics[width=\textwidth]{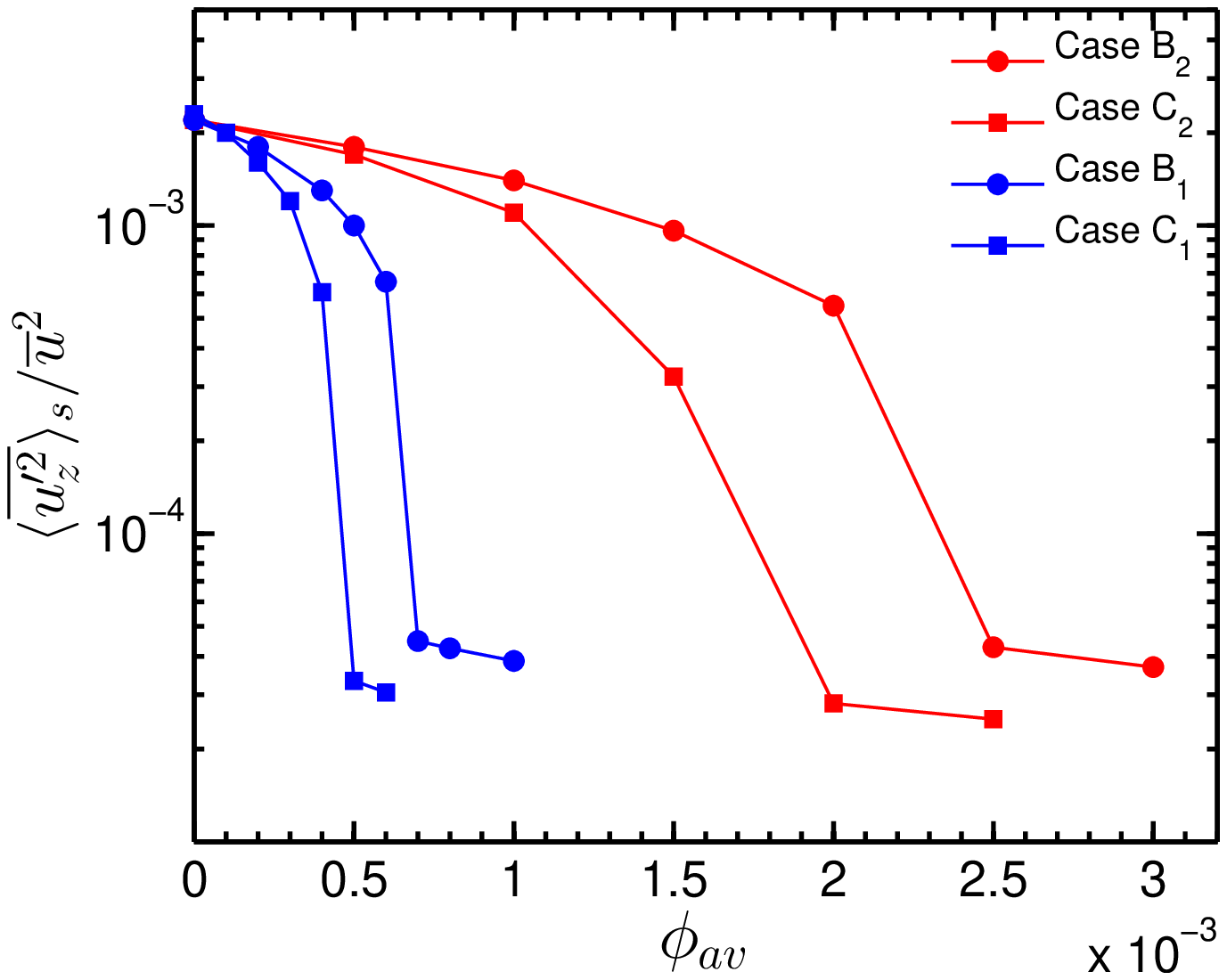}
	\caption{}
	\endminipage 
	\end{subfigure} 
	\caption{The average fluid fluctuations across the channel width are plotted over a range of volume fractions for cases $B_1$, $C_1$, $B_2$, and $C_2$. The fluid fluctuations are normalized by the fluid bulk velocity ($ \overline{u} $). Fig.~(a)  Streamwise, (b) Cross-stream, (c) Wall-normal, and (d) Spanwise fluid fluctuations. }
	\label{avg_u_Re395}
\end{figure*}

Since the extent of variation of second-order velocity fluctuations is a function of wall-normal distance, we compute a channel averaged value of the moments to quantify the variation of the moments as a function of solid volume fraction. The average fluid fluctuations are plotted in Fig.~\ref{avg_u_Re395} for both the Reynolds numbers of 5600 and 13750, and for channel dimensions of $2\delta/d_p = 81$ and 117 over a range of particle volume fractions. The decrease in the fluid fluctuations is observed with an increase in particle volume loading. It is observed that there is almost a 40\% decrease in the streamwise fluid fluctuations and almost 60\% decrease in the cross-stream, wall-normal, and spanwise fluid fluctuations compared to unladen cases before the CPVL for both the channels and Reynolds numbers. However, for the constant $2\delta/d_p$, the CPVL for $Re_b = 13750$ is significantly higher than that for the $Re_b = 5600$.


To understand the dependence of turbulence modulation and its collapse on the channel dimension and fluid phase Reynolds number, we conduct momentum and energy balance for a range of solid volume fractions and discuss them in the following section.

\subsection{Momentum and energy balance}

This section discusses the terms of the momentum balance and energy budget for the mean fluid flow. The filtered momentum equations for streamwise and wall-normal components are written as,

\begin{equation}
- \frac{1}{\rho_f } \frac{\partial \widetilde{P}}{\partial x} -\frac{\partial (\overline{\widetilde{u_x^{'}} \widetilde{u_y^{'}}})}{\partial y} + (\nu +\nu_t) \frac{\partial^2 (\widetilde{U}_x )}{ \partial y \partial y} - \overline{ \frac{\rho_p f \phi_l}{ \rho_f \tau_p} (\widetilde{u}_x  - \widetilde{v}_x) } = 0,
\label{mom_x}
\end{equation}
and
\begin{equation}
- \frac{1}{\rho_f } \frac{\partial \widetilde{P}}{\partial y} -\frac{\partial (\overline{\widetilde{u_y^{'}} \widetilde{u_y^{'}}})}{\partial y}  - \overline{ \frac{\rho_p f \phi_l}{ \rho_f \tau_p} (\widetilde{u}_y  - \widetilde{v}_y) } = 0.
\label{mom_y}
\end{equation}

\begin{figure*}
	\begin{subfigure}[b]{1\textwidth}
	\centering
	\minipage{0.4\textwidth}
	\includegraphics[width=\textwidth]{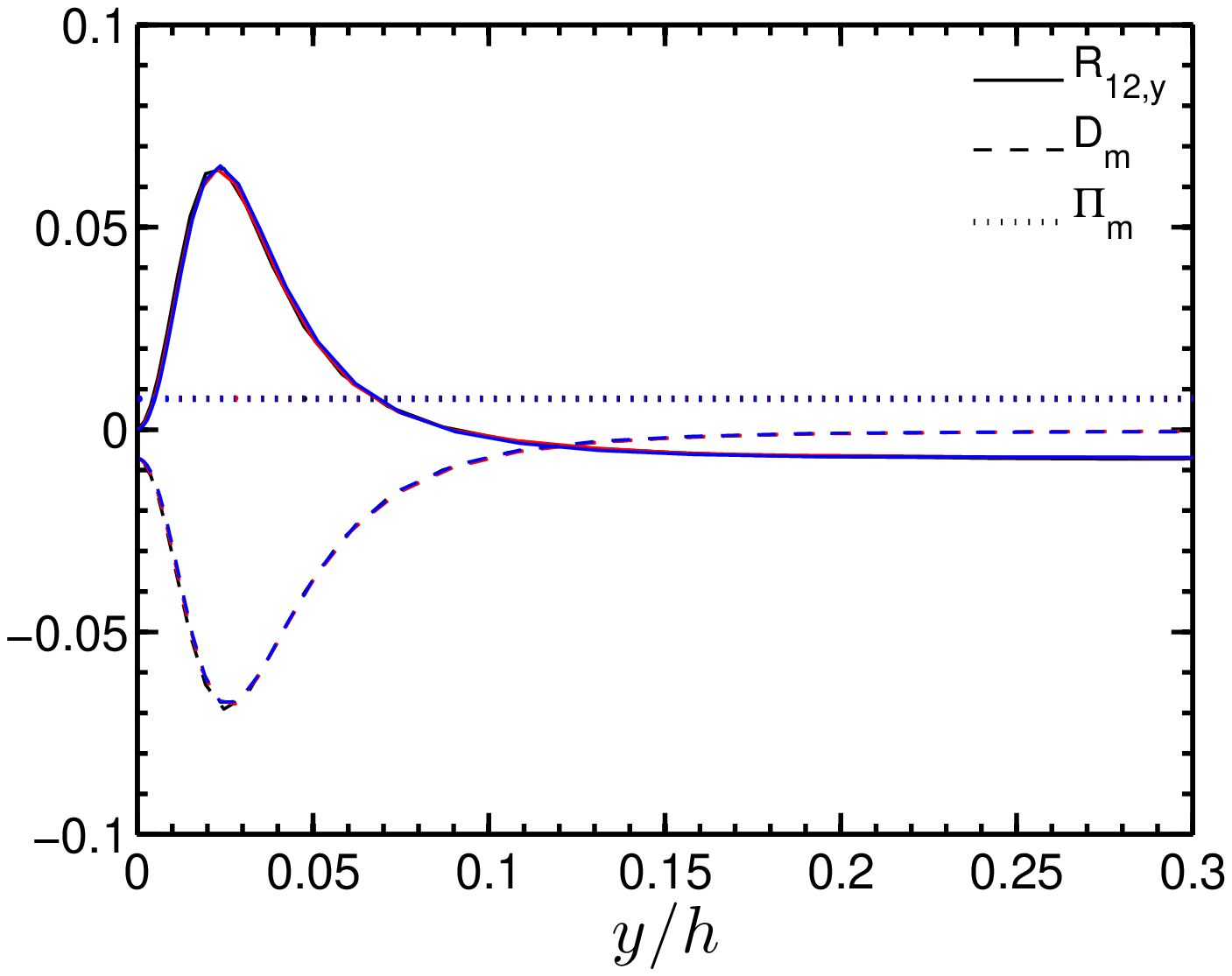}
	\caption{}
	\endminipage 
	\minipage{0.4\textwidth}
	\includegraphics[width=\textwidth]{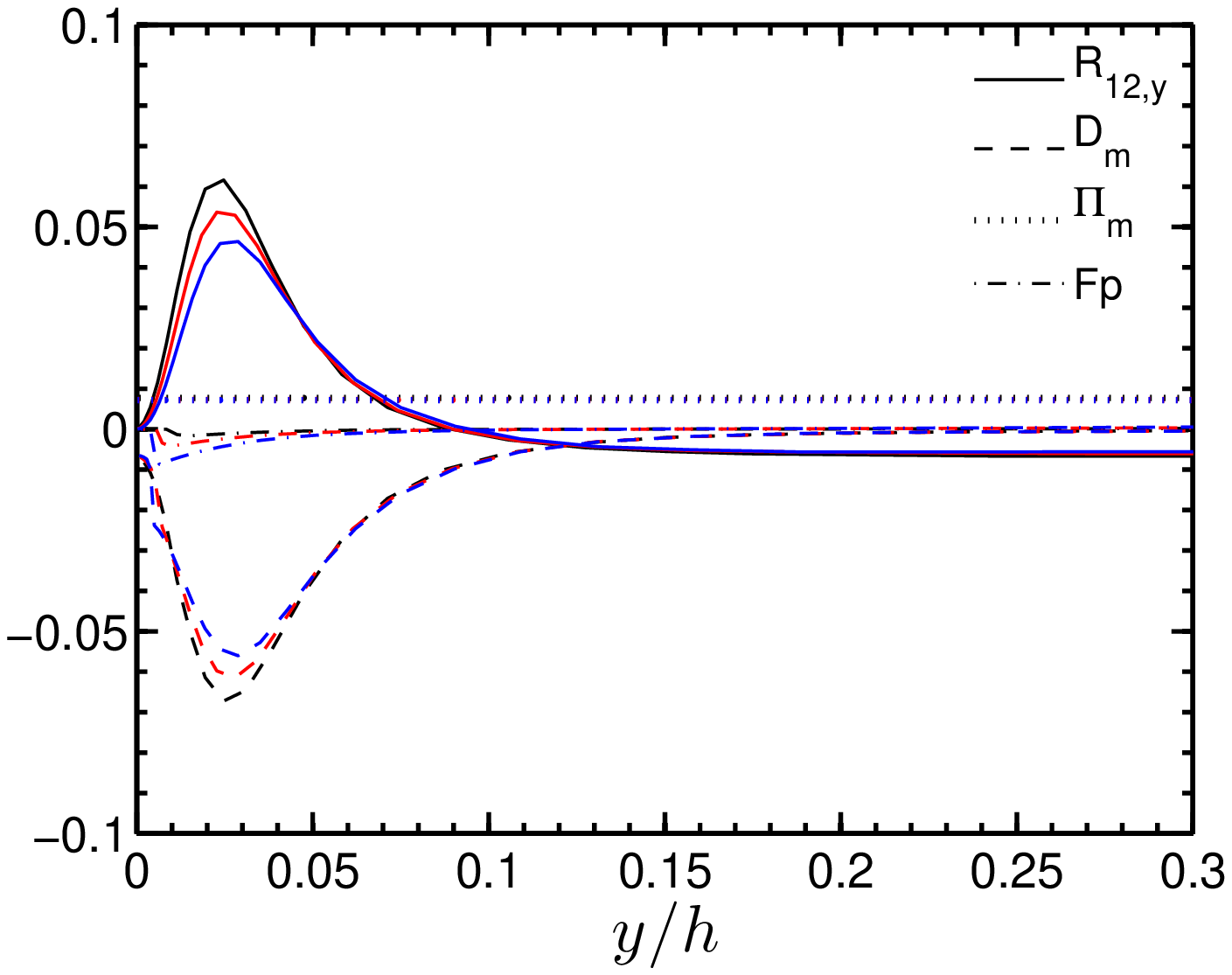}
	\caption{}
	\endminipage \\
	\minipage{0.4\textwidth}
	\includegraphics[width=\textwidth]{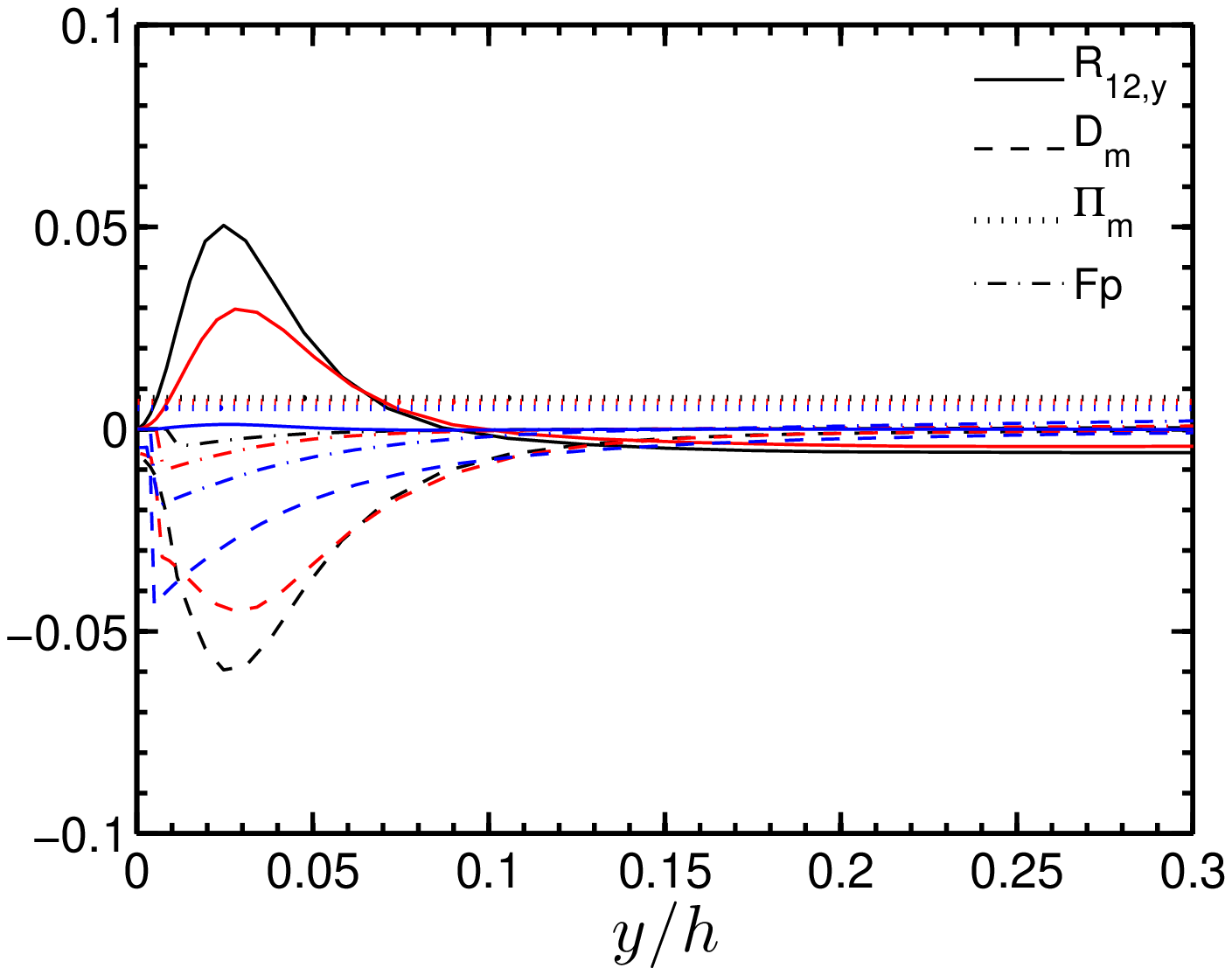}
	\caption{}
	\endminipage
	\end{subfigure} 
	\caption{The terms from Eqn.~\ref{mom_x} are plotted in the wall-normal direction for $Re_b = 5600$ at different volume fractions. (a) $\phi = 0$, (b) $\phi = 2\times10^{-4}$, and (c) $\phi = 5\times10^{-4}$. Solid line: Derivative of Reynolds stress ($R_{12,y}$), Dashed: viscous diffusion ( $D_m$), Dotted: Pressure gradient ($\Pi_m$), Dash dotted: particle feedback force ( $F_p$). Black lines are for the case $A_1$, red lines are for the case $B_1$, and blue lines are for the case $C_1$.}
	\label{mom_180_Ldp}
\end{figure*}

Here,  $\widetilde{p}$ and $\widetilde{u}_i$ are the fluid's instantaneous filtered pressure and velocity, respectively. $ \widetilde{P} $ and $\widetilde{U}_x$ are the mean filtered pressure and velocity, respectively. The filtered fluctuating velocity is denoted by $\widetilde{u'}_i$. The last term in both the above equations shows the feedback force exerted by the particles, where $\rho_p$ and $\rho_f$ are the particle and fluid density, $f (= 1 + 0.15 Re_p^{0.687})$ is the drag factor, $\tau_p$ is the particle relaxation time, and $\phi_l$ is the local volume fraction of the particle. The filtered Reynolds stress is represented by $\rho_f \overline{\widetilde{u^{'}_i} \widetilde{u^{'}_j}}$. The terms appearing in the streamwise filtered momentum equation are plotted in Fig.~\ref{mom_180_Ldp} as a function of wall-normal positions. All the terms are scaled with $\bar{u}^2/2\delta$. The wall-normal distance ($ y $) is normalized with channel width ($2\delta = h$). The terms of the momentum equation are plotted for different channel dimensions and volume fractions which present the effect of system size. Here, the derivative of Reynolds stress is denoted by $R_{12,y}$, diffusion term by $D_m$, pressure gradient by $\Pi_m$, and feedback force by $F_p$. For unladen cases, the terms for channels with different dimensions match with each other as expected (Fig.~\ref{mom_180_Ldp} a). Here, magnitude of $R_{12,y}$ is comparable to the $D_m$. However, the deviation is observed at $\phi = 2\times10^{-4}$ in Fig.~\ref{mom_180_Ldp} (b), and the peak values of $R_{12,y}$ and $D_m$ are smaller by approximately 20\% for case $C_1$ compared to the case $A_1$ (Table~\ref{Stokes_number}). A further increase in volume fraction to $\phi = 5\times 10^{-4}$ results a decrease in $D_m$, and $R_{12,y}$ almost becomes zero for case $C_1$ (higher $2\delta/d_p$) as shown in Fig.~\ref{mom_180_Ldp} (c). Fig.~\ref{mom_180_Ldp} (c) shows that $\langle \Pi_m \rangle_s$  decreases nearly 30\%, while $\langle F_p \rangle_s$ increases almost by 400\% for case $C_1$ compared to case $A_1$. Here, $\langle . \rangle_s$ denotes the averaging of a quantity across the channel width as defined by Eqn.~\ref{avg_fluid_fluc}. Here, a significant increase in the feedback force is observed with an increase in $2\delta/d_p$.

\begin{figure*}[htb]
	\begin{subfigure}[b]{1\textwidth}
	\centering
	\minipage{0.4\textwidth}
	\includegraphics[width=\textwidth]{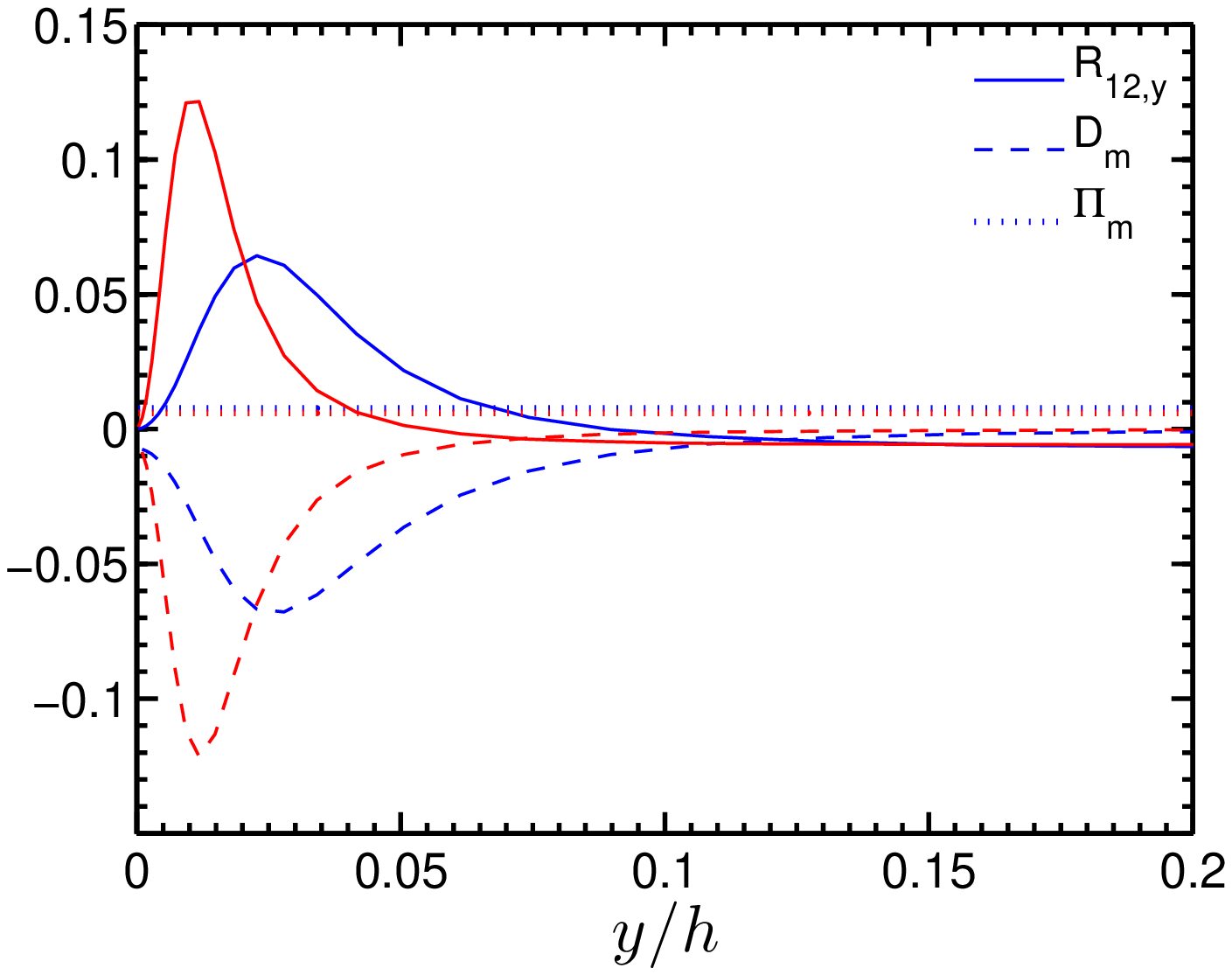}
	\caption{}
	\endminipage 
	\minipage{0.4\textwidth}
	\includegraphics[width=\textwidth]{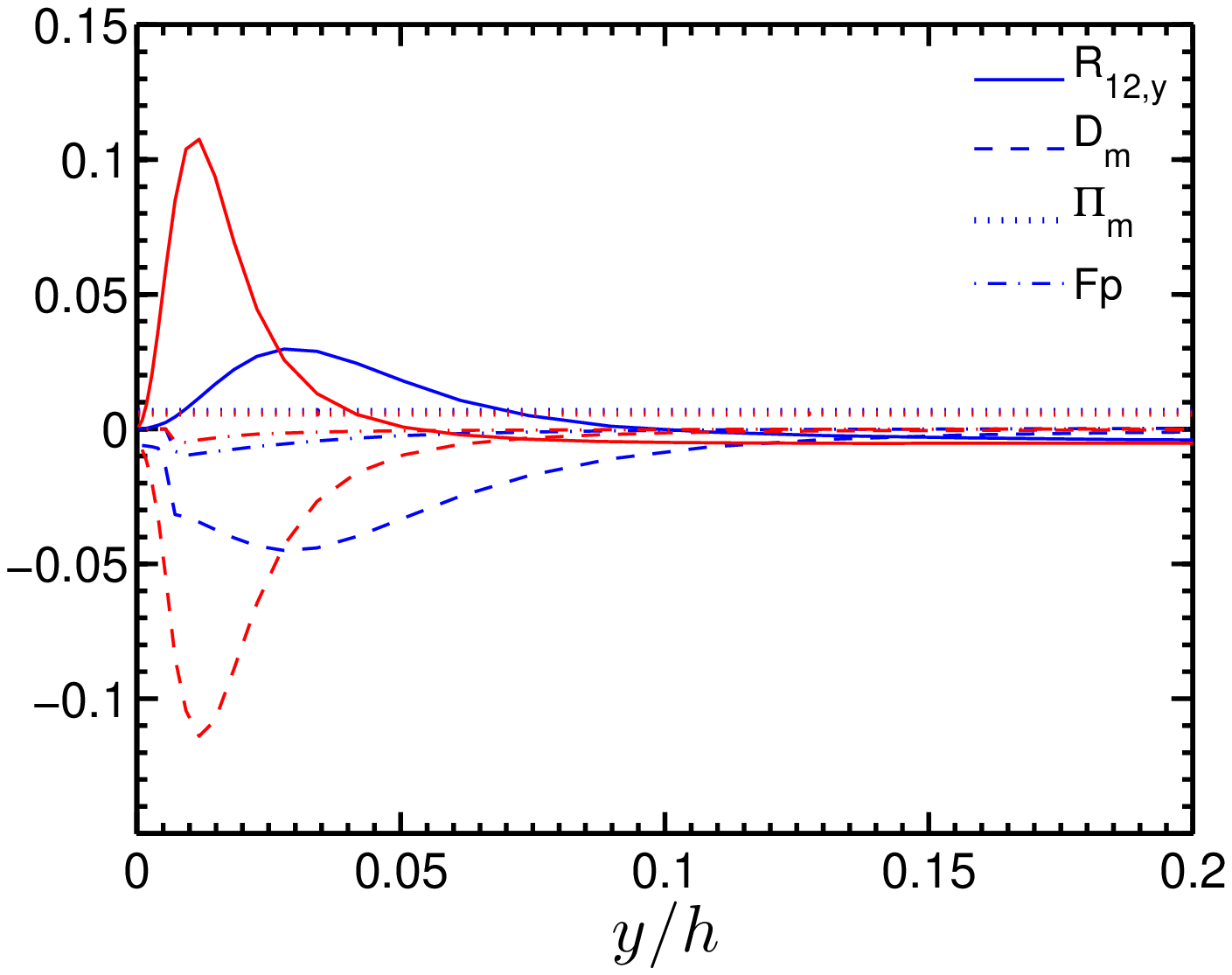}
	\caption{}
	\endminipage \\
	\minipage{0.4\textwidth}
	\includegraphics[width=\textwidth]{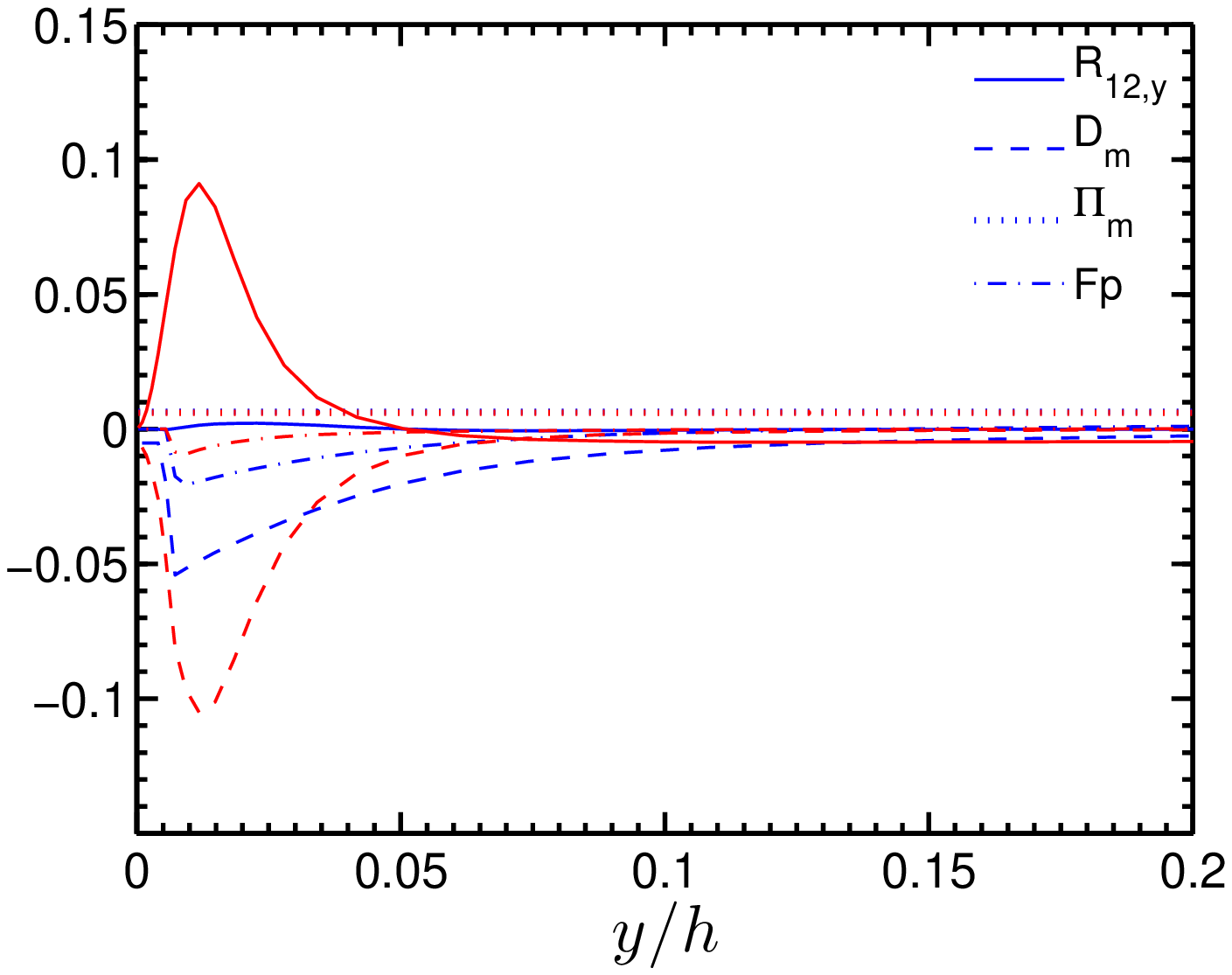}
	\caption{}
	\endminipage
	\end{subfigure} 
	\caption{The terms from Eqn.~\ref{mom_x} are plotted in the wall-normal direction for cases $B_1$ and $B_2$ at different volume fractions. (a) $\phi = 0$, (b) $\phi = 5\times10^{-4}$, and (c) $\phi = 1\times10^{-3}$. Solid line: Derivative of Reynolds stress ($R_{12,y}$), Dashed: viscous diffusion ( $D_m$), Dotted: Pressure gradient ($\Pi_m$), Dash dotted: particle feedback force ( $F_p$). Here, red lines are for the case $B_2$,  and blue lines are for the case $B_1$.}
	\label{mom_164_Re}
\end{figure*}

To quantify the effect of the Reynolds number, we have chosen a channel with $2\delta/d_p = 81$, and the terms of the momentum equation are plotted for $Re_b = 5600$ and 13750. In the case of the unladen flow, the peak location of $R_{12,y}$ and $D_m$ are closer to the wall, and magnitude of peak values are almost twice for $Re_b = 13750$ as shown in  Fig.~\ref{mom_164_Re} (a). It is interesting to note that the scaled pressure gradient is lowered by approximately 25\% for the case $B_2$ ($Re_b = 13750$) than case $B_1$ ($Re_b = 5600$), Fig.~\ref{mom_164_Re} (a). A decrease of nearly 25\% in $\langle \Pi_m \rangle_s$ and $ \langle D_m \rangle_s$ is observed, while there is no variation of $  \langle R_{12,y} \rangle_s$ in case of $B_2$ compared to $B_1$. The terms are shown for $\phi = 5\times 10^{-4}$ and $10^{-3}$ in  Fig.~\ref{mom_164_Re} (b and c). $R_{12,y}$ and $D_m$ decrease with an increase in volume fraction. However, the extent of the decrease is much lower for $Re_b = $13750 than $Re_b = 5600$. $R_{12,y}$ is almost zero for $Re_b = 5600$ at $\phi = 10^{-3}$, which is due to the collapse of fluid turbulence. In Fig.~\ref{mom_164_Re} (b and c), $\langle F_p \rangle_s$ in simulation $B_2$ is almost 65\% lower compared to $B_1$. Thus, it is observed that a significant decrease in the feedback force (scaled with fluid bulk velocity ($\bar{u}$) and channel width ($2\delta$)) occurs with an increase in the Reynolds number.

The mean filtered kinetic energy equation for the mean flow is written as,
\begin{multline}
(\nu +\nu_t) \frac{\partial}{\partial y}(\widetilde{U}_x \frac{\partial \widetilde{U_x}}{\partial y}) - \frac{\partial (\overline{\widetilde{u_x^{'}} \widetilde{u_y^{'}}} \widetilde{U}_x)}{\partial y} + \overline{\widetilde{u_x^{'}} \widetilde{u_y^{'}}} \frac{\partial ( \widetilde{U}_x)}{\partial y} \\
 - (\nu +\nu_t) \frac{\partial \widetilde{U}_x}{\partial y} \frac{\partial \widetilde{U}_x}{\partial y}  - \widetilde{U}_x \overline{ \frac{\rho_p f \phi}{ \rho_f \tau_p} (\widetilde{u}_x  - \widetilde{v}_x) }  - \widetilde{U}_x \frac{1}{\rho } \frac{\partial \widetilde{P}}{\partial x}   = 0.
\label{energy}
\end{multline}
In Eqn.~\ref{energy}, on the left-hand side, the first two terms are the energy flux terms due to molecular viscosity, eddy viscosity, and Reynolds stress. The third term is the energy utilized for the turbulent production, the fourth term is the dissipation of turbulent mean kinetic energy, the fifth term is the dissipation caused by the particle feedback, and the last term is the energy input due to pressure work. The spatial average of Eqn.~\ref{energy} in the wall-normal direction makes the flux term zero due to no-slip condition at the walls, and the equation becomes,
\begin{multline}
\langle \overline{\widetilde{u_x^{'}} \widetilde{u_y^{'}}} \frac{ \partial ( \widetilde{U}_x)}{\partial y}   \rangle_s  -  (\nu +\nu_t) \langle  \frac{ \partial \widetilde{U}_x}{\partial y} \frac{\partial \widetilde{U}_x}{\partial y} \rangle_s \\ - \langle \widetilde{U}_x \overline{ \frac{\rho_p f \phi_c}{ \rho_f \tau_p} (\widetilde{u}_x  - \widetilde{v}_x) } \rangle_s - \widetilde{U}_x \frac{1}{\rho_f } \frac{\partial \widetilde{P}}{\partial x}    = 0 
\label{energy_avg_x}
\end{multline}

\begin{figure*}[htb]
	\begin{subfigure}[b]{1\textwidth}
	\centering
	\minipage{0.4\textwidth}
	\includegraphics[width=\textwidth]{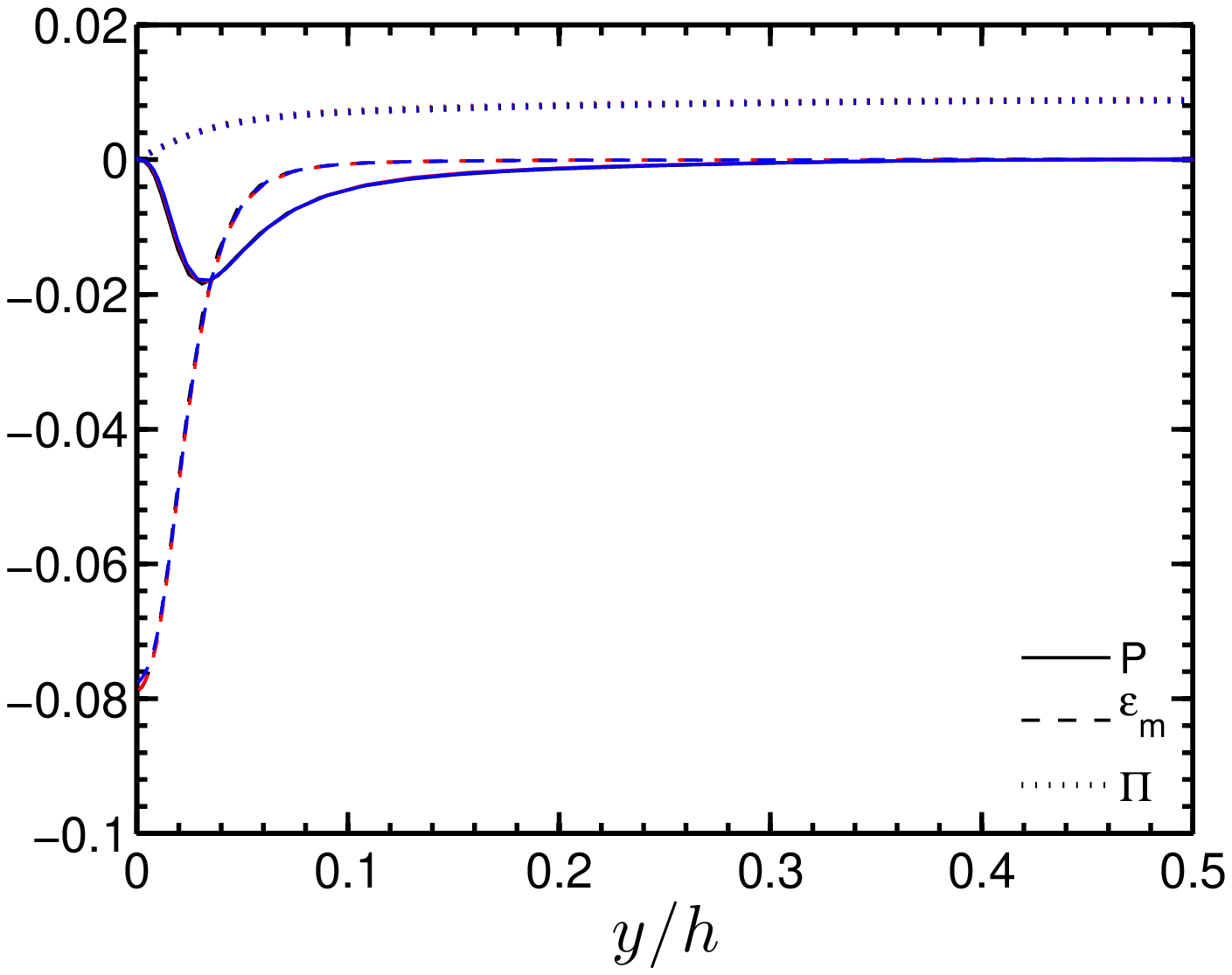}
	\caption{}
	\endminipage 
	\minipage{0.4\textwidth}
	\includegraphics[width=\textwidth]{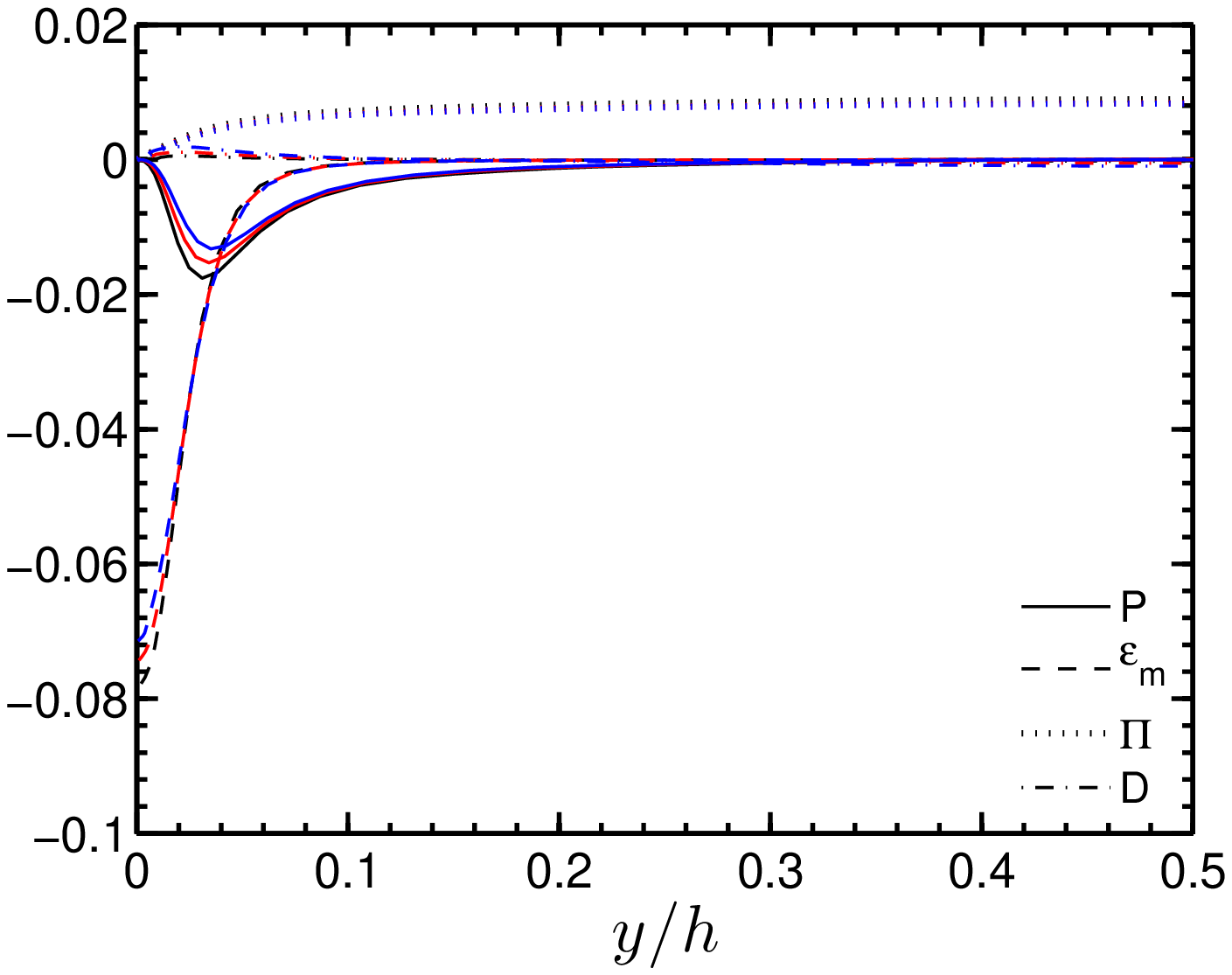}
	\caption{}
	\endminipage \\
	\minipage{0.4\textwidth}
	\includegraphics[width=\textwidth]{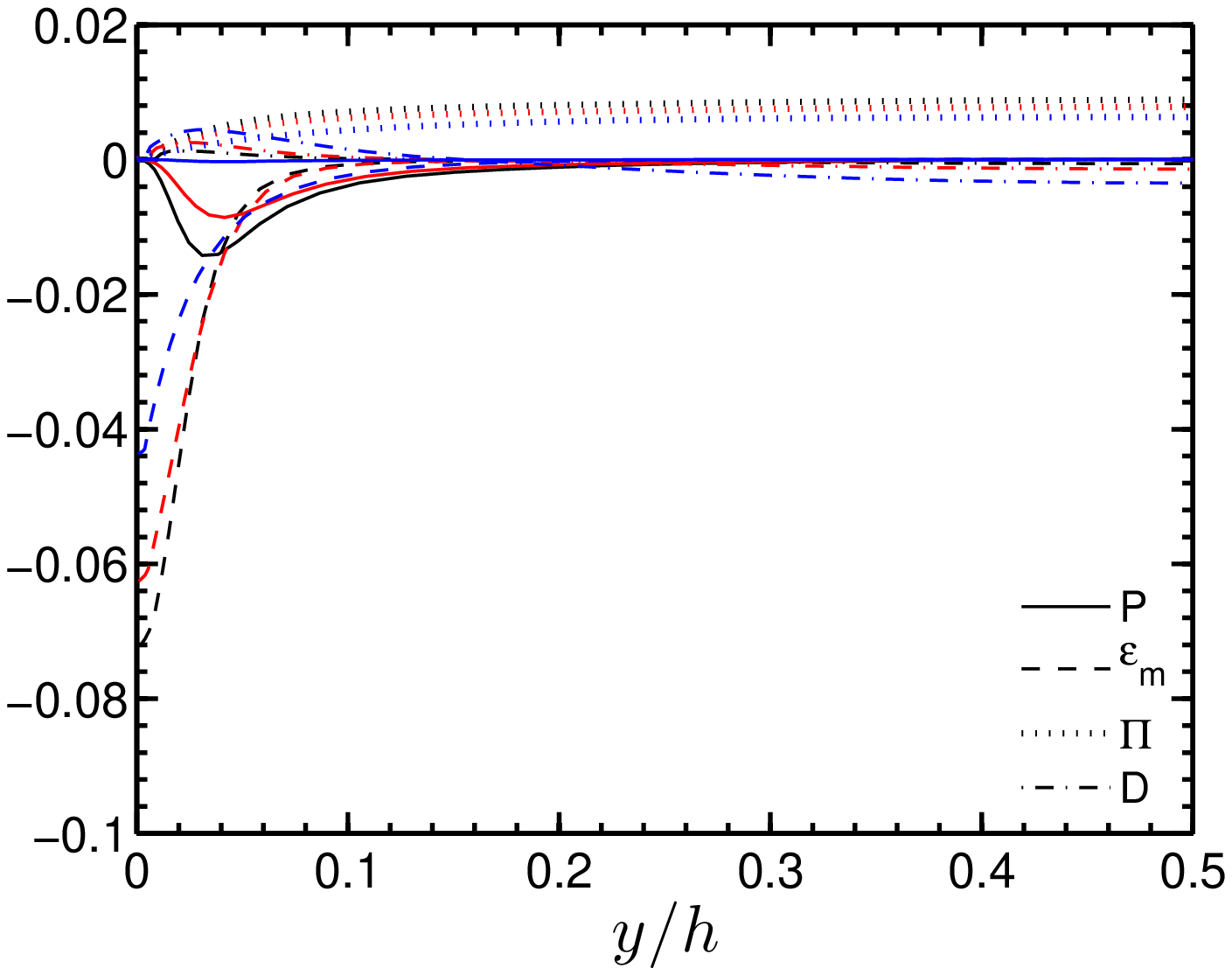}
	\caption{}
	\endminipage
	\end{subfigure} 
	\caption{The terms from Eqn.~\ref{energy_avg_x} are plotted in the wall-normal direction for $Re_b = 5600$ at different volume fractions. (a) $\phi = 0$, (b) $\phi = 2\times10^{-4}$, and (c) $\phi = 5\times10^{-4}$. Solid line: production term ($P$), Dashed: viscous dissipation ($\epsilon_m$), Dotted: Pressure work ($\Pi$), Dash dotted: dissipation due to particle feedback ($ D $). Black lines are for the case $A_1$, red lines are for the case $B_1$, and blue lines are for the case $C_1$.}
	\label{Energy_180_Ldp}
\end{figure*}

The terms in the Eqn.~\ref{energy_avg_x} are plotted in Fig.~\ref{Energy_180_Ldp} for different channels and volume fractions for $Re_b = 5600$. All the terms in Fig.~\ref{Energy_180_Ldp} are normalized by respective $\bar{u}^3/2\delta$. The turbulent production term is represented by $P$, mean viscous dissipation by $\epsilon_m$, pressure work by $\Pi$, and dissipation due to particles by $D$. The wall-normal distance ($y$) is scaled by channel width ($h$). For the unladen case in Fig.~\ref{Energy_180_Ldp} (a), all predicted terms for all the channels with different dimensions overlap as expected. However, for a particle volume fraction of $2\times 10^{-4}$ in Fig.~\ref{Energy_180_Ldp} (b), the deviation is observed in all the terms. The peak of turbulent production terms for the case $C_1$ is lowered by nearly 30\%, viscous dissipation in the near-wall is lower by 8\%, and pressure work in the channel center is lower by 10\% compared to the case $A_1$. However, $\langle D \rangle_s$ is higher by nearly 400\% for the case $C_1$ compared to the case $A_1$. For $\phi = 5\times 10^{-4}$ in Fig.~\ref{Energy_180_Ldp} (c), the turbulence collapse in case of $C_1$, and the turbulent production term becomes zero. The viscous dissipation in the near-wall is lower by approximately 40\%, and the pressure work is also reduced by 30\%. The dissipation due to particles is higher by nearly 400\% for the case $C_1$ than the case $A_1$.

\begin{figure*}
	\begin{subfigure}[b]{1\textwidth}
	\centering
	\minipage{0.4\textwidth}
	\includegraphics[width=\textwidth]{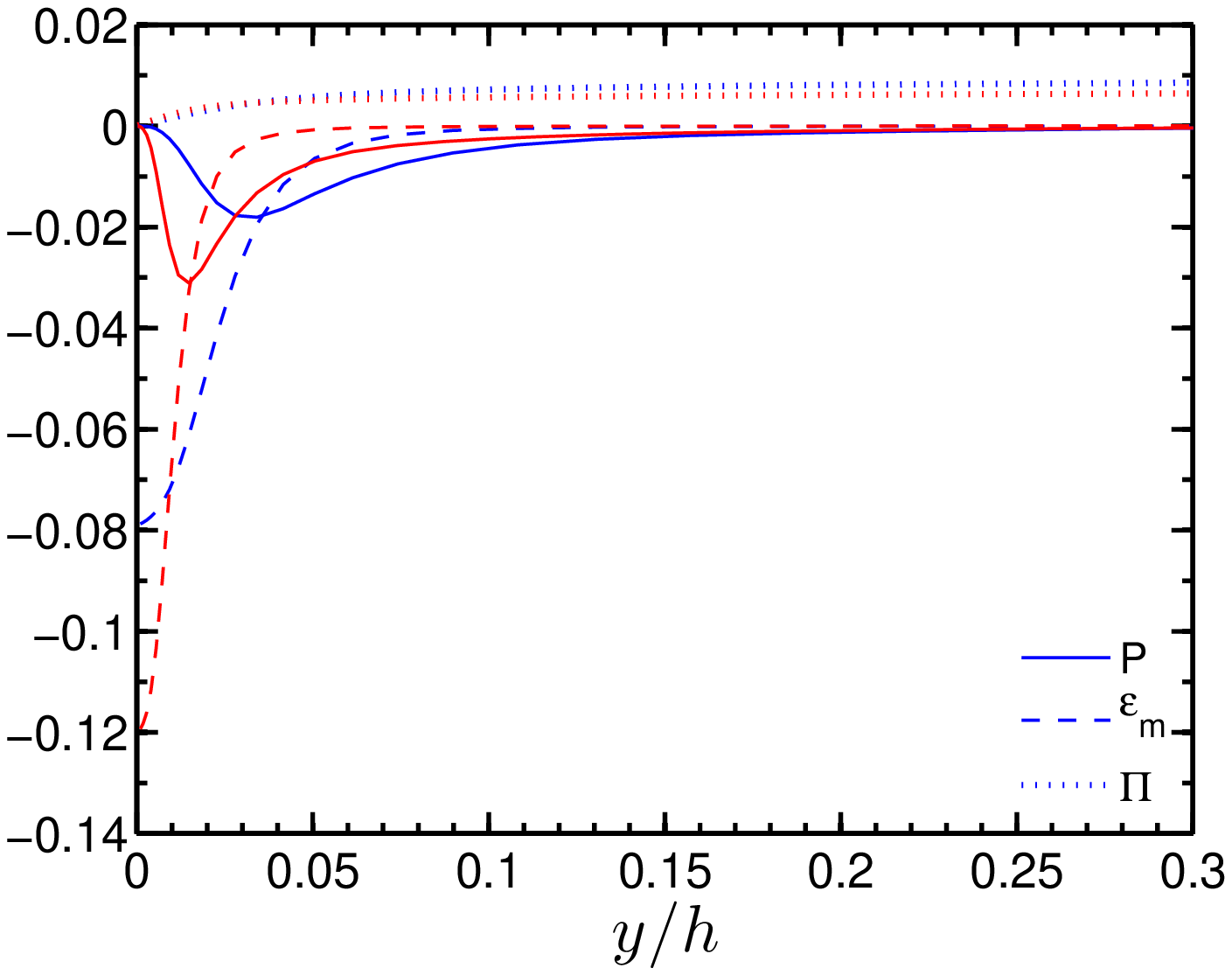}
	\caption{}
	\endminipage 
	\minipage{0.4\textwidth}
	\includegraphics[width=\textwidth]{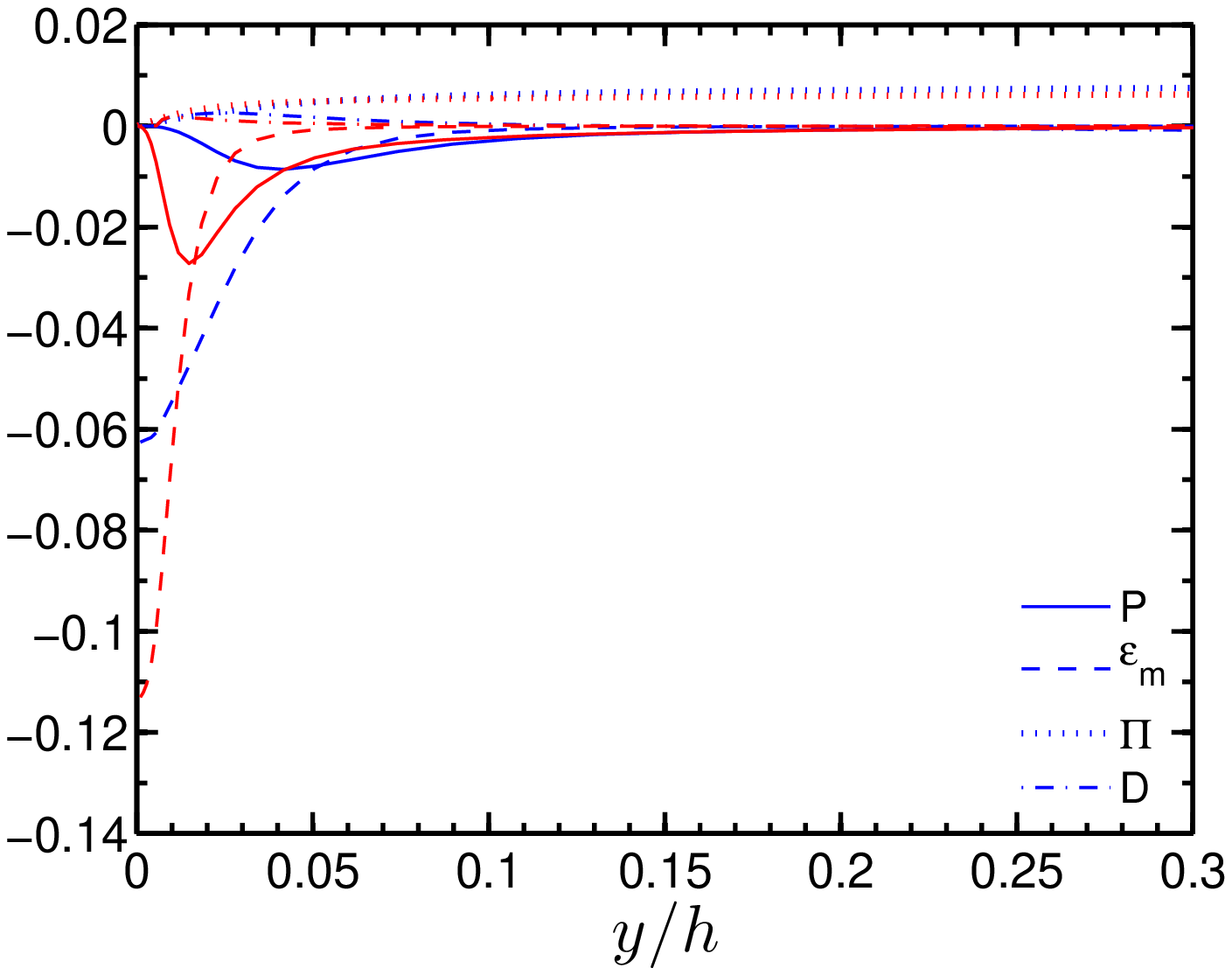}
	\caption{}
	\endminipage \\
	\minipage{0.4\textwidth}
	\includegraphics[width=\textwidth]{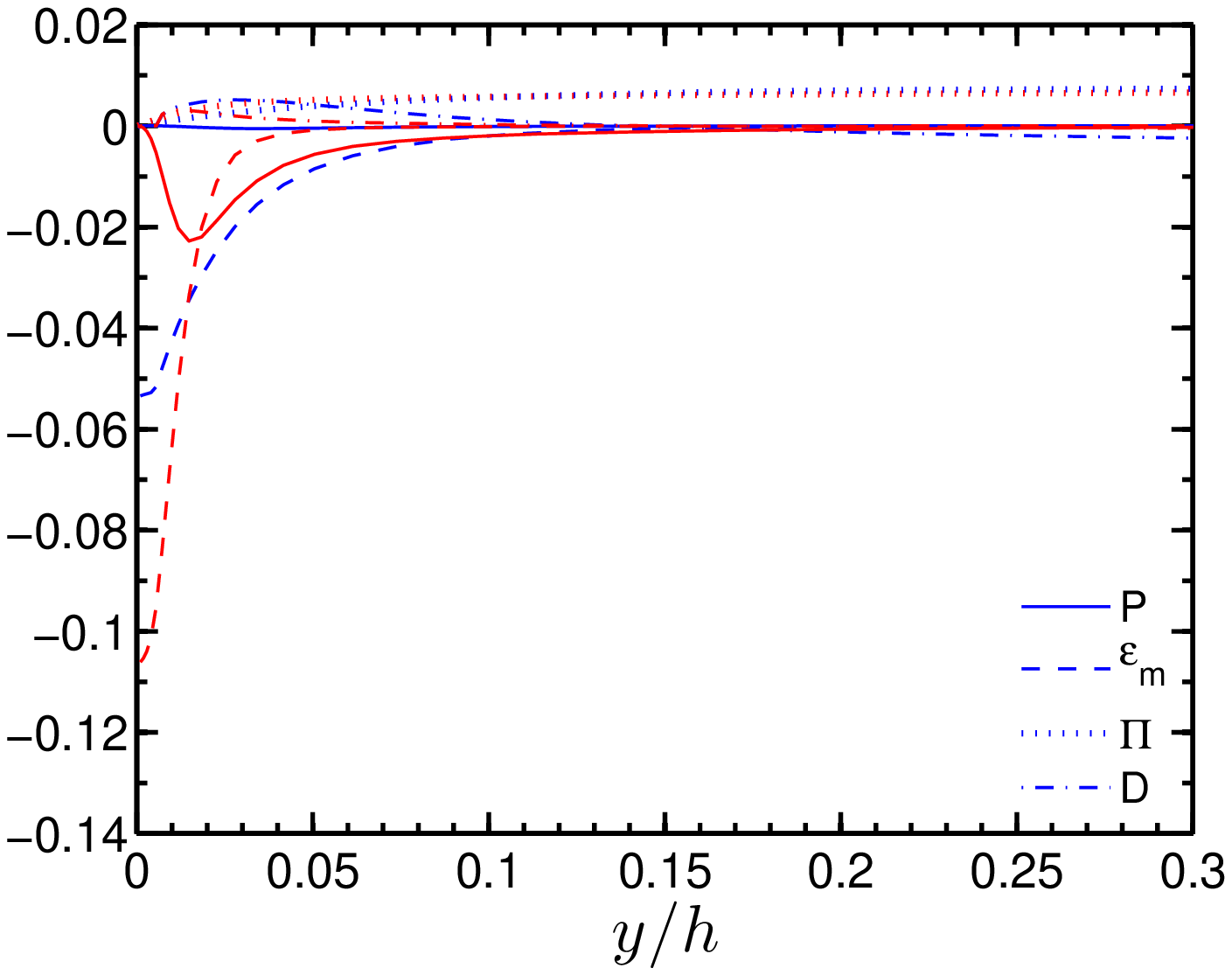}
	\caption{}
	\endminipage
	\end{subfigure} 
	\caption{The terms from Eqn.~\ref{energy_avg_x} are plotted in the wall-normal direction for $Re_b = 5600$ at different volume fractions. (a) $\phi = 0$, (b) $\phi = 5\times10^{-4}$, and (c) $\phi = 1\times10^{-3}$. Solid line: production term ($P$), Dashed: viscous dissipation ($\epsilon_m$), Dotted: Pressure work ($\Pi$), Dash dotted: dissipation due to particle feedback ($ D $). Here, red lines are for the case $B_2$, and blue lines are for the case $B_1$.}
	\label{Energy_Re}
\end{figure*}

In Fig.~\ref{Energy_Re}, the terms from Eqn.~\ref{energy_avg_x} are plotted for cases $B_1$ and $B_2$, and for different volume fractions to analyze the effect of Reynolds number. Fig.~\ref{Energy_Re} (a) shows that for unladen flow, the peak values of turbulent production and viscous dissipation (in the near-wall region) are higher by nearly 50\% for the case $B_2$ ($Re_b = 13750$) compared to the case $B_1$ ($Re_b = 5600$). However, $\langle \Pi \rangle_s$ is approximately 25\% less for $Re_b = 13750$ than $Re_b = 5600$. This decrease in $\langle \Pi \rangle_s$ is compensated by the decrease in $\langle \epsilon_m \rangle_s$, while there is almost no variation in the $\langle P \rangle_s$. The peak value of turbulent production is much closer to the wall for a higher Reynolds number. In Fig.~\ref{Energy_Re} (b) for $\phi = 5 \times 10^{-4}$, the peak value of production term and viscous dissipation in the near-wall region is almost 50\% for $Re_b = 5600$ compared to $Re_b = 13750$. However, $\langle \Pi \rangle_s$ is nearly 20\% and $\langle D \rangle_s$ is nearly 80\% lower for $Re_b = 13750$ compared to $Re_b = 5600$. In Fig.~\ref{Energy_Re} (c) for $\phi = 10^{-3}$, the mean production term almost becomes zero, and the mean viscous dissipation in the near-wall region is 50\% lower for the case $B_1$ compared to the case $B_2$. The scaled $\langle \Pi \rangle_s$ is nearly 5\%, while $\langle D \rangle_s$ is nearly one order of magnitude lower for $B_2$ compared to $B_1$. From the above analysis, it is clear that the variation in particle feedback term is more significant than the others before the critical loading for any of the cases. The effect of particle feedback increases with an increase in $2\delta/d_p$ for fixed Reynolds numbers, and decreases with an increase in fluid Reynolds number while keeping $2\delta/d_p$ fixed.

\subsection{Discussion}
In the above section, we observed that the particle-induced drag force depends on the channel dimension and the fluid phase Reynolds number. Such an increase in drag and drag-induced dissipation is the source of different extents in turbulence modulation when $\phi$ is less than CPVL. In this section, we discuss this dependence in detail. From Eqn.~\ref{LES eqn}, the scaled feedback force on the fluid due to particles can be written as,
\begin{equation}
F_p =  \frac{\rho_p f \phi_l}{ \rho_f \tau_p} (\widetilde{u}_x  - \widetilde{v}_x) \frac{2\delta}{\overline{u}^2} ,
\label{Feedback}
\end{equation}
\begin{eqnarray}
F_p =  \frac{\rho_p f \phi_l}{ \rho_f (\rho_p d_p^2/18 \mu)} (\widetilde{u}_x  - \widetilde{v}_x) \frac{2\delta}{\overline{u}^2} ,
\end{eqnarray}
\begin{equation}                
F_p = 18 f \phi_l  \frac{Re_p }{ Re_b^2 }     \left(\frac{2\delta}{d_p }\right)^3.
\label{scaled_Fp}
\end{equation}

\begin{figure*}
	\begin{subfigure}[b]{1\textwidth}
	\centering
	\minipage{0.45\textwidth}
	\includegraphics[width=\textwidth]{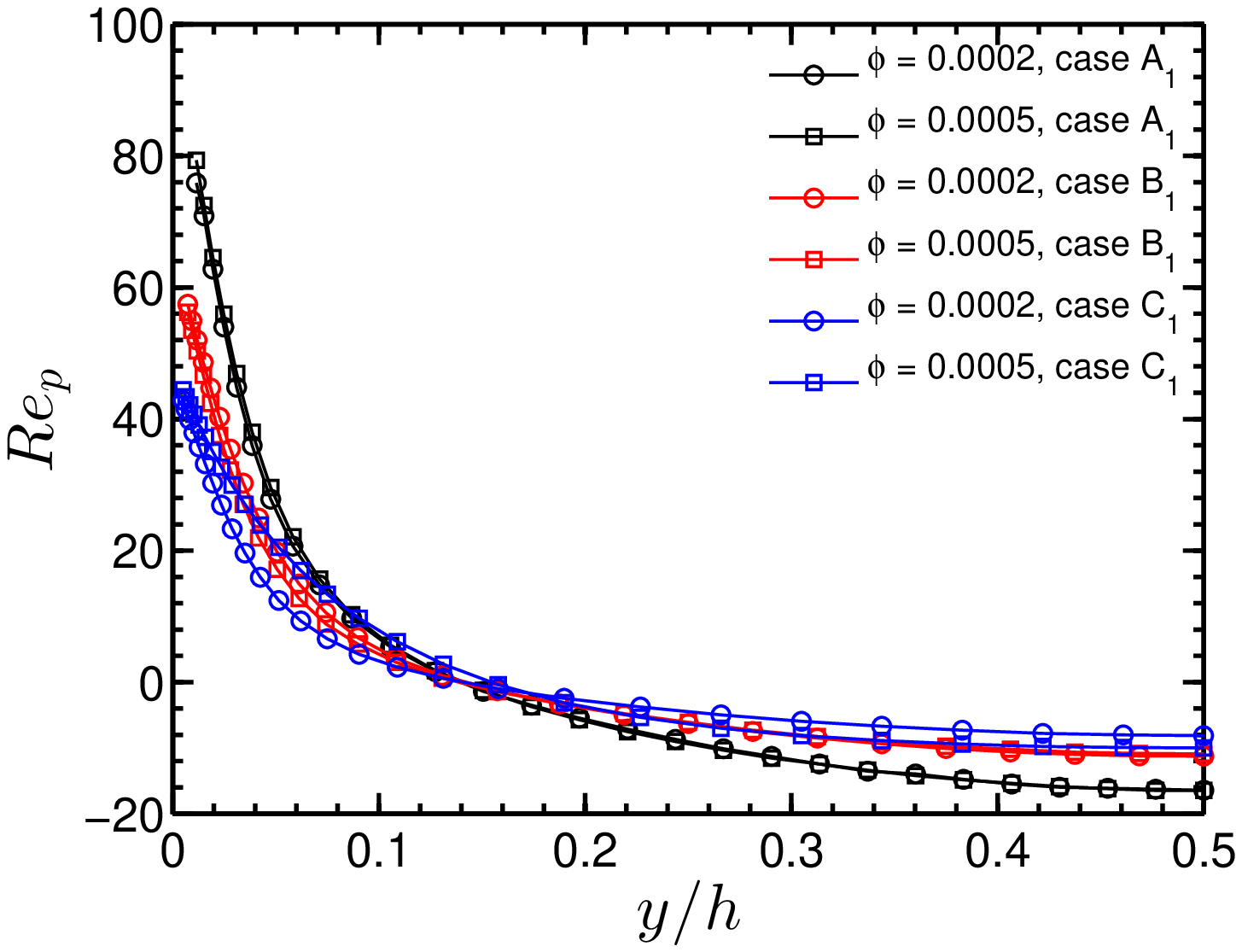}
	\caption{}
	\endminipage
	\minipage{0.45\textwidth}
	\includegraphics[width=\textwidth]{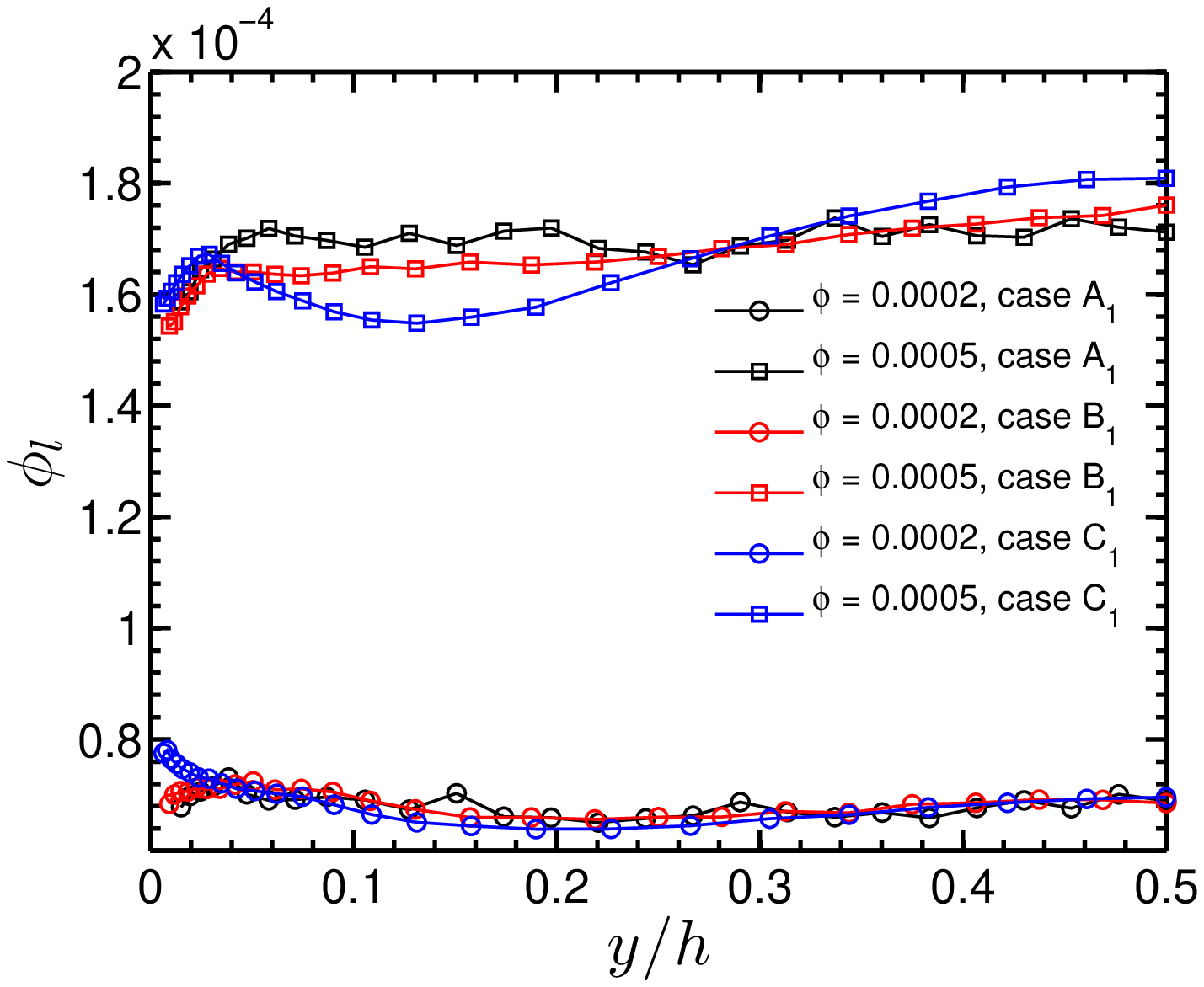}
	\caption{}
	\endminipage	
	\end{subfigure} 
	\caption{(a) The particle Reynolds number and (b) local particle volume fraction in the wall-normal direction for different system sizes and volume fractions for $Re_b = 5600$.}
	\label{reason_Ldp}
\end{figure*}

Here, $\phi_l$ is the local volume fraction of the particle, $f$ is the inertial correction factor, $Re_p$ is the particle Reynolds number, $Re_b$ is the fluid bulk Reynolds number, $\delta$ is the half-channel width, and $d_p$ is the particle diameter. In the case of fixed $Re_b$ and $d_p$, when $2\delta/d_p$ is modified by varying $\delta$, it is observed that there is no significant variation in $Re_p$, $\phi_l$, and $f$ over a range of volume fractions as shown in Fig.~\ref{reason_Ldp}. Under those conditions, we can write,

\begin{figure*}[htb]
	\begin{subfigure}[b]{1\textwidth}
	\centering
	\minipage{0.45\textwidth}
	\includegraphics[width=\textwidth]{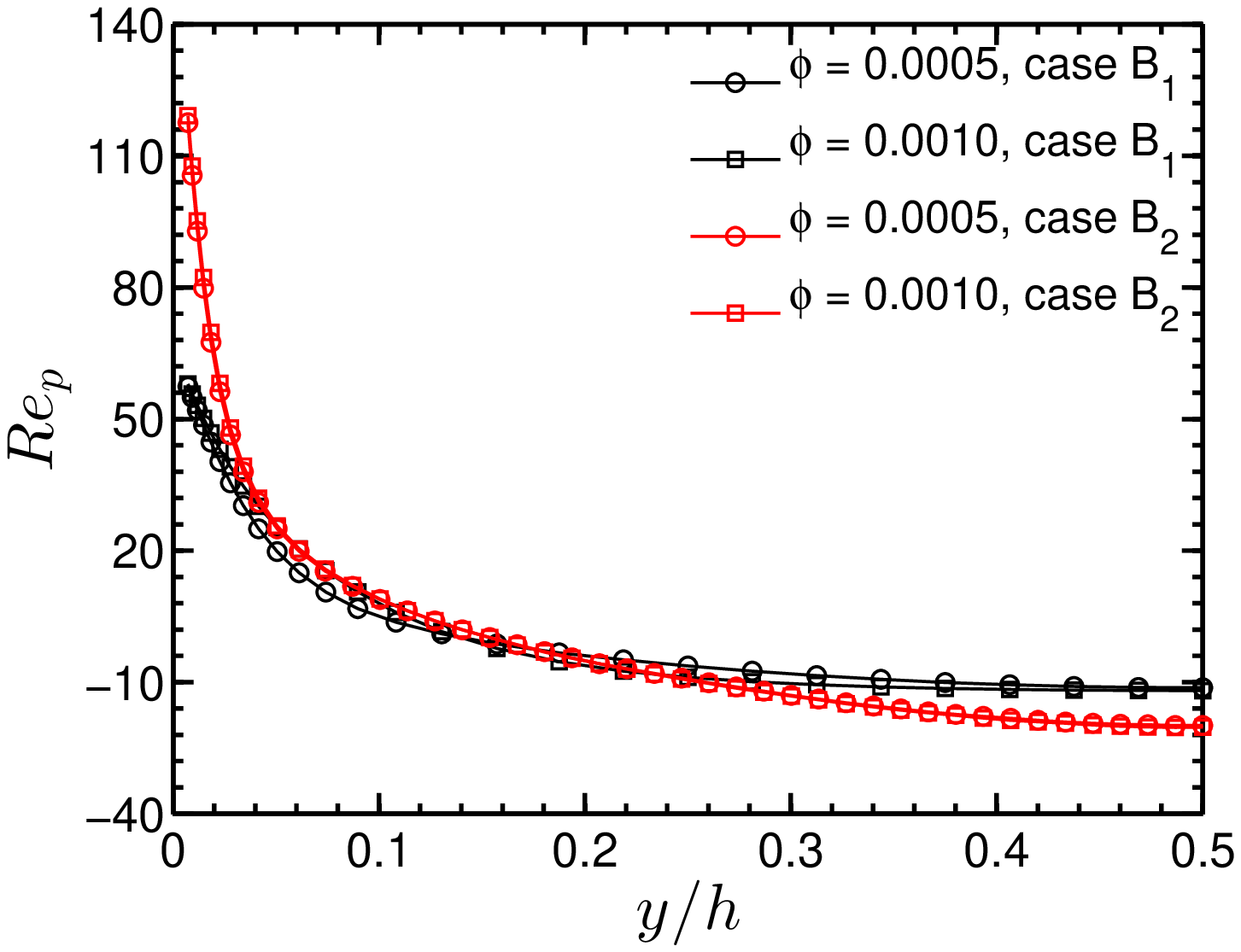}
	\caption{}
	\endminipage 
	\minipage{0.45\textwidth}
	\includegraphics[width=\textwidth]{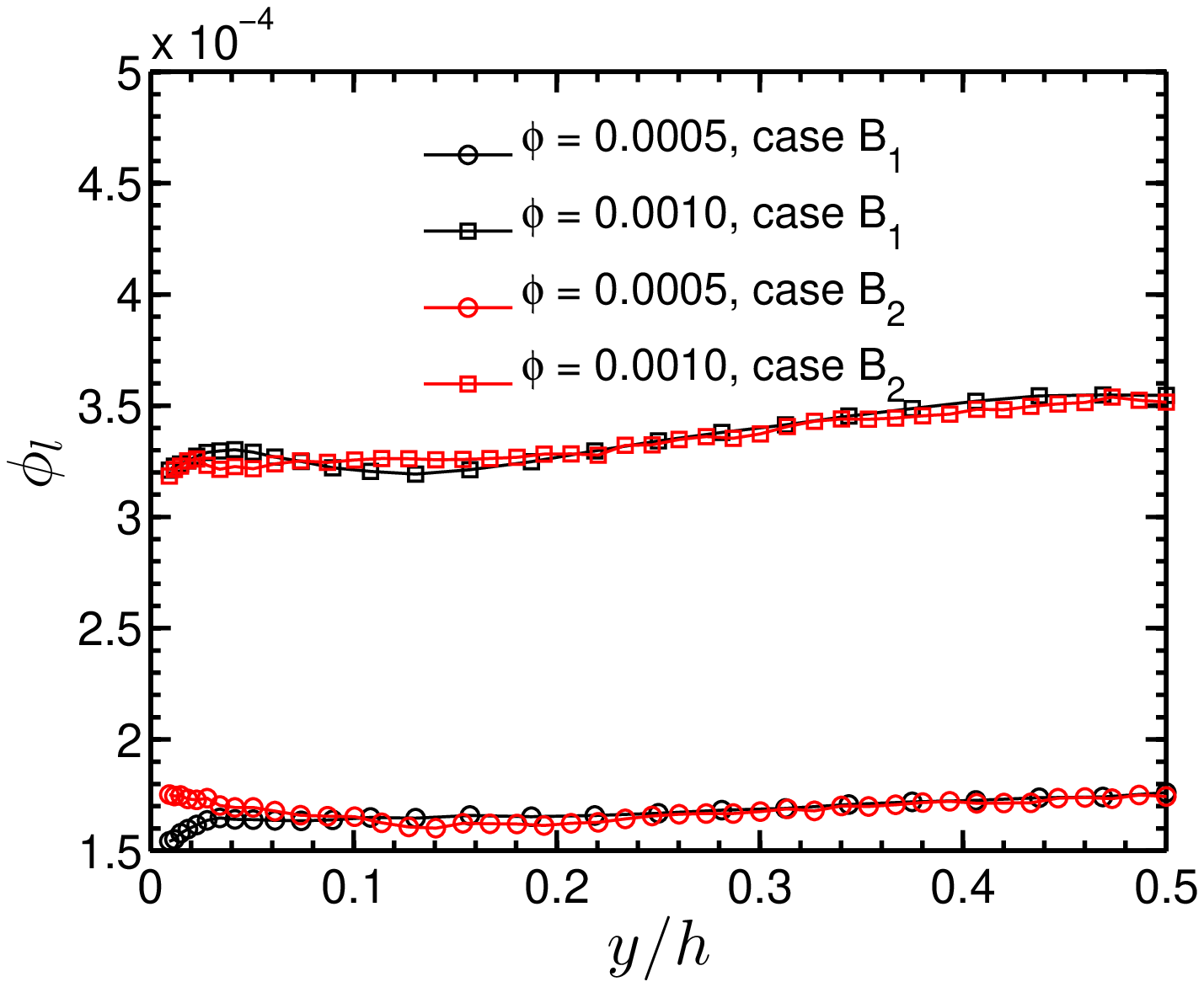}
	\caption{}
	\endminipage
	\end{subfigure} 
	\caption{(a) The particle Reynolds number and (b) local particle volume fraction in the wall-normal direction for cases $B_1$ and $B_2$ at different volume fractions.}
	\label{reason_Re}
\end{figure*}
\begin{equation}
F_p \propto \delta^3.
\end{equation}
Thus, a change in $2\delta/d_p$ from 54 to 117 results in almost one order of magnitude change in the $(2\delta/d_p)^3$, which is much higher than the variation of other parameters. While in the case of constant $2\delta/d_p$ when $Re_b$ is changed, there is no significant variation in $Re_p$, $\phi_l$, and $f$ over a range of volume fractions (Fig.~\ref{reason_Re}). Then,   using Eqn.~\ref{scaled_Fp} we can write, 
\begin{equation}
F_p \propto \frac{1}{Re_b^2}.
\end{equation}
Here, an increase in $Re_b$ from 5600 to 13750 leads to a nearly six-fold decrease in the $1/Re_b^2$ term, which is higher than the variation of other parameters. Therefore, the above formulation indicates that $2\delta/d_p$ and $Re_b$ are the dominating factors in controlling the extent of turbulence modulation compared to the other parameters.

\subsection{Two point statistics: Spatial correlation}

The spatial correlation of the fluid fluctuations qualitatively describes the length scale associated with the structure of the fluid phase. The spatial correlation of fluid fluctuation in the streamwise ($ x $) direction is defined as,
\begin{equation}
r_{ij}(r,t) = \overline{u'_i(x,t)u'_j(x+r,t)}.
\end{equation}
Where $r$ is the distance between the two points, and $i$ and $j$ are the components of fluctuations. The overbar denotes the ensemble averaging. The spatial correlation coefficient is defined as,
\begin{equation}
R_{ij}(r,t) = \frac{\overline{u'_i(x,t)u'_j(x+r,t)}}{\overline{u'_i(x,t)u'_j(x,t)}}.
\end{equation}

\begin{figure*}
	\begin{subfigure}[b]{1\textwidth}
	\centering
	\minipage{0.4\textwidth}
	\includegraphics[width=\textwidth]{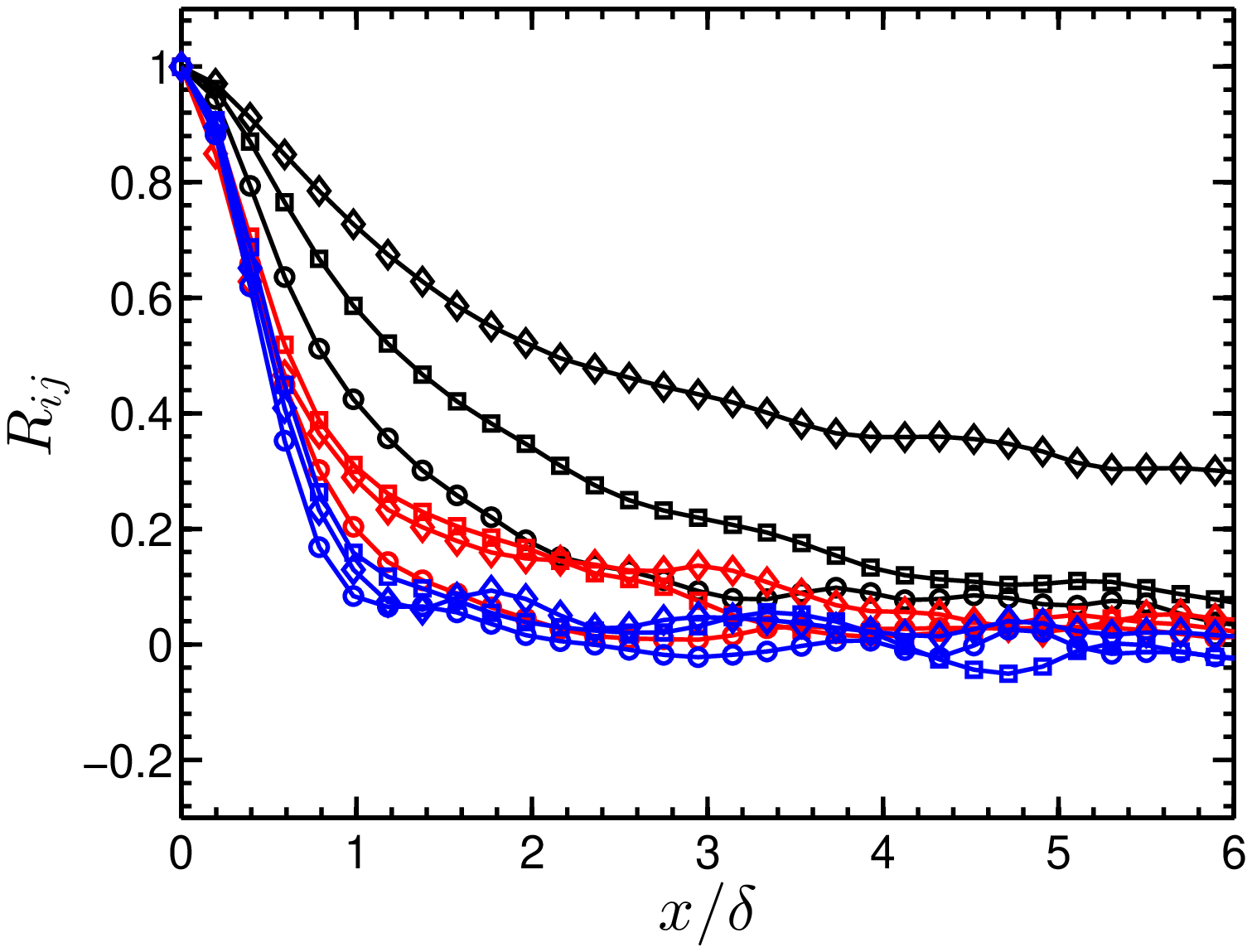}
	\caption{}
	\endminipage
	\minipage{0.4\textwidth}
	\includegraphics[width=\textwidth]{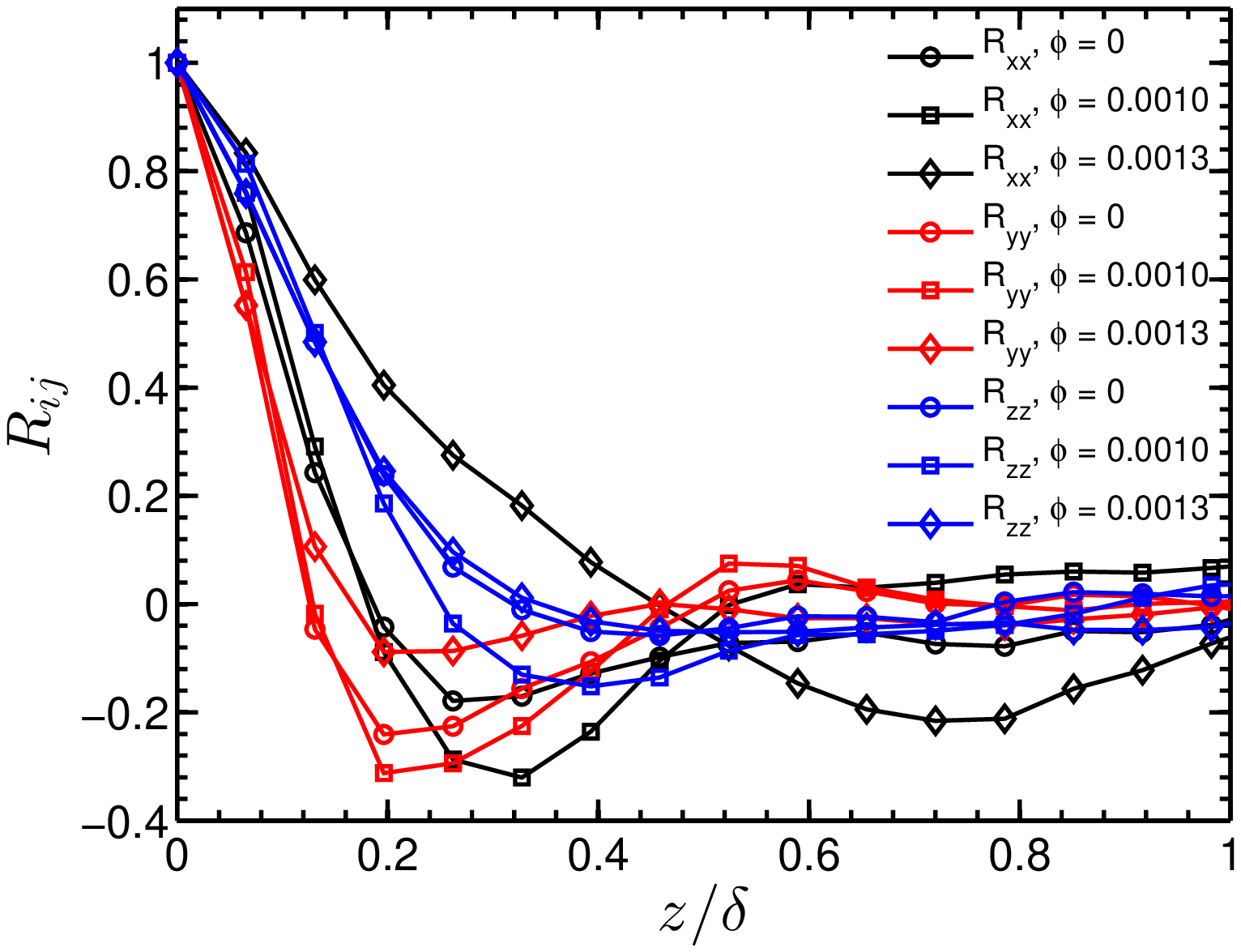}
	\caption{}
	\endminipage \\
	\minipage{0.4\textwidth}
	\includegraphics[width=\textwidth]{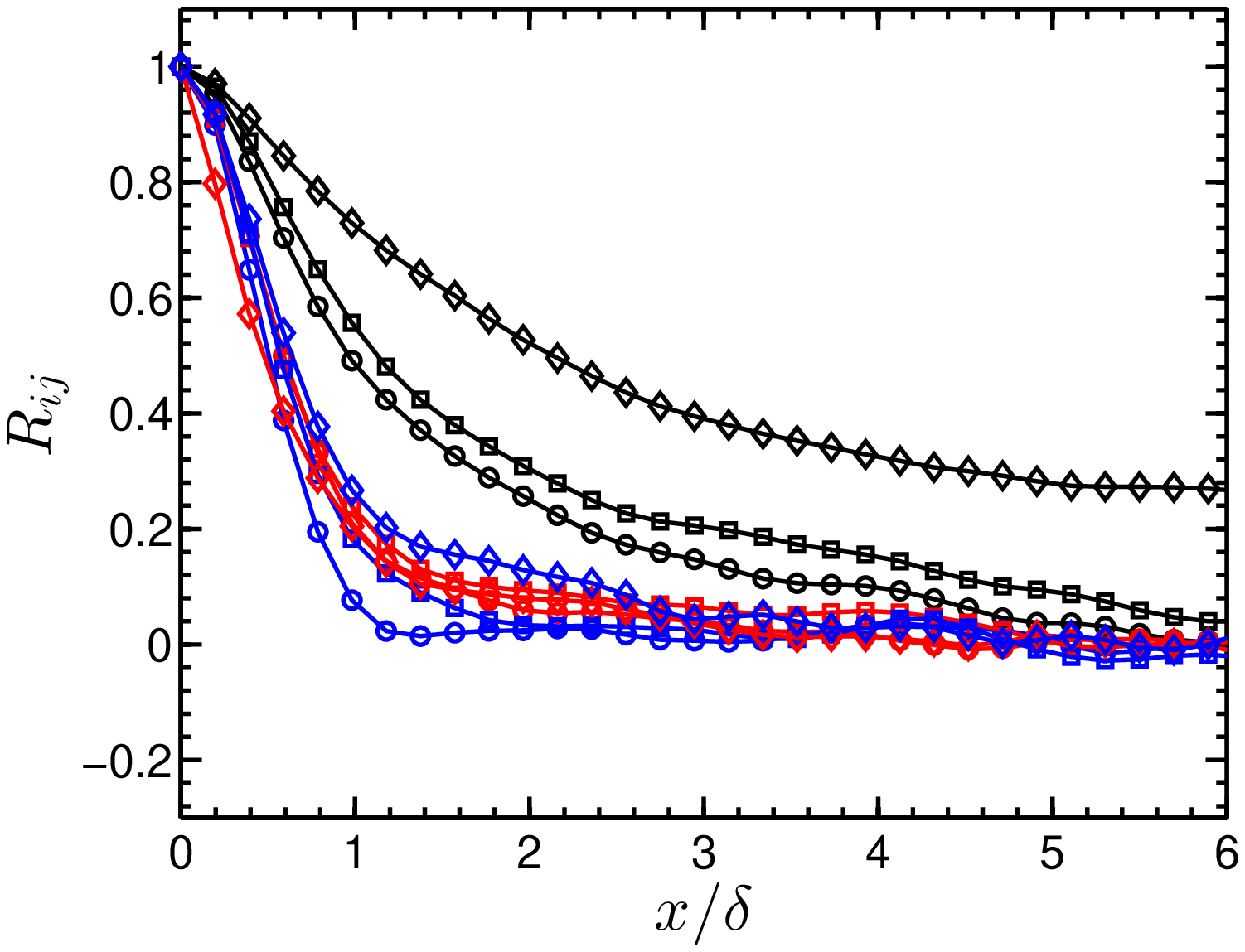}
	\caption{}
	\endminipage
	\minipage{0.4\textwidth}
	\includegraphics[width=\textwidth]{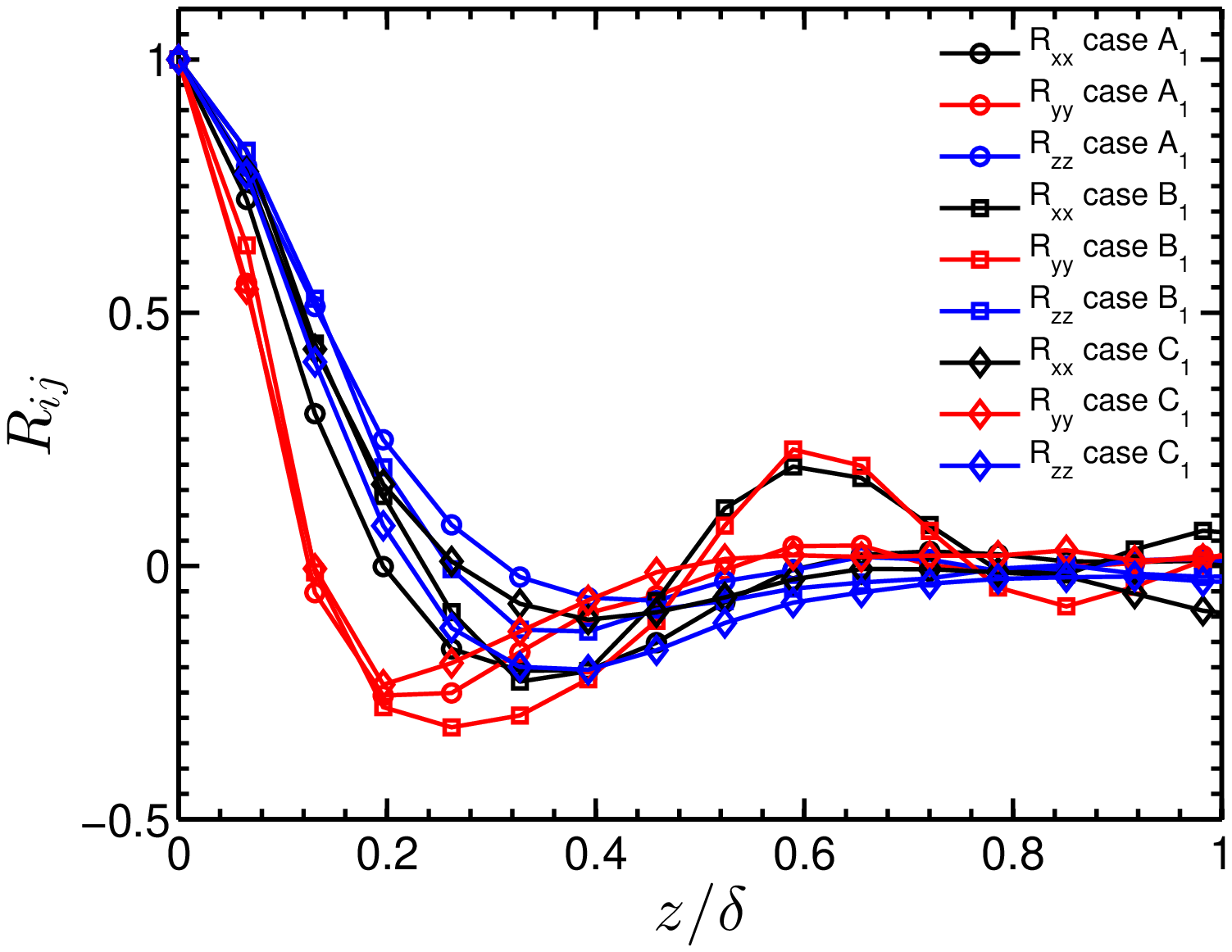}
	\caption{}
	\endminipage\\
	\minipage{0.4\textwidth}
	\includegraphics[width=\textwidth]{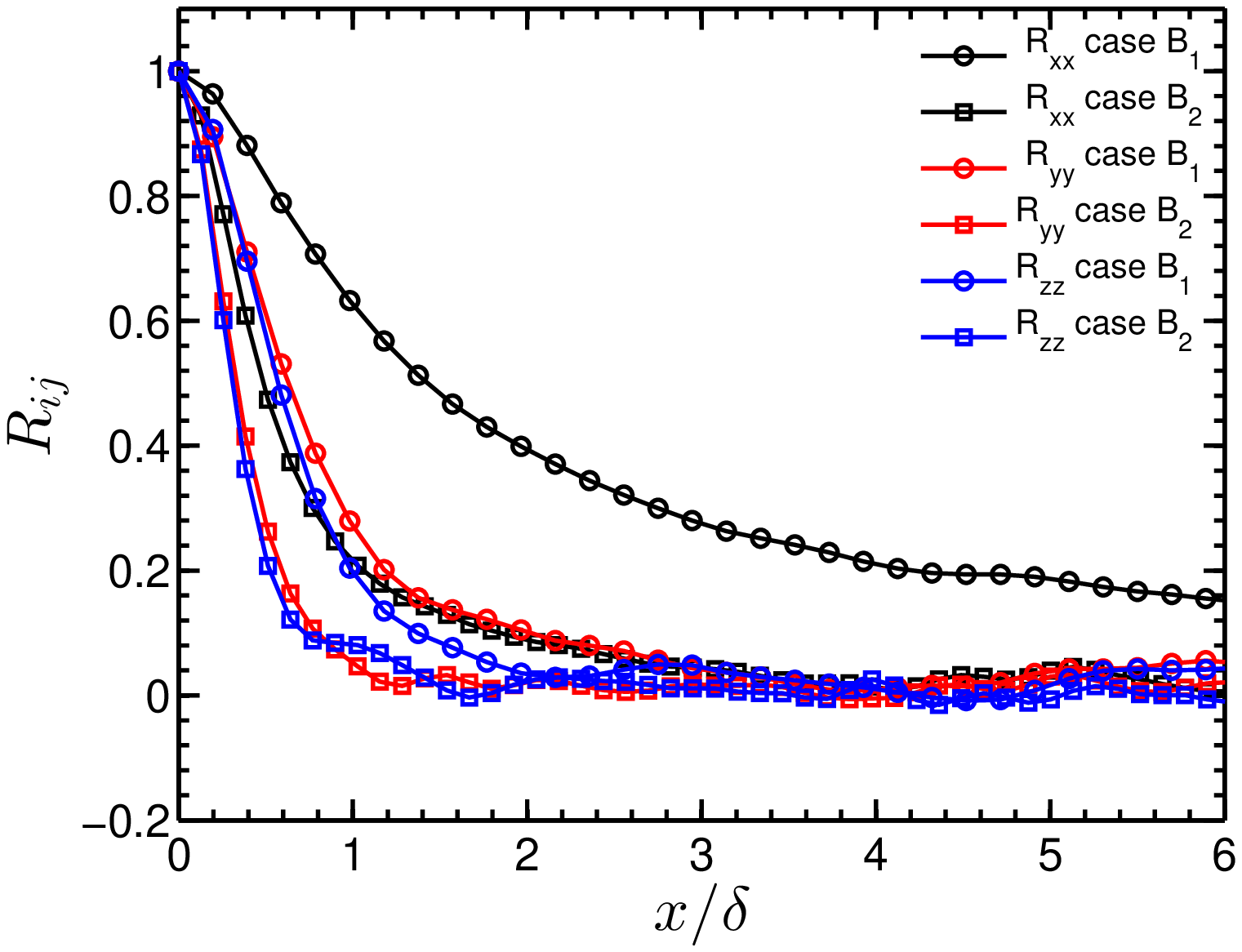}
	\caption{}
	\endminipage
	\minipage{0.4\textwidth}
	\includegraphics[width=\textwidth]{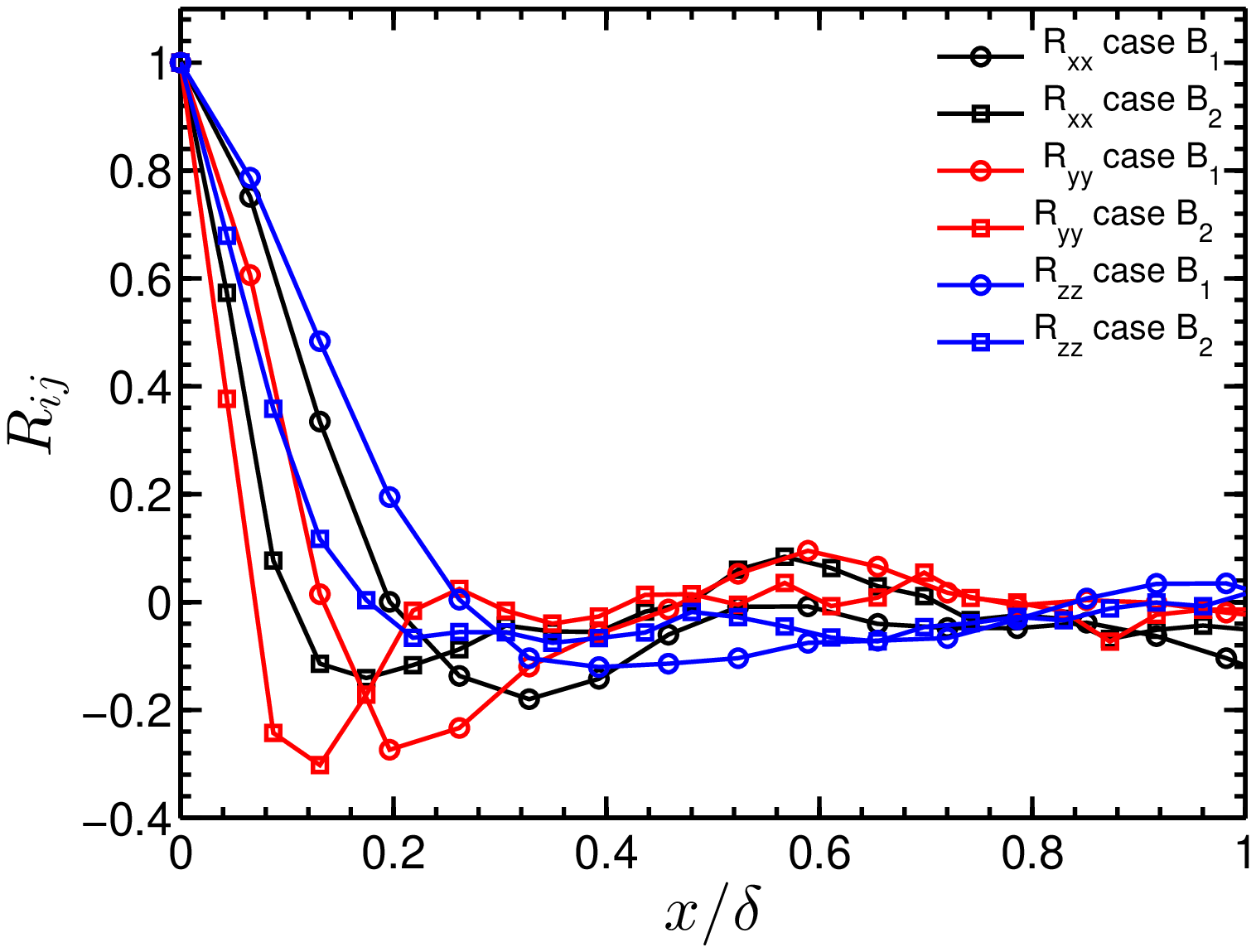}
	\caption{}
	\endminipage
	\end{subfigure} 
	\caption{The spatial correlation coefficient is plotted in the streamwise direction (Fig.~(a, c, and e)) and spanwise direction (Fig.~(b, d, and f). (a and b) $R_{ij}$ is plotted for different volume fractions for case $A_1$. (c and d) $R_{ij}$ is plotted for cases $A_1$, $B_1$ and $C_1$ for $\phi = 4\times 10^{-4}$. (e and f) $R_{ij}$ is plotted for cases $B_1$ and $B_2$ for $\phi = 5\times 10^{-4}$. The legends for Fig.~(a) are the same as in (b), and for Fig.~(c) are the same as in (d).}
	\label{spatial_corre}
\end{figure*}

$R_{ij}$ is plotted for streamwise and spanwise directions in the near-wall ($y^+ = 15$) regime for a range of particle volume fractions and shown in Fig.~\ref{spatial_corre}. In Fig.~\ref{spatial_corre} (a and b), the correlation coefficient is plotted for $\phi = 0$, $1.0 \times 10^{-3}$, and $1.3\times 10^{-3}$ for the case $A_1$. For the streamwise direction, the correlation coefficient of the streamwise component ($R_{xx}$) decays slower than the $R_{yy}$ and $R_{zz}$ as shown Fig.~\ref{spatial_corre} (a). The decay of the correlation coefficient significantly decreases with an increase in the volume fraction from $\phi = 0$ to $1.3\times 10^{-3}$ for each of the components. However, the variation in decay rate is more pronounced for the streamwise component. Such an observation indicates that the turbulent structures become lengthier with an increase in volume fraction. A decrease in the correlation coefficient for all the components is observed when $\phi = 1.5 \times 10^{-3}$ (for the case $A_1$) as the turbulence has collapsed for this volume loading (not shown here). For the spanwise spatial correlation coefficient in Fig.~\ref{spatial_corre} (b), no significant variation is observed in the correlation coefficient for wall-normal and spanwise components. However, an increase is observed for the streamwise component with an increase in volume loading, which signifies that the vortical structures become wider in the spanwise direction.

In Fig.~\ref{spatial_corre} (c and d), the effect of channel dimension on the correlation coefficient is analyzed for $Re_b = 5600$ and $\phi = 5\times 10^{-4}$. In the streamwise direction for streamwise and spanwise velocity fluctuations, the correlation decay becomes slower with an increase in $2\delta/d_p$, Fig.~\ref{spatial_corre} (c). Such an observation is a signature of the lengthier structure with increased channel dimensions for a fixed Reynolds number. However, in the case of correlations in the spanwise direction, the wall-normal fluctuations decay faster, and there is almost no significant modification of the dimension of turbulent streaks. The effect of Reynolds number on the correlation coefficients for the cases $B_1$ and $B_2$ are plotted in Fig.~\ref{spatial_corre} (e and f). Interestingly, the correlation coefficient decay faster for the higher Reynolds number of $Re_b = 13750$ compared to the $Re_b = 5600$ for both the directions and all the components of velocity fluctuations. This suggests that the vortical structures are smaller in dimensions at a high Reynolds number. In this context, it is worth mentioning that for $2\delta/d_p = 20$, $\phi = 2.36\times10^{-3}$ and $Re_p <50$, \citet{yu2021modulation} have reported less turbulence attenuation for higher Reynolds number of $Re_b = 12000$ than the $Re_b = 5746$. They have commented that at a high Reynolds number, the large-scale vortices, although smaller in size, are stronger than those of a low Reynolds number. \citet{zhou2020non} have also reported that the vortical structures become weaker and fewer with an increase in mass loading. The authors also found that the spacing between the streaks increases with the addition of particles.

\begin{figure*}
	\begin{subfigure}[b]{1\textwidth}
	\centering
	\minipage{0.5\textwidth}
	\includegraphics[width=\textwidth]{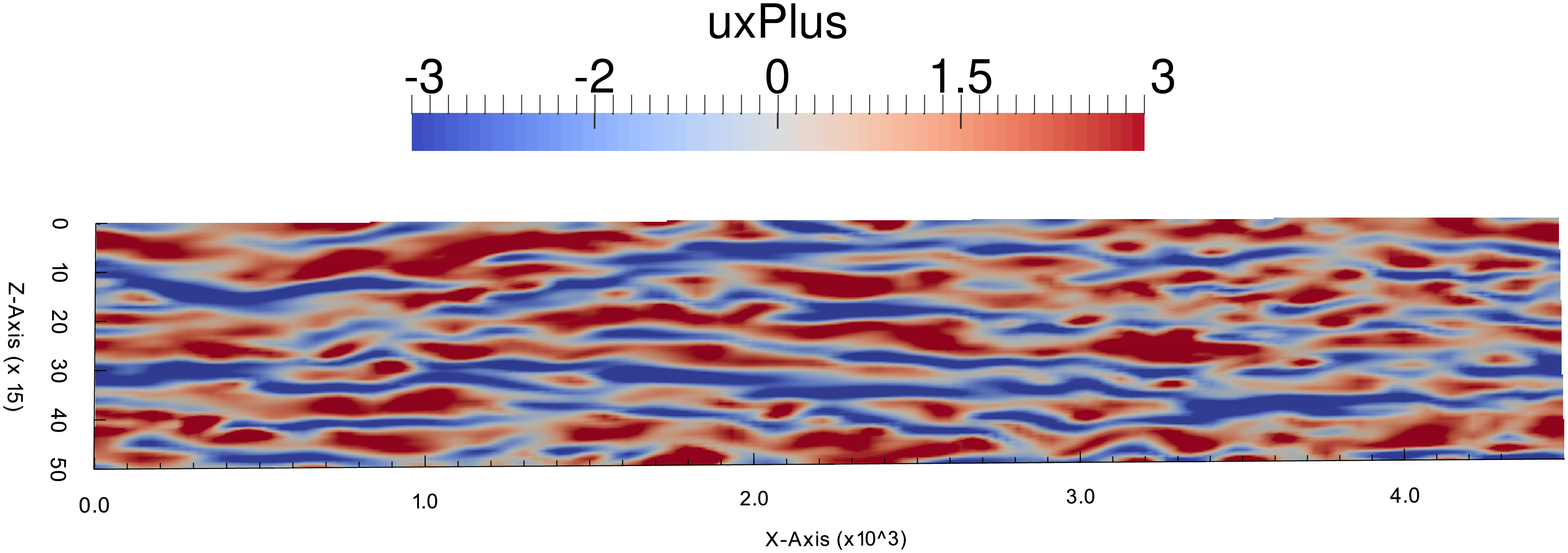}
	\vspace{-2cm}
	\caption{}
	\endminipage
	\minipage{0.5\textwidth}
	\includegraphics[width=\textwidth]{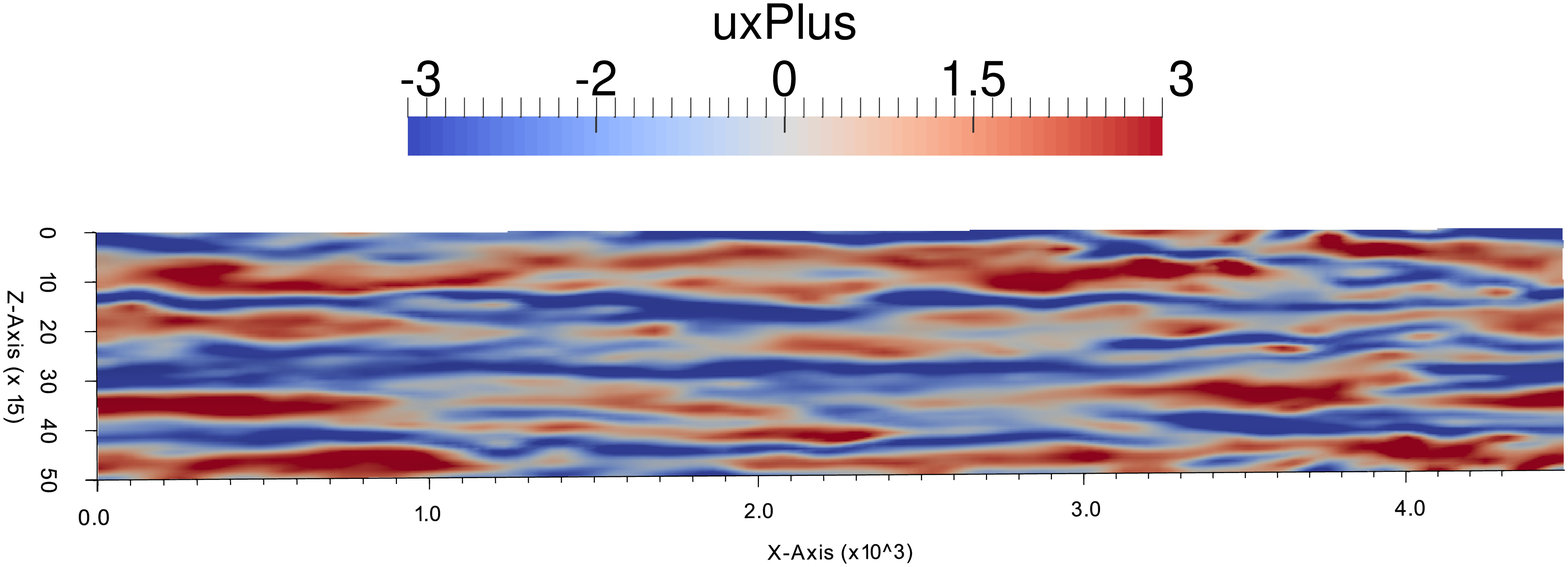}
	\vspace{-2cm}
	\caption{}
	\endminipage
	\end{subfigure} 
	\caption{ The contours for instantaneous streamwise fluctuations (uxPlus is  $\widetilde{u}'_x/u_\tau$) plotted for case $A_1$ at (a) $\phi = 4\times10^{-4}$, and (b) $\phi = 10^{-3}$ in x-z plane at $y^+ = 15$. The streamwise and spanwise distances are normalized with viscous scales.}
	\label{streaks_Ldp}
	\end{figure*}
	\begin{figure*}
	\begin{subfigure}[b]{1\textwidth}
	\centering
	\minipage{0.49\textwidth}
	\includegraphics[width=\textwidth]{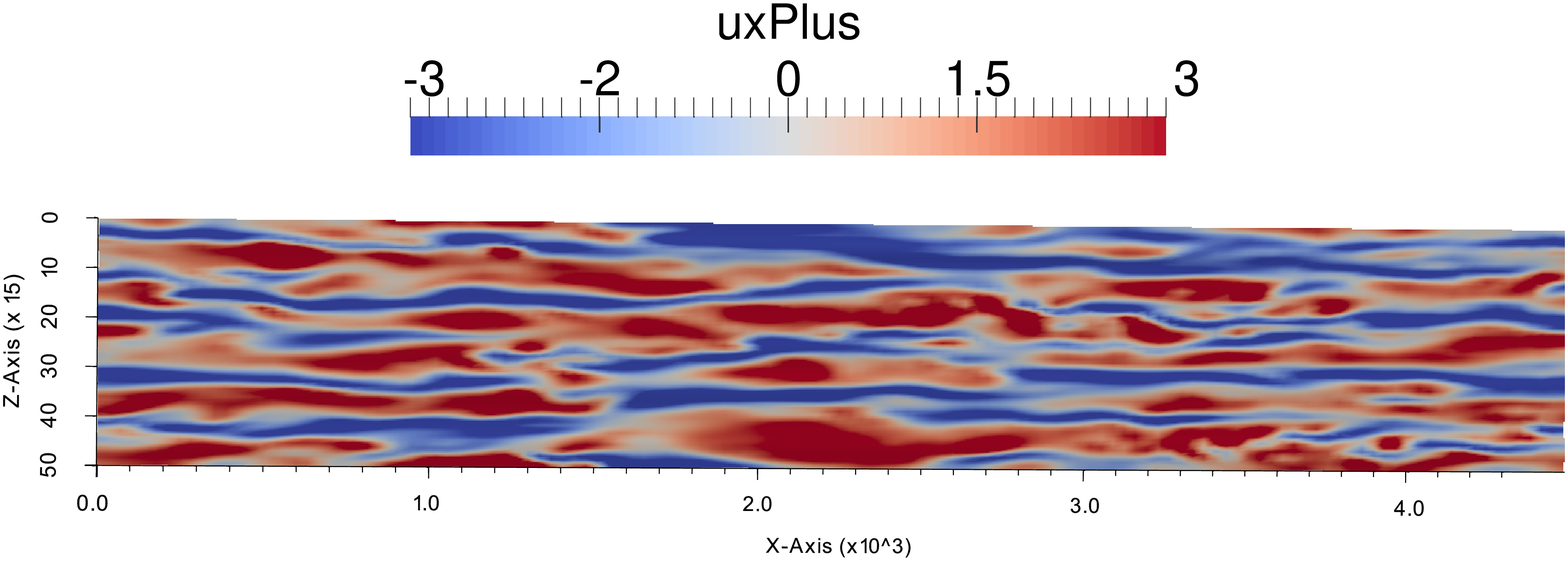}
	\vspace{-2cm}
	\caption{}
	\endminipage
	\minipage{0.49\textwidth}
	\includegraphics[width=\textwidth]{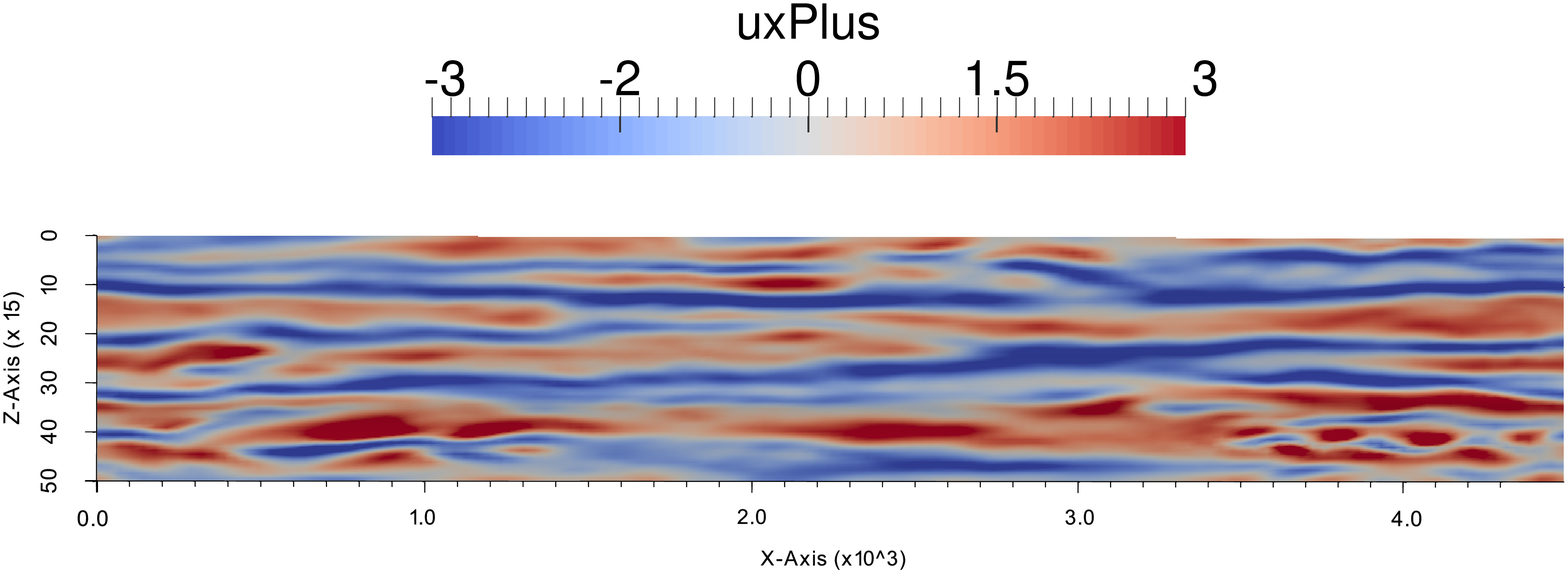}
	\vspace{-2cm}
	\caption{}
	\endminipage 
	\end{subfigure} 
	\caption{ The contours for instantaneous streamwise fluctuations (uxPlus is  $\widetilde{u}'_x/u_\tau$) plotted for $\phi = 4\times10^{-4}$ (a) case $B_1$, and (b) case $C_1$ in x-z plane at $y^+ = 15$. The streamwise and spanwise distances are normalized with viscous scales.}
	\label{streaks_Ldp2}
\end{figure*}

In Figs.~\ref{streaks_Ldp}-\ref{streaks_Ldp2}, the contours are plotted for instantaneous streamwise velocity fluctuations ($\widetilde{u}'_x$) in the near-wall region ($y^+ = 15$) to depict the size and strength of the vortical structures. The streamwise fluctuations are scaled with respective unladen friction velocity ($u_\tau$). In Fig.~\ref{streaks_Ldp} (a), the low and high-speed streaks are shown in the $ x-z $ plane for the case $A_1$ at lower volume fraction ($ 4 \times 10^{-4} $). The streamwise and spanwise dimensions are provided in viscous units. It is observed that the length of the high-speed streaks (red regime) is around 2000 viscous units, whereas low-speed streaks (blue regime) are extended up to complete channel length. For the same channel in Fig.~\ref{streaks_Ldp} (b), the streaks become fewer and wider as the volume fraction is increased to $10^{-3}$. The high-speed streaks become more elongated and extend up to 3500 viscous units, while there is no significant change in the lengths of low-speed streaks. The effect of $2\delta/d_p$ on the high-low speed streaks for a volume fraction of $\phi =4 \times 10^{-4}$ is shown in Figs.~\ref{streaks_Ldp}~(a) and \ref{streaks_Ldp2}. It is observed that the high-speed streaks become wider and fewer as the $2\delta/d_p$ is increased while keeping $\phi$ constant. The length of high-speed streaks extends up to 3500 viscous units from case $A_1$ to $C_1$.

\begin{figure*}
	\begin{subfigure}[b]{1\textwidth}
	\centering
	\minipage{0.5\textwidth}
	\includegraphics[width=\textwidth]{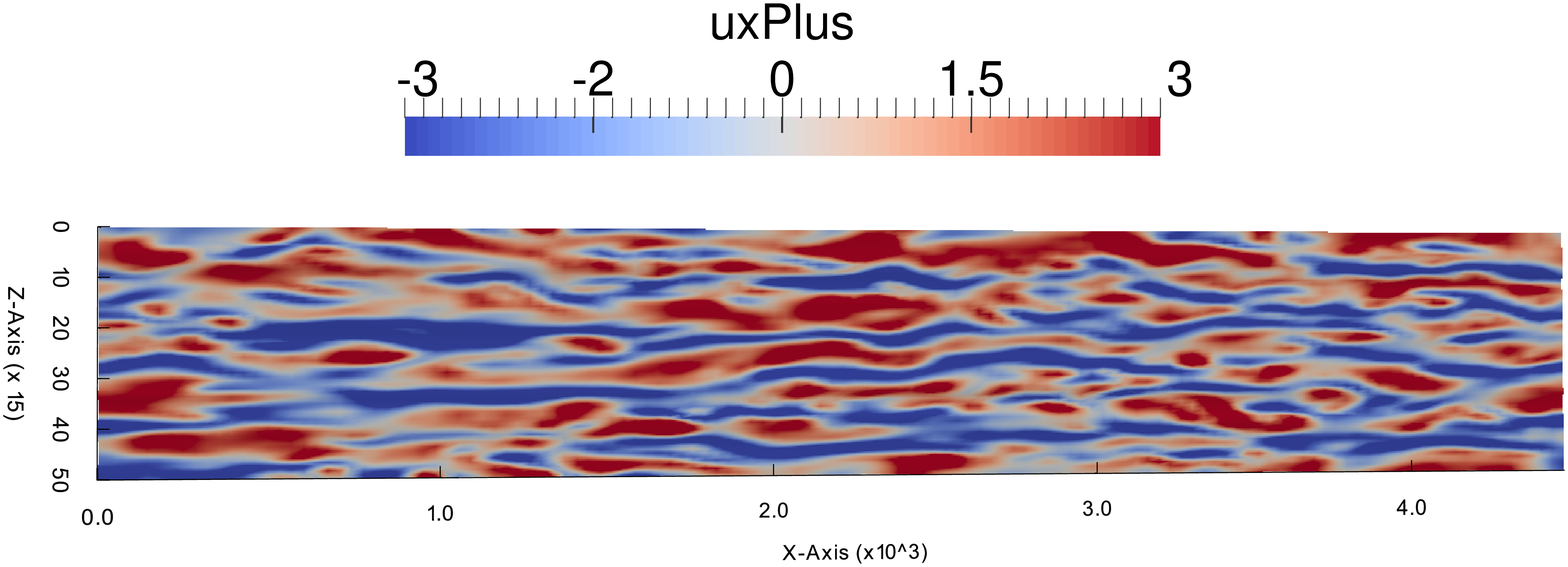}
	\vspace{-2cm}
	\caption{}
	\endminipage
	\minipage{0.5\textwidth}
	\includegraphics[width=\textwidth]{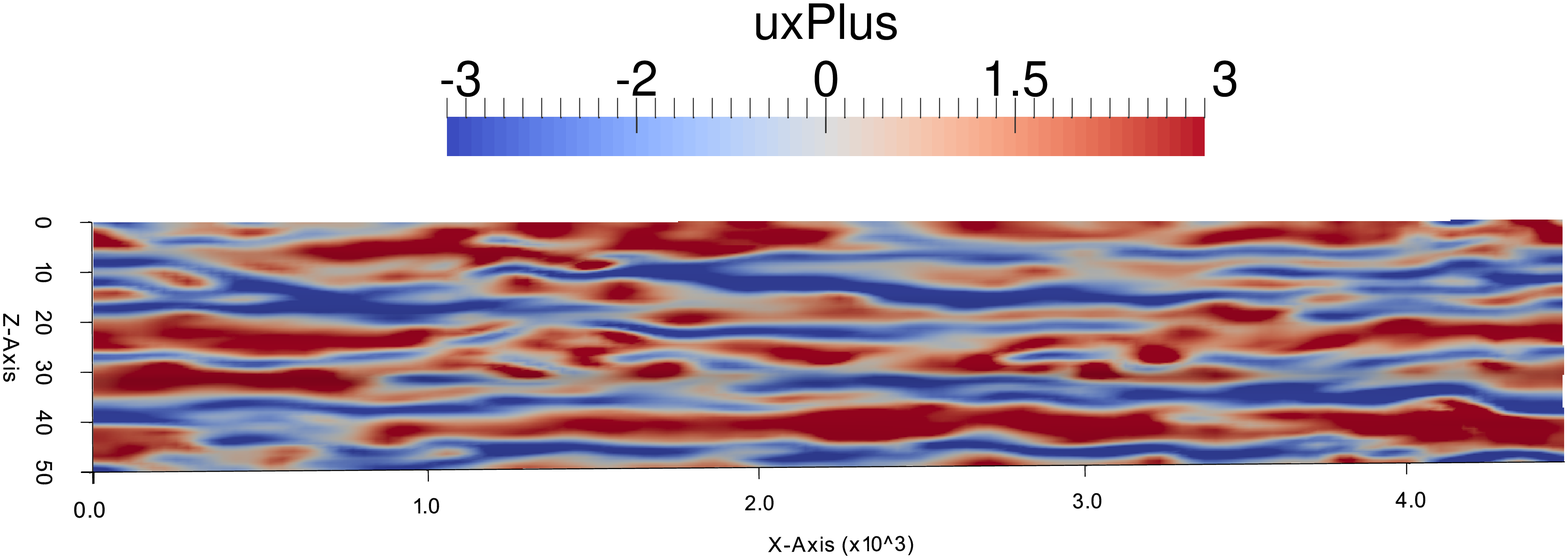}
	\vspace{-2cm}
	\caption{}
	\endminipage \\
	\minipage{0.49\textwidth}
	\includegraphics[width=\textwidth]{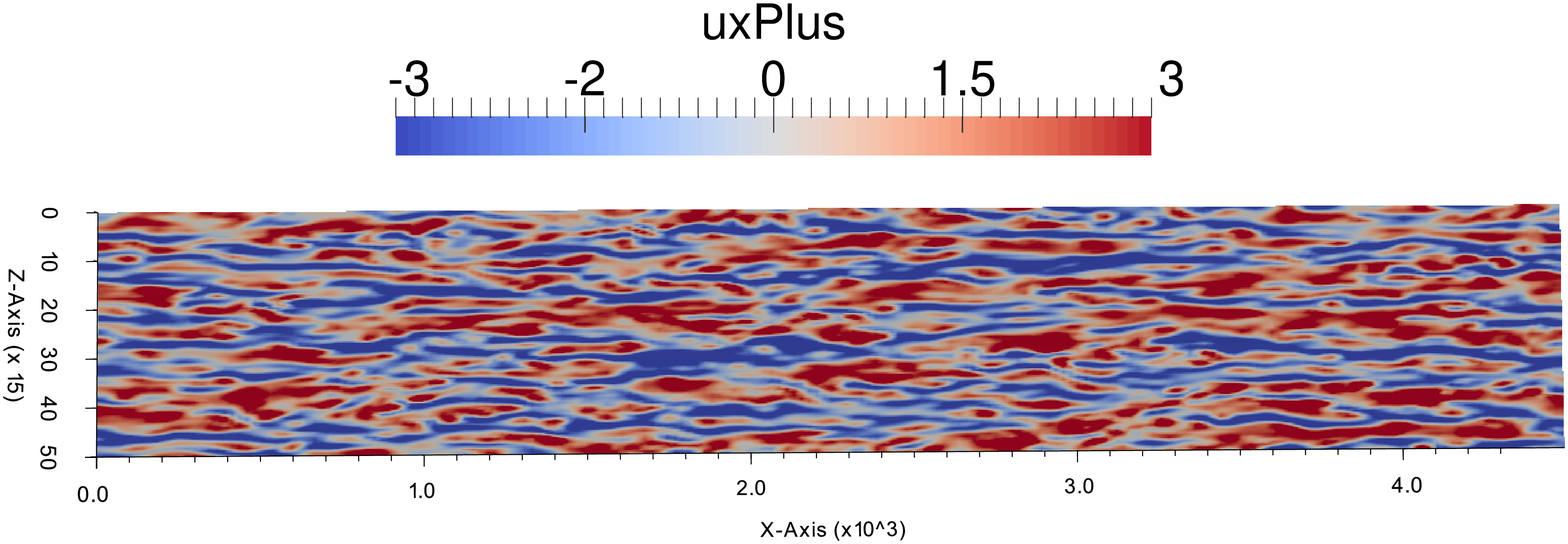}
	\vspace{-2cm}
	\caption{}
	\endminipage
	\minipage{0.49\textwidth}
	\includegraphics[width=\textwidth]{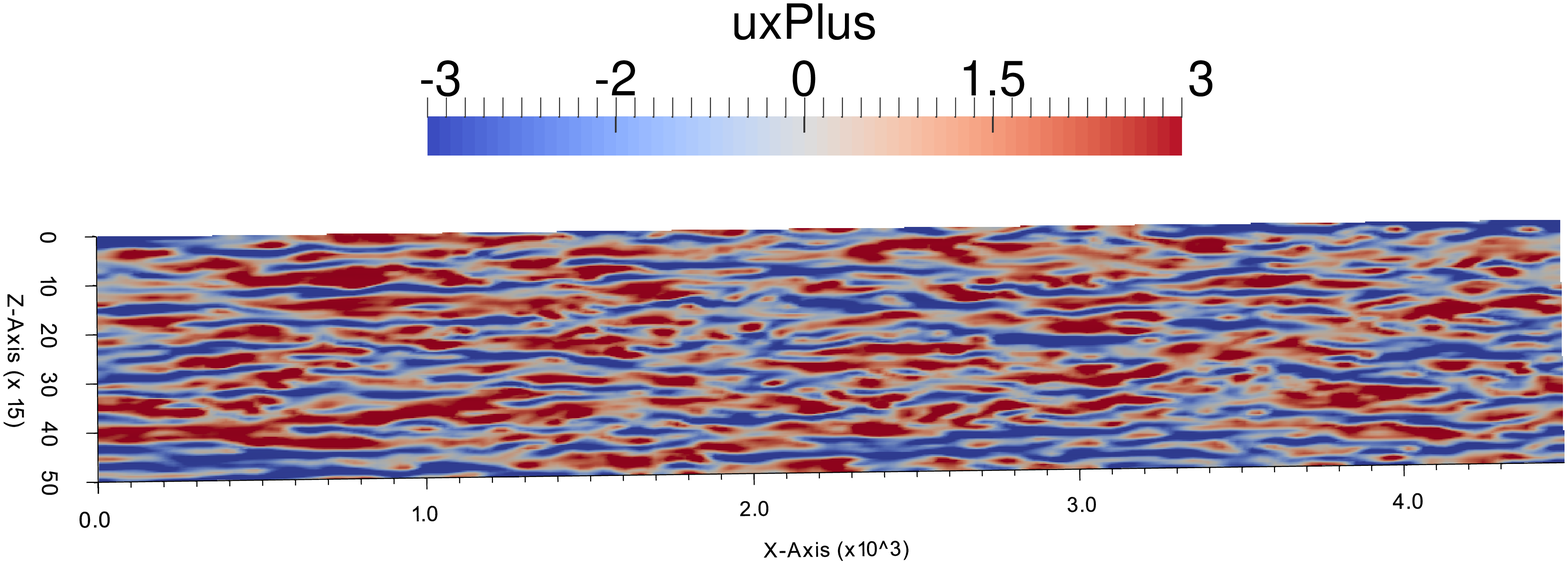}
	\vspace{-2cm}
	\caption{}
	\endminipage 
	\end{subfigure} 
	\caption{ The contours for instantaneous streamwise fluctuations (uxPlus is $\widetilde{u}'_x/u_\tau$) plotted for (a) case $B_1$, $\phi = 0$, (b) case $B_1$, $\phi = 5\times 10^{-4}$, (c) case $B_2$, $\phi = 0$, and (d) case $B_2$, $\phi = 5\times 10^{-4}$ in x-z plane at $y^+ = 15$. The streamwise and spanwise distances are normalized with viscous scales.}
	\label{streaks_Re}
\end{figure*}

\begin{figure*}[htb]
	\begin{subfigure}[b]{1\textwidth}
	\centering
	\minipage{0.45\textwidth}
\includegraphics[width=\textwidth]{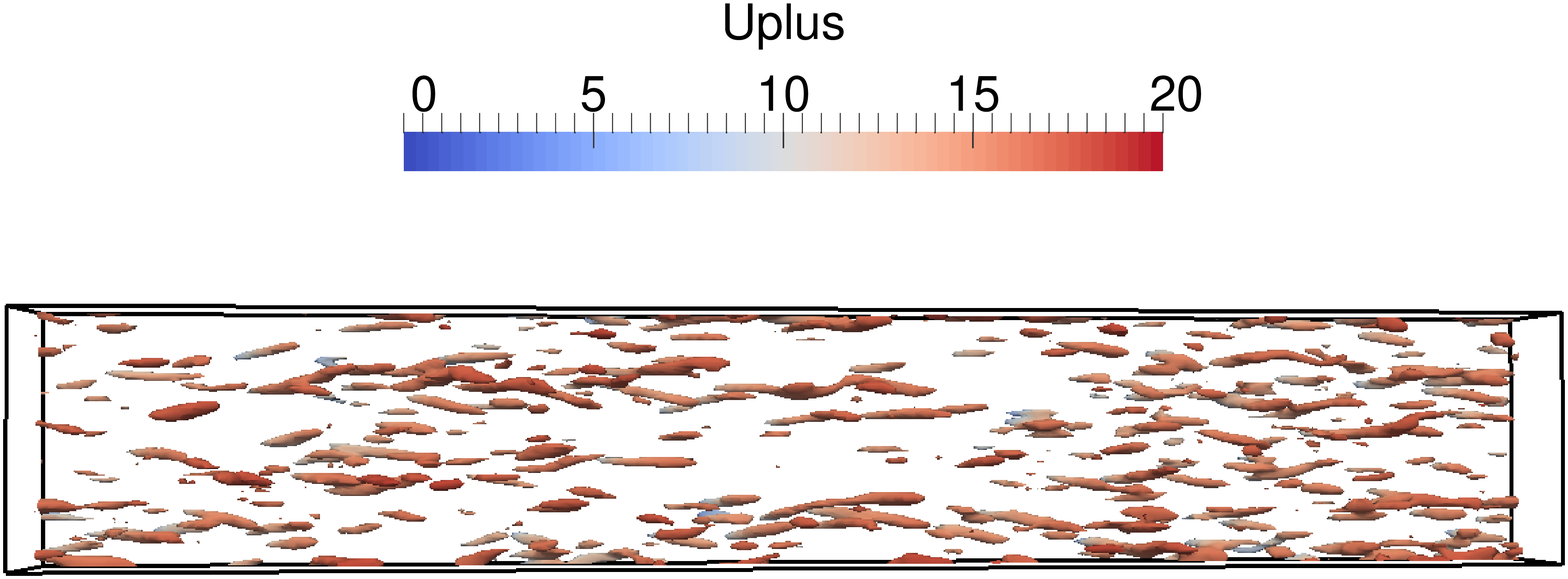}
	\vspace{-2cm}
	\caption{}
	\endminipage \\
	\minipage{0.45\textwidth}
\includegraphics[width=\textwidth]{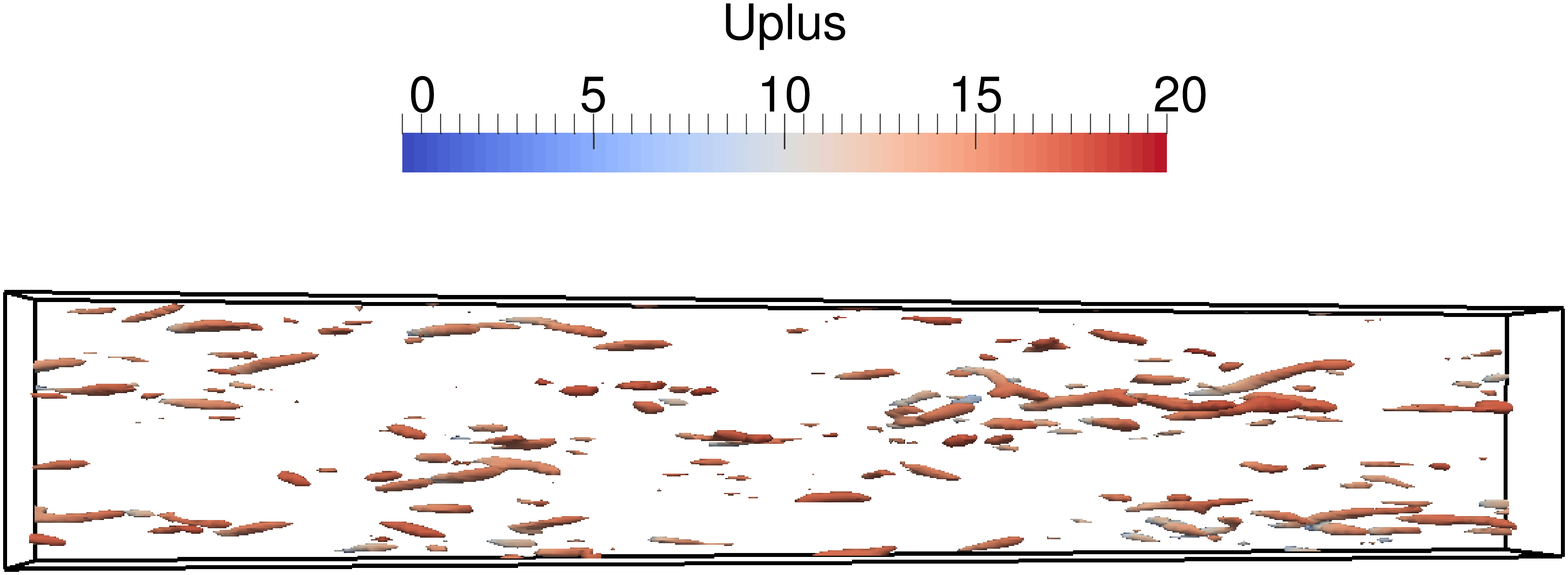}
	\vspace{-2cm}
	\caption{}
	\endminipage
	\minipage{0.45\textwidth}
\includegraphics[width=\textwidth]{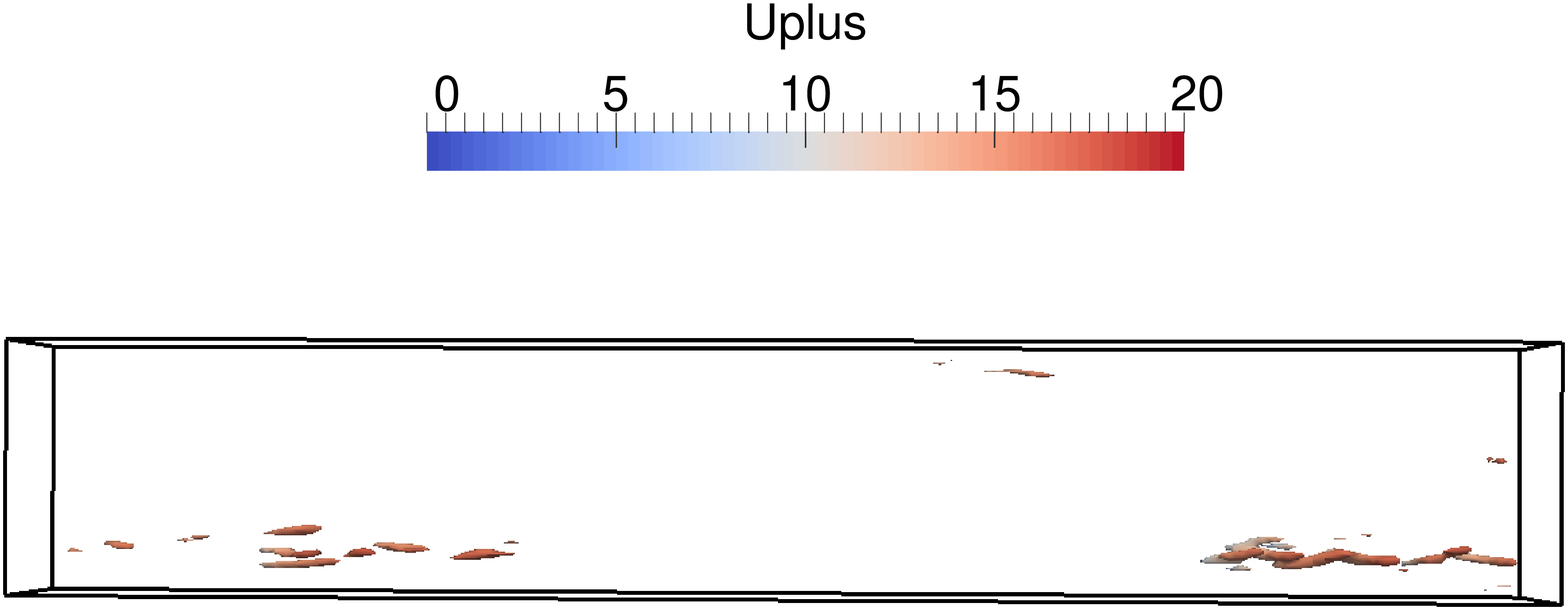}
	\vspace{-2cm}
	\caption{}
	\endminipage
	\end{subfigure} 
	\caption{The $-\lambda_2^+ = 0.0025$ isocontours for cases (a) $A_1$, (b) $B_1$, and (c) $C_1$ at $\phi = 4\times10^{-4}$ in the near-wall x-z plane. The colors show the magnitude of the mean fluid velocity (Uplus is $\overline{u}/u_\tau$) within the contours.}
	\label{Lambda2_Ldp}
\end{figure*}

In Fig.~\ref{streaks_Re}, the normalized contours for instantaneous streamwise fluctuations are shown for cases $B_1$ and $B_2$. The contours are plotted in the $ x-z $ plane at the near-wall region ($y^+ = 15$) for unladen flow and a volume fraction of $5\times10^{-4}$ while channel dimensions are kept constant at $2\delta/d_p = 81$. In Fig.~\ref{streaks_Re}~(a and b) for the case $B_1$, it is observed that the streaks become fewer, lengthier, and fatter with an increase in particle volume loading. However, the streaks are smaller, thinner, and closely spaced for $Re_b = 13750$ when compared to similar cases in Fig.~\ref{streaks_Re} (a and c) and Fig.~\ref{streaks_Re} (b and d). It is interesting to note that there is almost no change in the streaks spacing and lengths in case of higher Reynolds number, Fig.~\ref{streaks_Re} (c and d). While streaks become fewer, fatter and the spacing becomes more for lower $Re_b = 5600$ when volume fraction is increased from 0 to $5\times10^{-4}$.

\begin{figure*}
	\begin{subfigure}[b]{1\textwidth}
	\centering
	\minipage{0.45\textwidth}
\includegraphics[width=\textwidth]{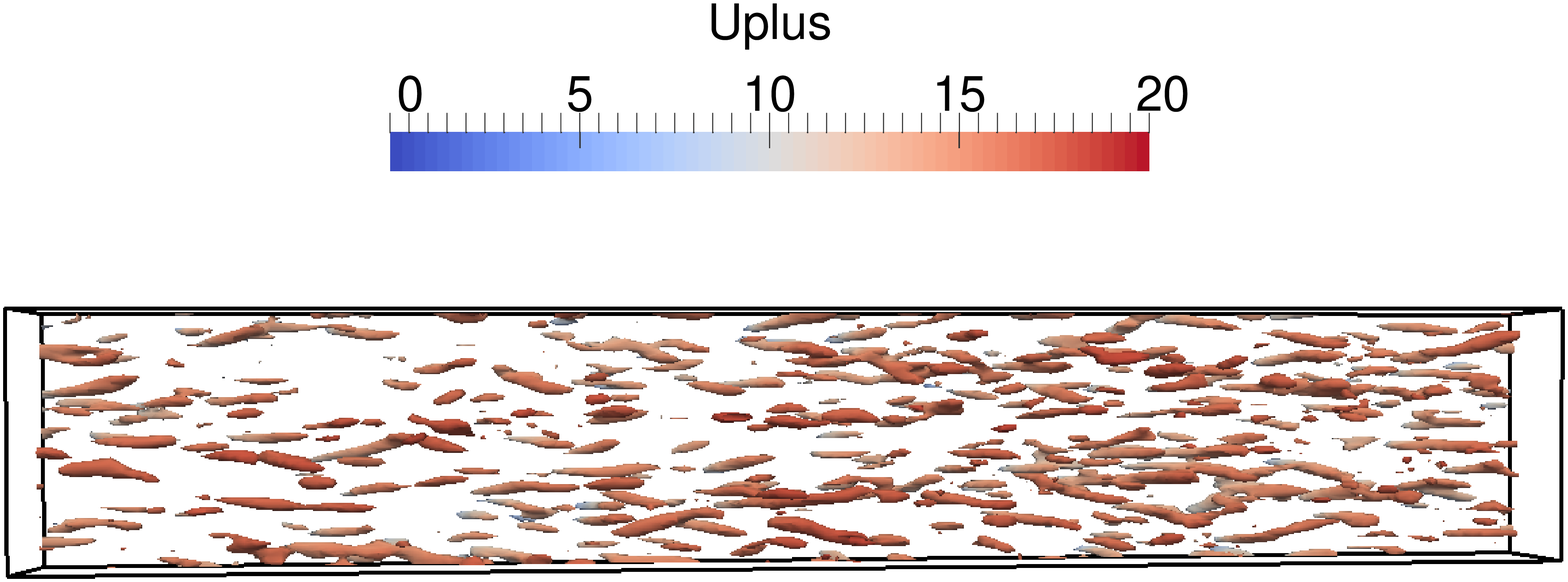}
	\vspace{-2cm}
	\caption{}
	\endminipage
	\minipage{0.45\textwidth}
\includegraphics[width=\textwidth]{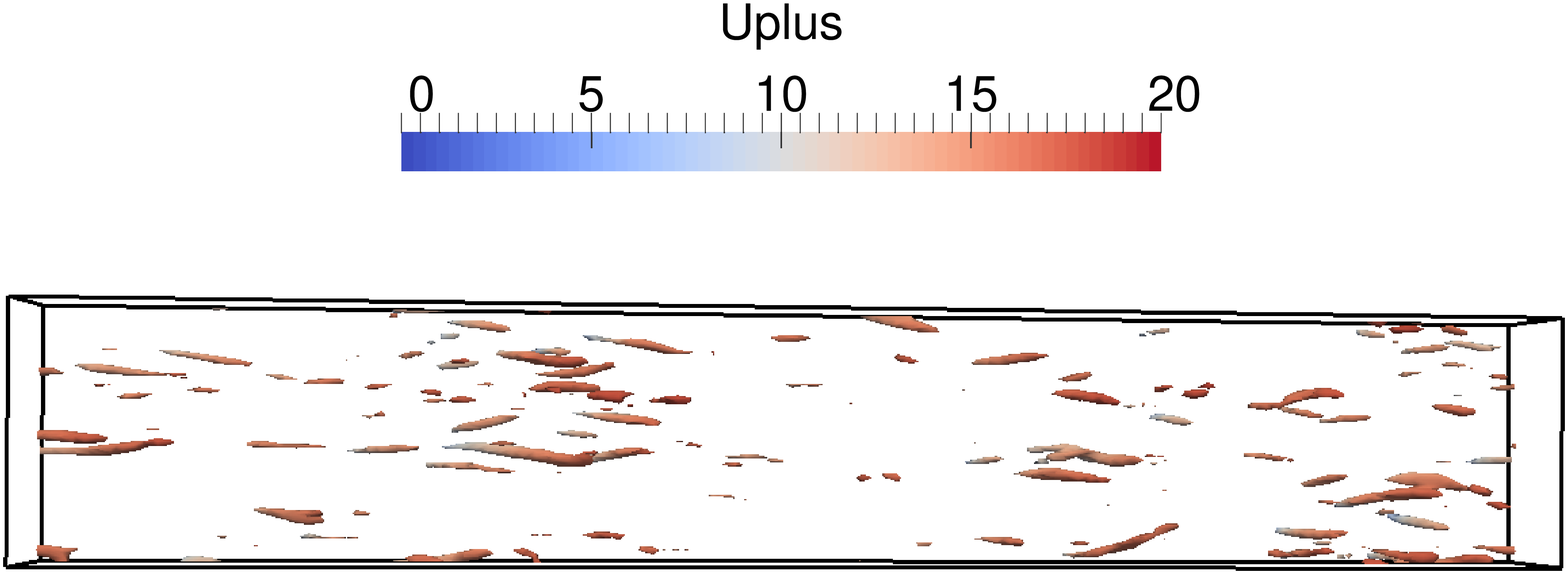}
	\vspace{-2cm}
	\caption{}
	\endminipage \\
	\minipage{0.45\textwidth}
\includegraphics[width=\textwidth]{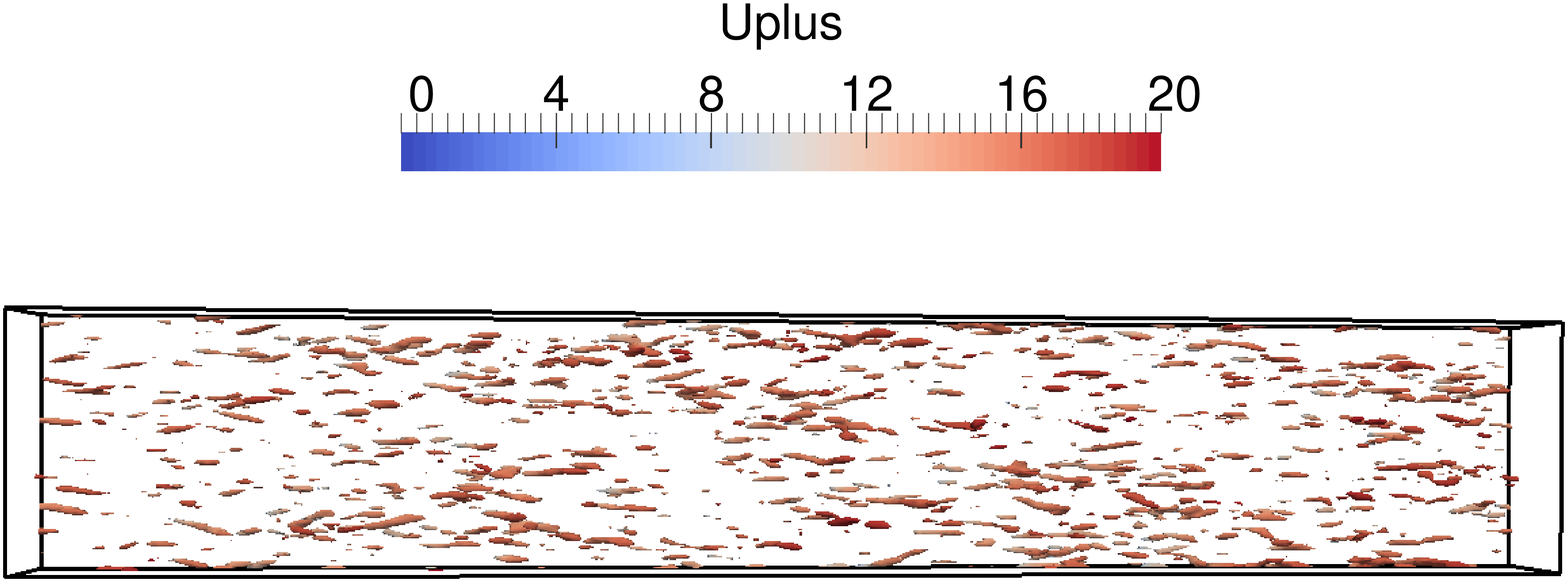}
	\vspace{-2cm}
	\caption{}
	\endminipage
	\minipage{0.45\textwidth}
\includegraphics[width=\textwidth]{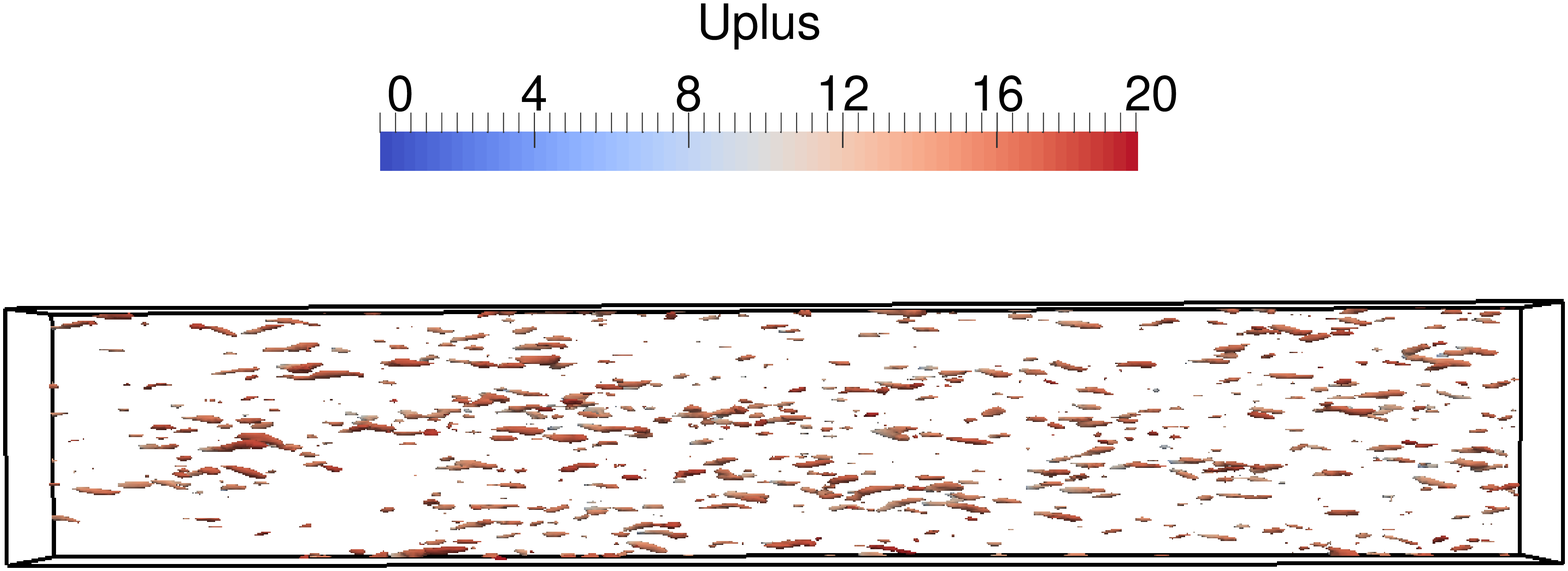}
	\vspace{-2cm}
	\caption{}
	\endminipage
	\end{subfigure} 
	\caption{The $-\lambda_2^+ = 0.0025$ isocontours for (a) case $B_1$, $\phi = 0$, (b) case $B_1$, $\phi = 5\times 10^{-4}$, (c) case $B_2$, $\phi = 0$, and (d) case $B_2$, $\phi = 5\times 10^{-4}$ in the near-wall x-z plane. The colors show the magnitude of the mean fluid velocity (Uplus is $\overline{u}/u_\tau$) within the contours.}
	\label{Lambda2_Re}
\end{figure*}

The coherent structures responsible for the turbulent production are analyzed for different $2\delta/d_p$, Reynolds numbers and volume fractions. Different methods have been reported in the literature to analyze the turbulent structures \cite{chong1990general, jeong1995identification, chakraborty2005relationships, dritselis2011large, yu2021investigation}. In the present work, we report the contours of $-\lambda_2$ which is the second largest eigenvalue of $S^2 + \Omega^2$. Here, $S$ is the strain rate tensor, and $\Omega$ is the rotational tensor of the velocity gradient. This method was proposed and discussed in detail by \citet{jeong1995identification}. This method neglects the unsteady straining which can cause minimum pressure without the presence of vortical structures and viscous effects that may eliminate minimum pressure in vortical flows\cite{jeong1995identification}. The isocontours for $-\lambda_2^+ = 0.0025$ shows the effect of volume fraction, system size, and Reynolds numbers, Fig.~\ref{Lambda2_Ldp} and Fig.~\ref{Lambda2_Re}. The (+) symbol indicates that the values are reported in viscous units. The contours are colored according to the fluid mean velocity ($\bar{u}$) normalized with respective unladen friction velocity ($u_\tau$). In Fig.~\ref{Lambda2_Ldp} (a) for case $A_1$, the $-\lambda_2^+ = 0.0025$ structures are shown for lower volume fraction of $4\times10^{-4}$. Here, the length and width of the structures are approximately 200 and 50 viscous units, respectively. However, the $-\lambda_2^+$ structures vanish as the volume fraction reaches near CPVL (not shown here). In Fig.~\ref{Lambda2_Ldp} (a, b and c), It is observed that the $-\lambda_2^+$ structures become fewer as channel dimension is changed from case $A_1$ to case $C_1$ while volume fraction and Reynolds number are kept constant. In Fig.~\ref{Lambda2_Re}, $-\lambda_2^+$ contours are plotted for both the unladen and particle-laden flows for cases $B_1$ and $B_2$. The $-\lambda_2^+ = 0.0025$ contours become smaller and thinner as the Reynolds number is increased from 5600 to 13750 for unladen flows. When particle volume fraction is $5\times10^{-4}$, the number of $-\lambda_2^+$ structures becomes significantly smaller for low Reynolds number.

\subsection{Particle statistics}

\begin{figure*}
	\begin{subfigure}[b]{1\textwidth}
	\centering
	\minipage{0.4\textwidth}
	\includegraphics[width=\textwidth]{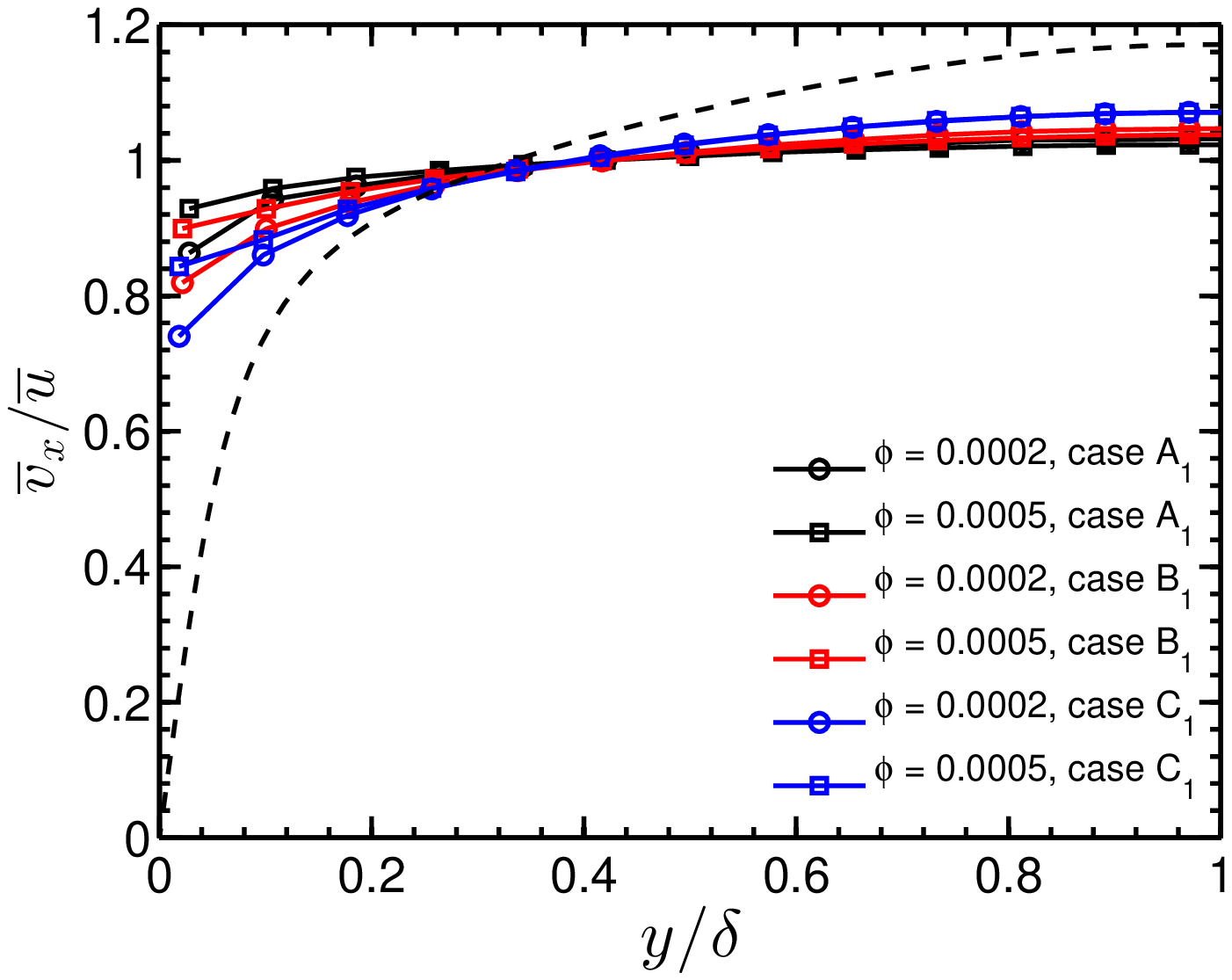}
	\caption{}
	\endminipage 
	\minipage{0.4\textwidth}
	\includegraphics[width=\textwidth]{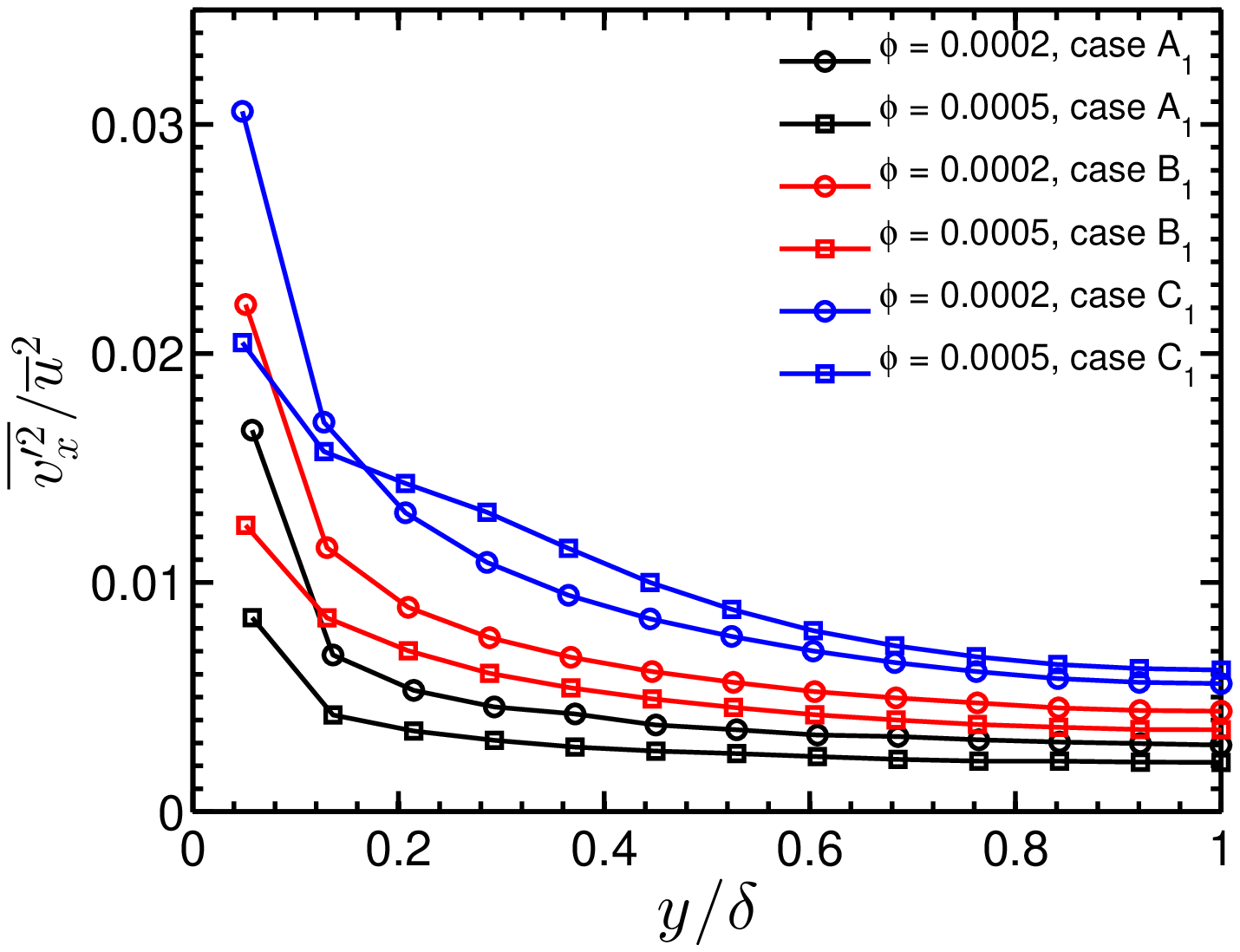}
	\caption{}
	\endminipage \\
	\minipage{0.4\textwidth}
	\includegraphics[width=\textwidth]{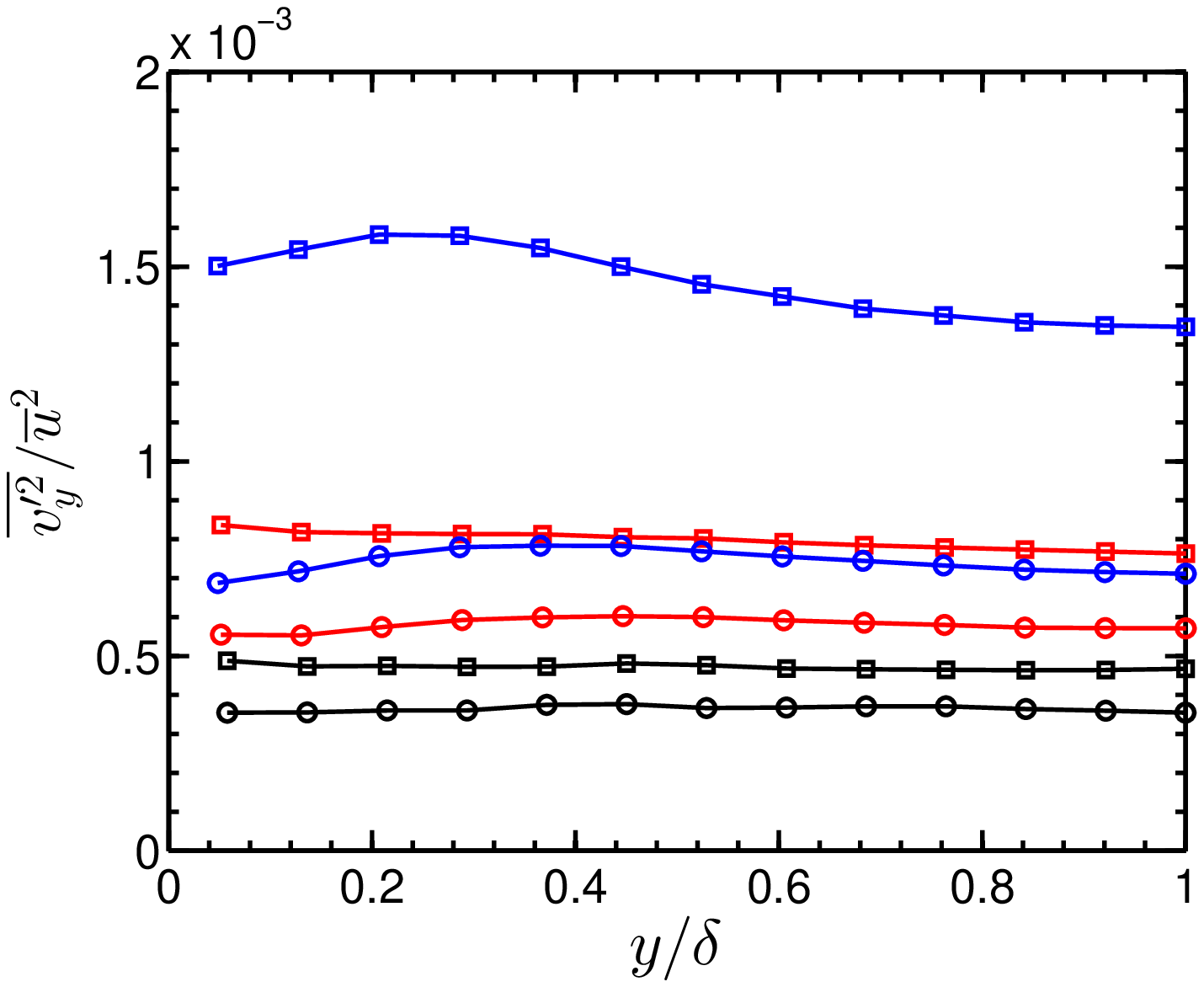}
	\caption{}
	\endminipage	
	\minipage{0.4\textwidth}
	\includegraphics[width=\textwidth]{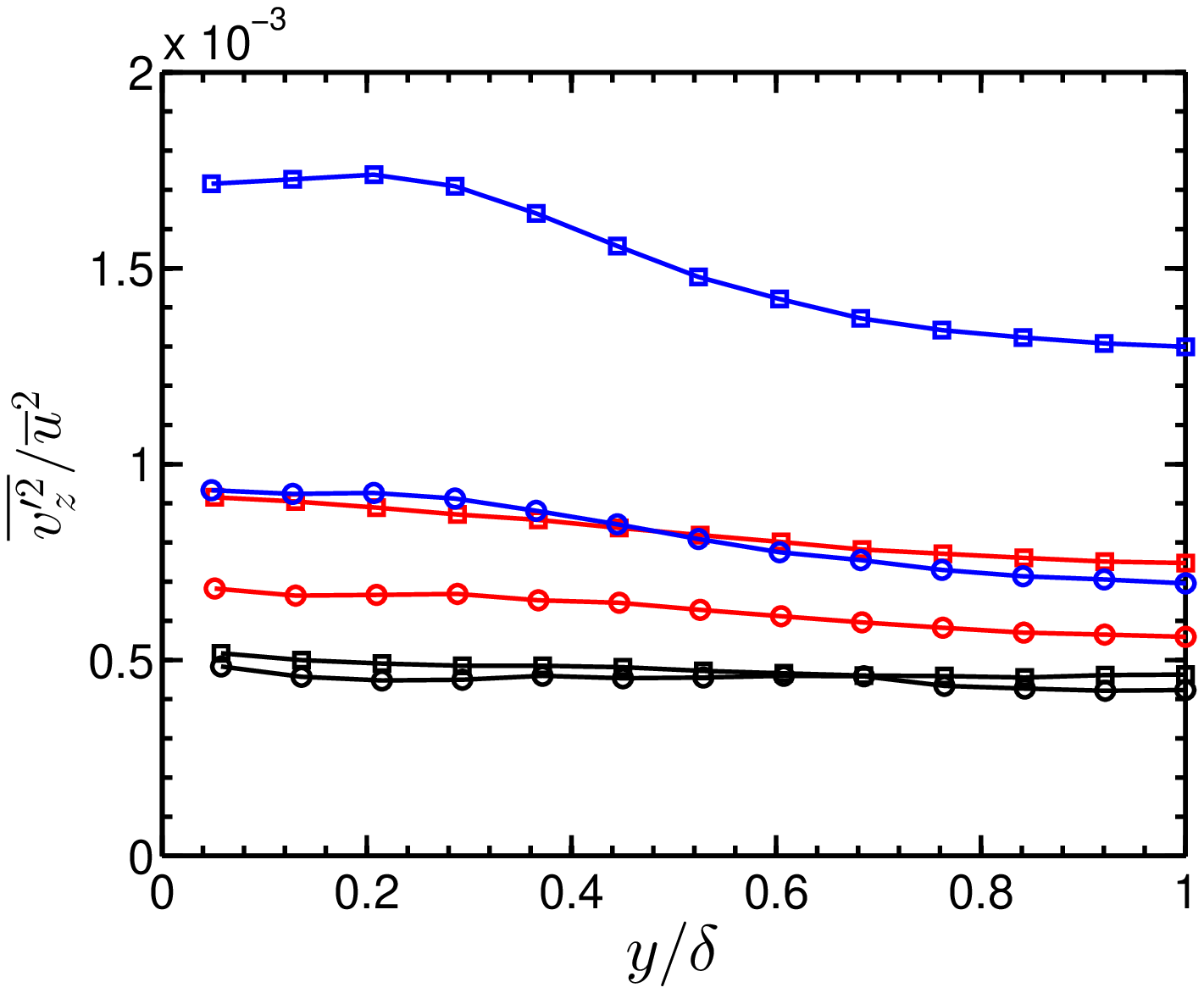}
	\caption{vz rms}
	\endminipage \\	
	\minipage{0.4\textwidth}
	\includegraphics[width=\textwidth]{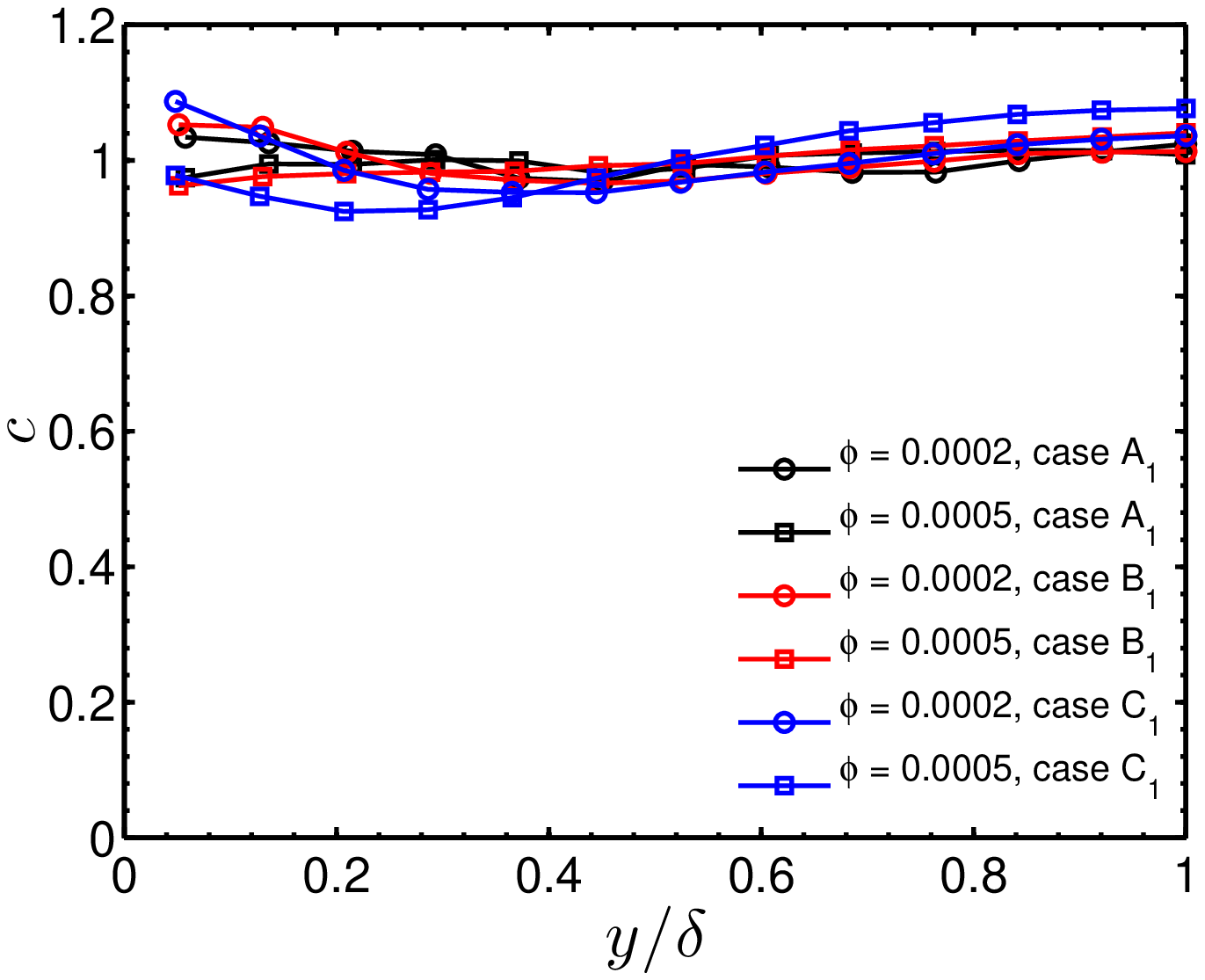}
	\caption{}
	\endminipage
	\minipage{0.4\textwidth}
	\includegraphics[width=\textwidth]{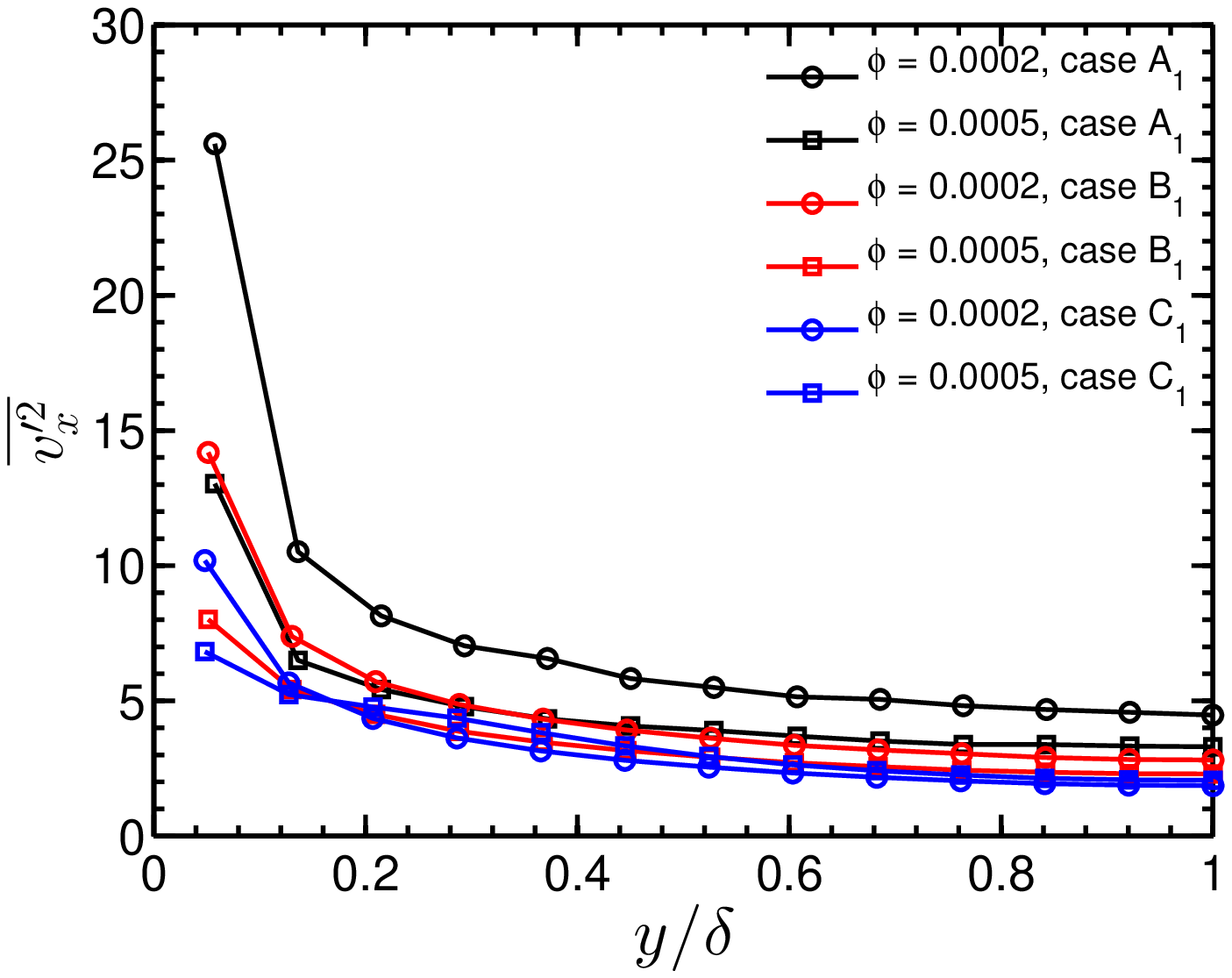}
	\caption{}
	\endminipage
	\end{subfigure} 
	\caption{The particle properties are plotted in the wall-normal direction for cases $A_1$, $B_1$, and $C_1$ for different volume fractions. The particle velocities (except for Fig.~(f)) are scaled with fluid bulk velocity ($\overline{u}$), and wall-normal distance is scaled with channel width ($h$). (a) Mean velocity, (b) second moments of streamwise, (c) wall-normal, (d) spanwise fluctuations, (e) particle concentration, and (f) second moments of streamwise fluctuations. The particle concentration is normalized with an average concentration across the channel width. }
	\label{particle_stat_Ldp}
\end{figure*}

\begin{figure*}
	\begin{subfigure}[b]{1\textwidth}
	\centering
	\minipage{0.4\textwidth}
	\includegraphics[width=\textwidth]{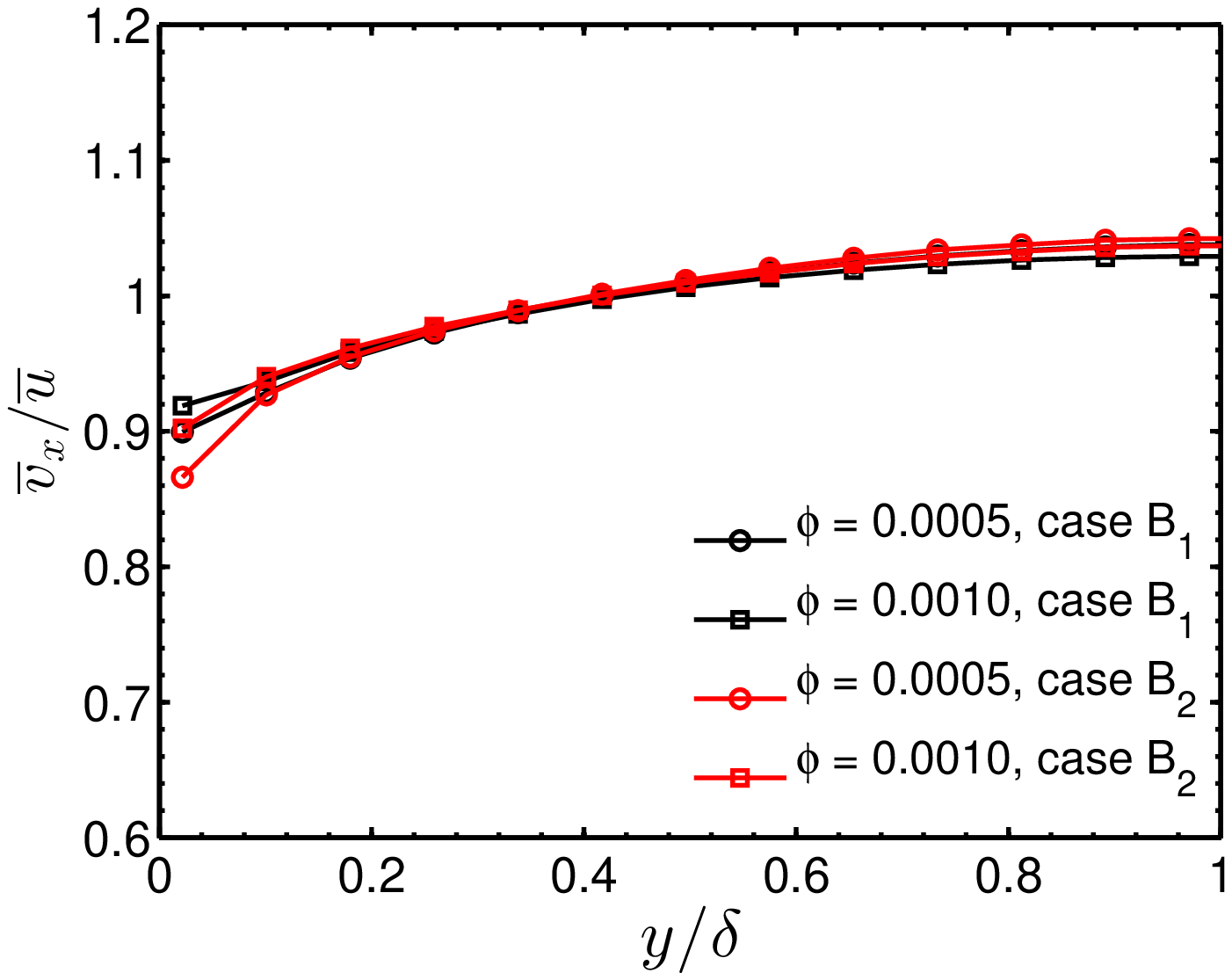}
	\caption{}
	\endminipage 
	\minipage{0.4\textwidth}
	\includegraphics[width=\textwidth]{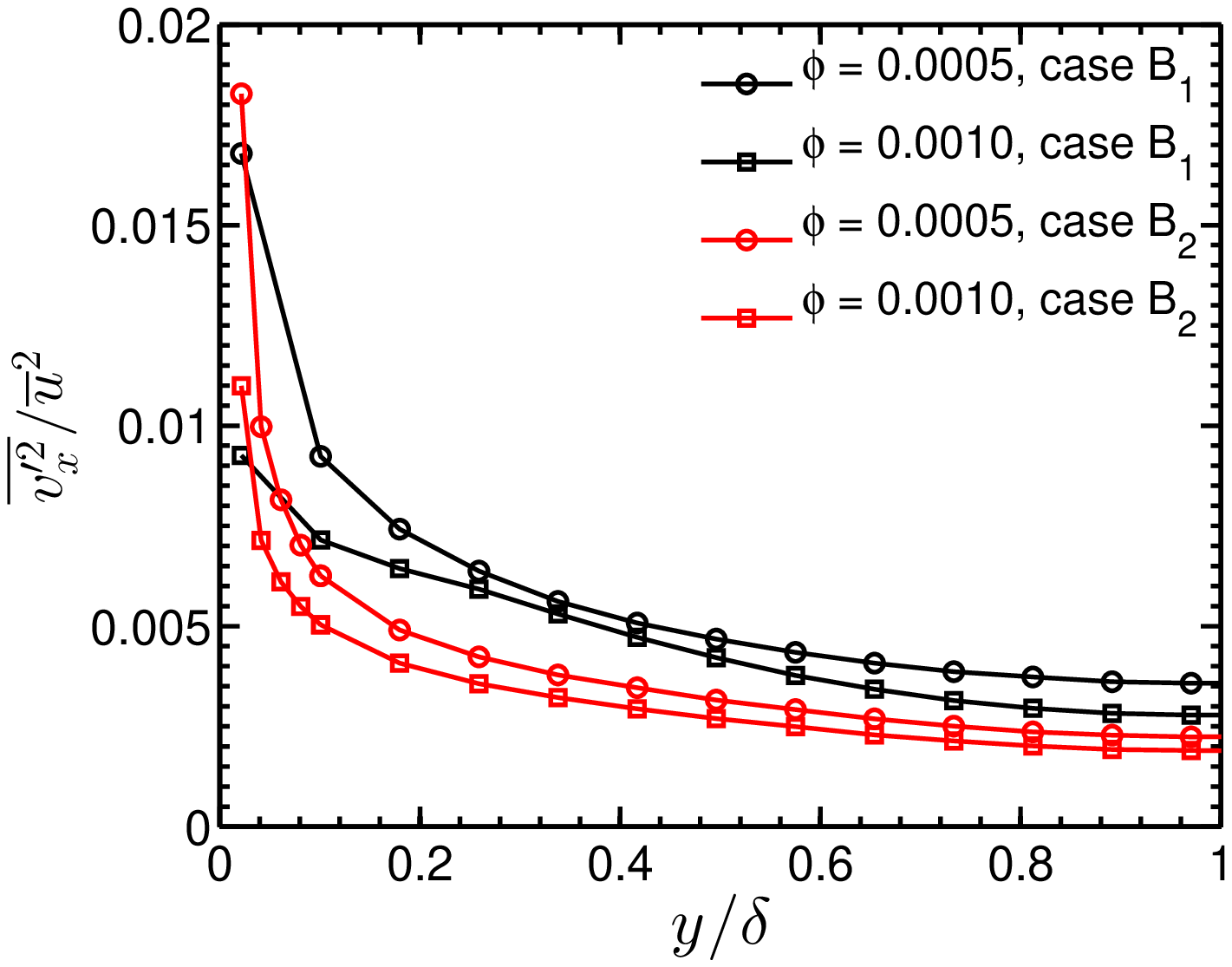}
	\caption{}
	\endminipage \\
	\minipage{0.4\textwidth}
	\includegraphics[width=\textwidth]{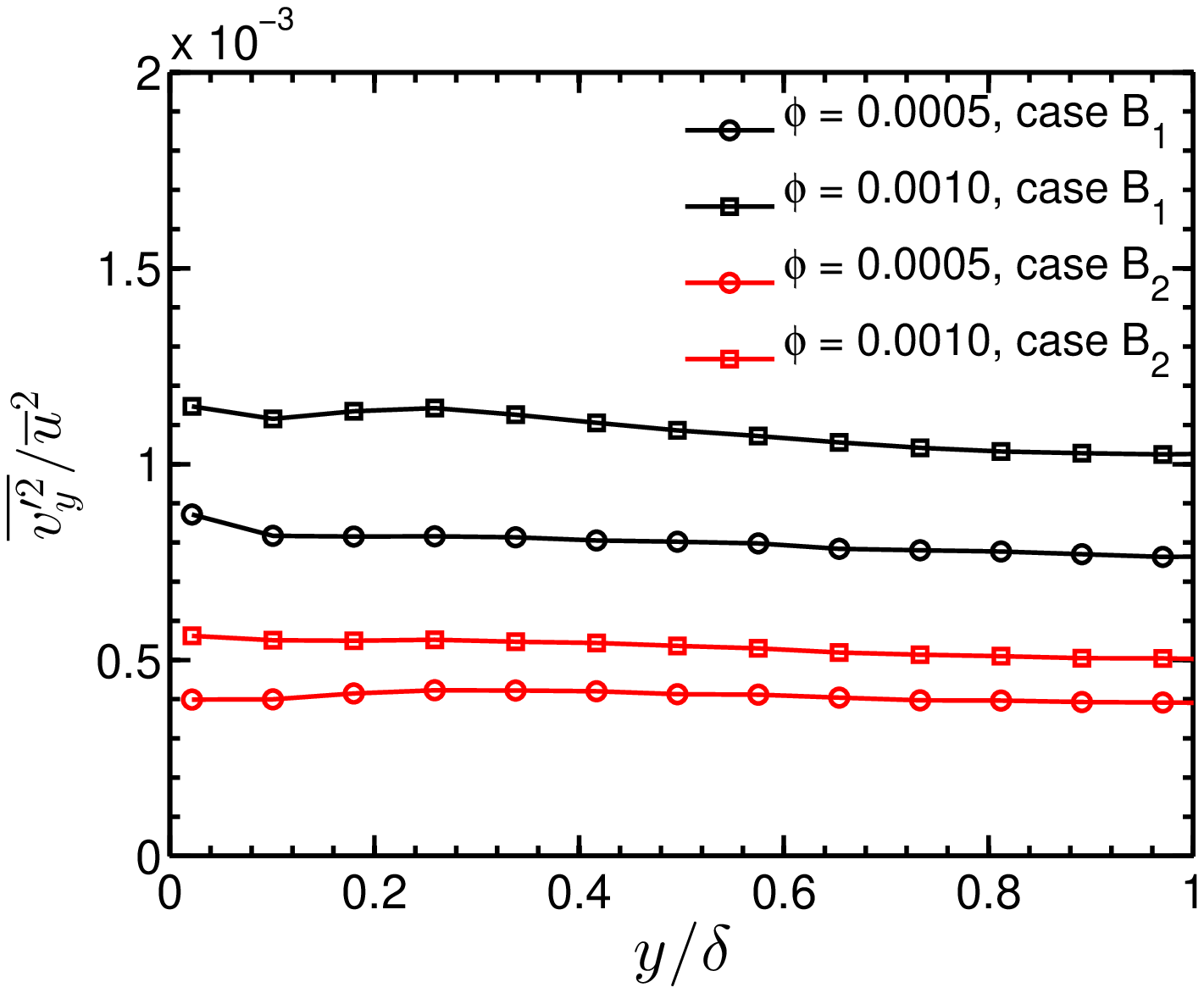}
	\caption{}
	\endminipage	
	\minipage{0.4\textwidth}
	\includegraphics[width=\textwidth]{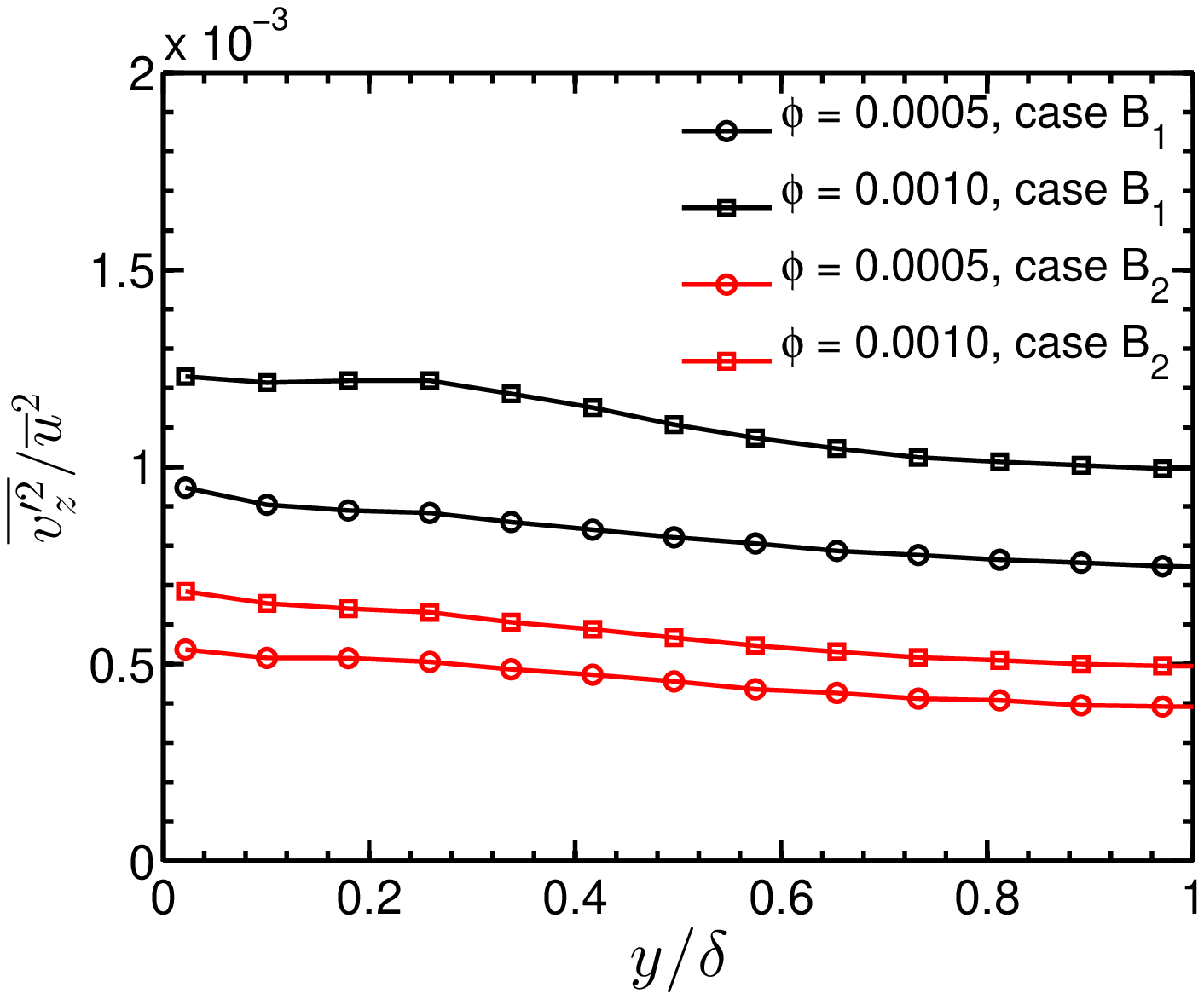}
	\caption{}
	\endminipage \\	
	\minipage{0.4\textwidth}
	\includegraphics[width=\textwidth]{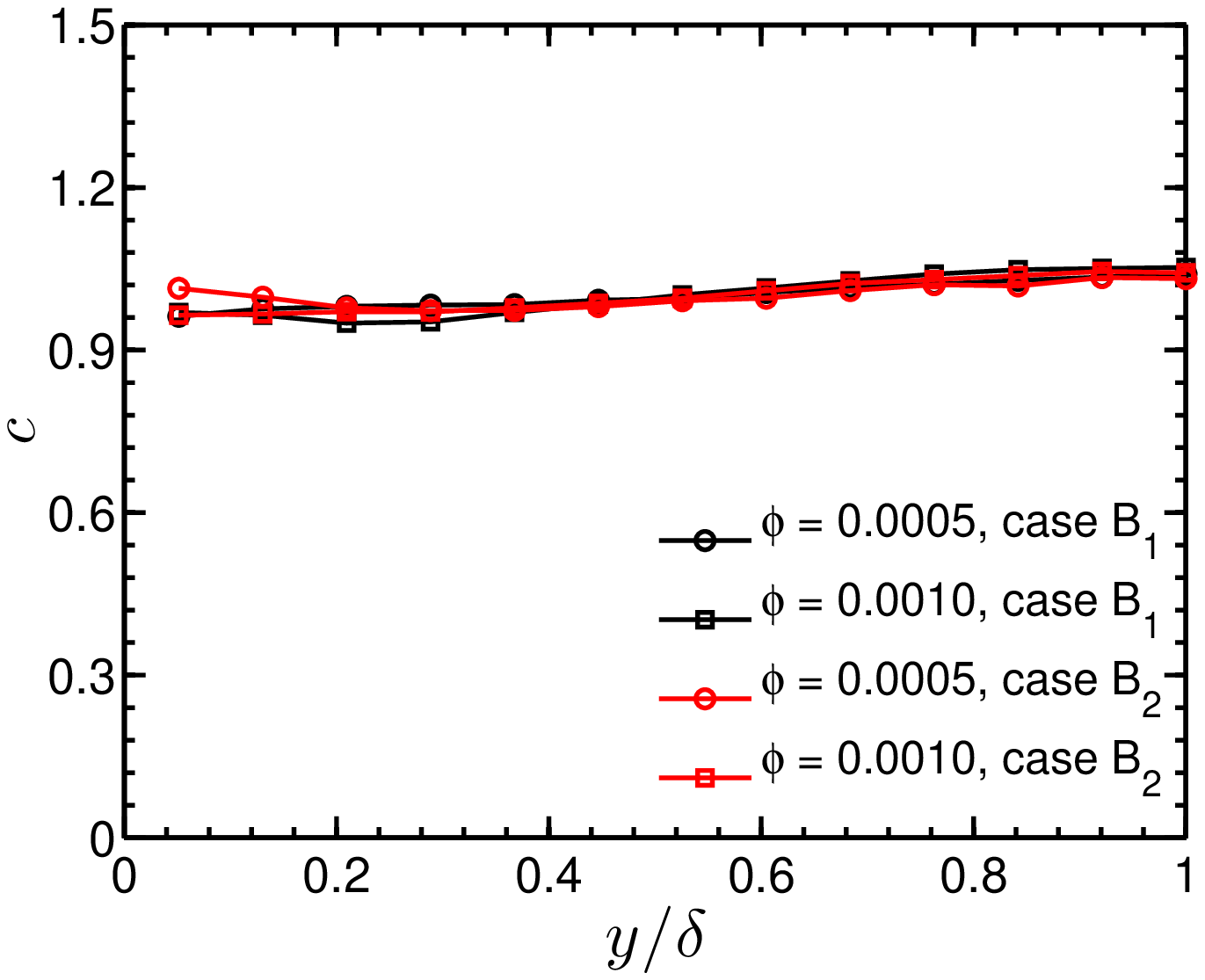}
	\caption{}
	\endminipage
	\minipage{0.4\textwidth}
	\includegraphics[width=\textwidth]{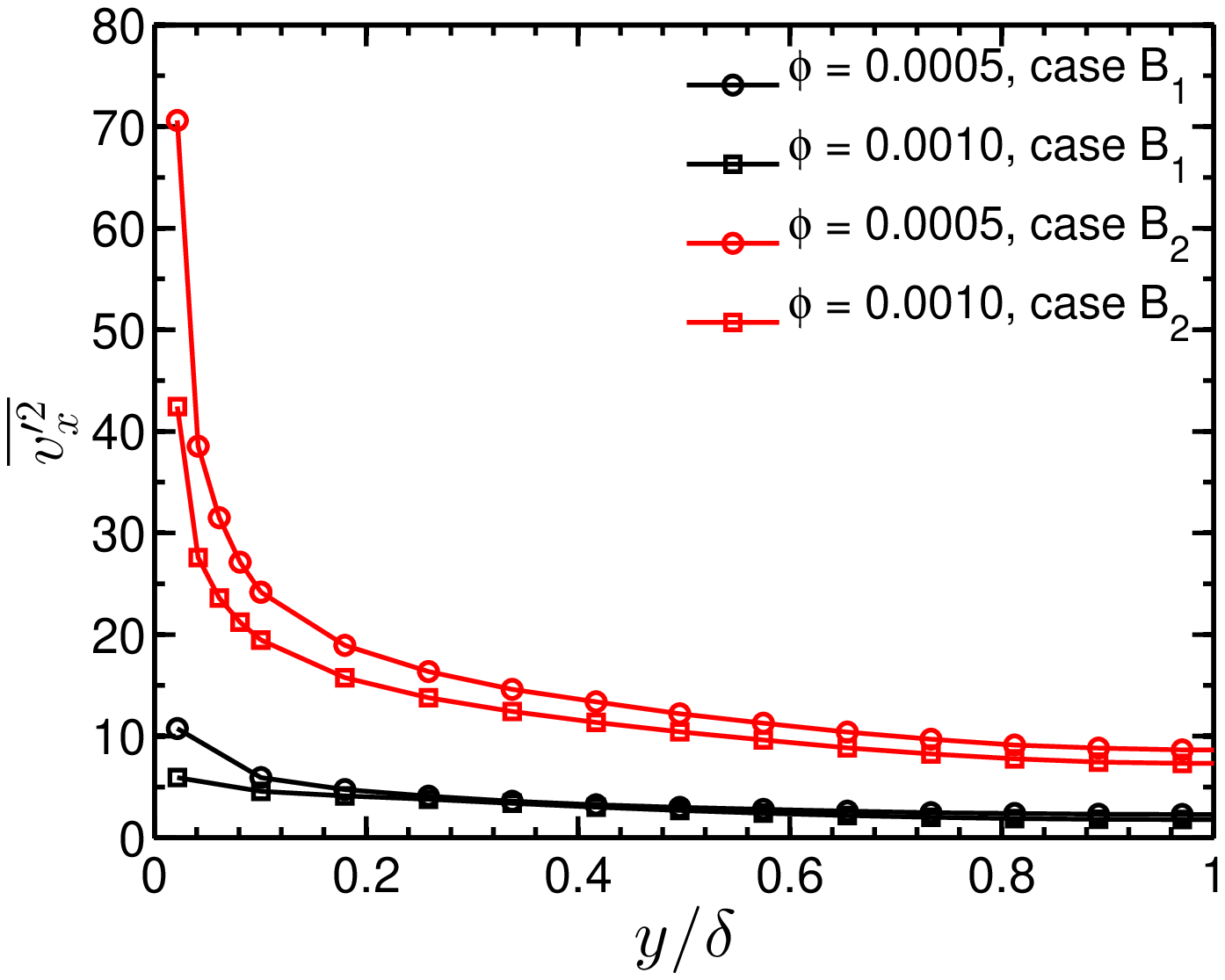}
	\caption{}
	\endminipage
	\end{subfigure} 
	\caption{The particle properties are plotted in the wall-normal direction for cases $B_1$ and  $B_2$ for different volume fractions. The particle velocities (except for Fig.~(f)) are scaled with fluid bulk velocity ($\overline{u}$), and wall-normal distance is scaled with channel width ($h$). (a) Mean velocity, second moments of (b) streamwise, (c) wall-normal, (d) spanwise fluctuations, (e) particle concentration, and (f) second moments of streamwise fluctuations. The particle concentration is normalized with an average concentration across the channel width. }
	\label{particle_stat_Re}
\end{figure*}

In the previous sections, the effect of $\phi $, $2\delta/d_p$, and $Re_b$ are studied on the fluid phase properties. It is observed that the turbulence attenuation increases with an increase in  $\phi $ up to a volume fraction, and a sudden turbulence collapse is observed at CPVL. However, with an increase in the $2\delta/d_p$ ratio, a higher extent of turbulence attenuation is observed for the fixed $\phi $ and Reynolds number. But, for the same volume fraction and fixed channel dimension, the extent of attenuation is low at a higher Reynolds number. This section discusses the modification of particle phase properties for different $\phi $, $2\delta/d_p$, and fluid phase Reynolds number.

The effect of the channel dimension ($2\delta/d_p$) on the particle phase properties at $Re_b = 5600$ is shown in Fig.~\ref{particle_stat_Ldp}. Mean velocities and second moments of the fluctuating velocities are plotted for cases $A_1$, $B_1$, and $C_1$ for two different volume fractions of $\phi = 2\times10^{-4}$ and $5\times10^{-4}$. The normalized particle mean velocities are plotted along with unladen fluid mean velocity, shown in Fig.~\ref{particle_stat_Ldp} (a). The particle mean velocity profile is almost flat across the channel width for the case $A_1$. However, the particle mean velocity is lower in the near-wall region for larger channel sizes of case $B_1$ and $C_1$. For case $C_1$, the particle mean velocity is less in the near-wall region for $\phi = 2\times10^{-4}$ than the other volume fraction cases. At the channel center, the particle mean velocity is higher for both the volume fractions for the case $C_1$ as the considered Stokes number is less here than the other two system sizes, Table~\ref{Stokes_number}. With an increase in solid volume fraction for a fixed channel dimension, near-wall particle velocity increases due to the transfer of momentum from the channel center to the wall through particle-particle interaction. With the increase in channel dimension, Stokes number of the particle decreases, leading to a decrease in the fluid-particle relative velocity near the wall.

The normalized particle fluctuations are plotted in Fig.~\ref{particle_stat_Ldp} (b - d). The streamwise particle fluctuations shown in Fig.~\ref{particle_stat_Ldp} (b) are higher in the near-wall region and decrease away from the wall. The streamwise particle fluctuations decrease as the volume fraction is increased from the $\phi = 2\times10^{-4}$ to $5\times10^{-4}$ for a fixed system size. The wall-normal and spanwise particle fluctuations increase with an increase in the volume fraction due to an increase in the collision frequency for a fixed system size \cite{YAMAMOTO2001}, Fig.~\ref{particle_stat_Ldp} (c and d). It is observed that the streamwise particle fluctuation increase with an increase in system size. Similar behavior is observed for the wall-normal and spanwise particle fluctuations. A more significant increase in wall-normal and spanwise particle fluctuations is observed for the case $C_1$ as the volume fraction is increased from $\phi = 2\times10^{-4}$ to $5\times10^{-4}$. The dimensional particle streamwise fluctuations are plotted in Fig.~\ref{particle_stat_Ldp} (f) for similar cases as in Fig.~\ref{particle_stat_Ldp} (b). Here, it is observed that the dimensional particle streamwise fluctuations decrease with an increase in system size for the same volume fraction. This is similar to the behavior observed for fluid bulk velocity which decreases with an increase in system size for a constant Reynolds number. However, it is interesting to note that the particle fluctuations increase relative to the fluid bulk velocity with an increase in channel dimensions, Fig.~\ref{particle_stat_Ldp} (b). 

Next, we have analyzed the mechanisms of the generation of particle streamwise fluctuations which originate from three sources~\cite{goswami2010particle, Goswami2011}. The first is due to the force exerted by the fluid fluctuations in the streamwise direction, which is proportional to the $D_{xx} \tau_p$ where $D_{xx}$ is the velocity space diffusion coefficient due to fluid fluctuations. Velocity space diffusion coefficient ($D_{ij}$) for the particle in turbulent field is defined as,
\begin{equation}
D_{ij} = \frac{\langle u'_i(0) u'_j(0) \rangle}{\tau_p^2} \int\limits_{0}^{\infty} dt' R_{ij}.
\label{Dij}
\end{equation}
Here, $R_{ij}$ is the Eulerian time correlation tensor of the fluid fluctuations, and $\int\limits_{0}^{\infty} dt' R_{ij}$ is the integral time scale of the fluid phase ($\tau_I$). Eqn.~\ref{Dij} can be written as,
\begin{equation}
D_{ij} = \frac{\langle u'_i(0) u'_j(0) \rangle}{\tau_p^2} \tau_I.
\end{equation}
In the above equation, $\tau_I$ may be replaced with the time scale defined as $2\delta/\bar{u} ( = \tau_f)$. Therefore, 
\begin{equation}
D_{ij} \propto \frac{\langle u'_i(0) u'_j(0) \rangle}{\tau_p^2} \tau_f.
\end{equation}
Normalizing the above relation with fluid average velocity($\bar{u}$), we get,
\begin{equation}
\frac{\tau_p D_{ij}}{\overline{u}^2} \propto \frac{\langle u'_i u'_j/\overline{u}^2 \rangle}{St}.
\end{equation}
Here, $St$ is the particle Stokes number based on the fluid integral time scale ($2\delta/\bar{u}$). The second contribution to the streamwise particle velocity fluctuations is due to particle migration across the streamlines due to wall-normal fluid fluctuations, which is proportional to the $(\tau_p D_{yy})St_\gamma^2$. The third source is due to collision induced by the mean velocity gradient of the particle phase, which is proportional to the $\phi_l (\dot{\gamma}d_p)^2 St_\gamma^3$. Here, $\phi_l$ is the local particle volume fraction, $\dot{\gamma}$ is the mean velocity gradient of particle phase, and $St_\gamma$ is the particle Stokes number based on the mean strain rate as $St_\gamma = \tau_p \dot{\gamma}$. Simplifying, $St_\gamma =  \tau_p (dv_p/dy) = \tau_p (dv_p^*/dy^*) (\bar{u}/\delta) = 2 \times St  (dv_p^*/dy^*)$, where $v_p^*$ is the normalized particle velocity, and $y^*$ is the normalized wall-normal distance. In the near-wall region, the streamwise particle fluctuations generated due to fluid streamwise fluctuations, which are inversely proportional to $St$, are low as the $St$ is large for cases $A_1$, $B_1$ and $C_1$. The contribution due to particle collisions is also low due to smaller local volume fraction ($\phi_l$). The significant contribution is due to the wall-normal migration of the particle, which is proportional to $St_\gamma^2$. In the channel center, the contribution due to wall-normal migration decreases due to low $St_\gamma$. However, it remains dominant compared to the other two sources. The streamwise particle fluctuations increase with an increase in system size (shown in  Fig.~\ref{particle_stat_Ldp} b) as $St_\gamma$ increases from case $A_1$ to $C_1$.

The wall-normal particle fluctuations are also generated due to three mechanisms. First, due to wall-normal fluid fluctuations, which is proportional to $(\tau_p D_{yy})$. The second mechanism is the collision of the particles having different mean velocities, which is proportional to the $\phi_l (\dot{\gamma}d_p)^2 St_\gamma$. The third contribution is due to the collision induced by the streamwise particle fluctuations, which is proportional to the $\phi_l(T^{1/2}_{xx} \tau_p/d_p)T_{xx}$. Here, $T_{xx} = \overline{v'^2_x}$ is the mean square streamwise particle fluctuations. Non-dimensionalising the $\phi_l(T^{1/2}_{xx} \tau_p/d_p)T_{xx}$ with the fluid bulk velocity and channel width ($2\delta$), it becomes $\phi_l (T^{*3/2}_{xx} \times St \times 2\delta/d_p)$. In the present study, the particle Stokes number is high, thus, the contribution from $\tau_p D_{yy}$ is low. The second contribution, $\phi_l (\dot{\gamma}d_p)^2 St_\gamma$, is also not significant as local volume fraction  ($\phi_l$) and $(\dot{\gamma}d_p)$ are low. The significant contribution is due to the third mechanism for $A_1$, $B_1$, and $C_1$. This dominant term increases (due to an increase in $T_{xx} $ and $2\delta/d_p$) with the increase in system sizes from $A_1$ to $C_1$ which is also observed in Fig.~\ref{particle_stat_Ldp} (c).

The normalized particle concentration is plotted in Fig.~\ref{particle_stat_Ldp} (e). The concentration profiles are almost flat for lower system sizes of $2\delta/d_p = 54$ and 81 for the range of volume fractions of $\phi = 2\times10^{-4}$ and $5\times10^{-4}$ considered in this study. However, the particle concentration is larger in the near-wall and channel center region for case $C_1$. This might be due to the decrease in effective particle inertia with the reduction of Stokes number (Table~\ref{Stokes_number}).

The effect of the Reynolds number on the particle properties is examined in Fig.~\ref{particle_stat_Re} for cases $B_1$ and $B_2$. The particle properties are plotted for two volume fractions of $\phi = 5\times 10^{-4}$ and $10^{-3}$. The particle mean velocity predicted for both the volume fractions are almost similar as the Stokes number is also the same as shown in Fig.~\ref{particle_stat_Re} (a). The particle mean velocity is almost 10\% lower in the near-wall region compared to the channel center. In case of streamwise particle fluctuations, the fluctuations are higher for $Re_b = 13750$ compared to the $Re_b = 5600$ for the same volume fractions near the wall. However, away from the wall, the streamwise fluctuations are lower for case $B_2$ ($Re_b = 13750$) than that in case of $B_1$ ($Re_b = 5600$). For particle wall-normal and spanwise fluctuations, it is observed that the fluctuations are lower for the case $B_2$ compared to the case $B_1$. The normalized particle concentration is almost flat and the same for all the considered volume fractions as shown in Fig.~\ref{particle_stat_Re} (e). The dimensional particle streamwise fluctuations are plotted in Fig.~\ref{particle_stat_Re} (f) for similar cases as in Fig.~\ref{particle_stat_Re} (b). Here, it is observed that the dimensional particle streamwise fluctuations increase with an increase in Reynolds number for the same volume fraction. The fluid bulk velocity also increases with an increase in Reynolds number for constant system size; however, the particle fluctuations scaled with the fluid bulk velocity decrease with an increase in Reynolds number, Fig.~\ref{particle_stat_Re} (b). 

When volume fraction is $5\times 10^{-4}$, the component $ (\tau_p D_{yy} ) St_\gamma^2$ increases in the near-wall region for the case of $B_2$ due to high $St_\gamma$. Thus, the streamwise particle fluctuations caused by wall-normal migration are higher in the near-wall region for the case $B_2$ compared to $B_1$ for $\phi = 5\times 10^{-4}$ (Fig.~\ref{particle_stat_Re} b).  $ (\tau_p D_{yy} ) St_\gamma^2$ decreases in channel center due to reduction in $\dot{\gamma}$. In the channel center, $ (\tau_p D_{yy} ) St_\gamma^2$ provides almost simialr values for the cases $B_1$ and $B_2$ as $St_\gamma (= 2 \times St  (dv_p^*/dy^*))$ is also similar for both the cases. However, it is not observed in  Fig.~\ref{particle_stat_Re} (b), and the particle streamwise fluctuations are lower for $Re_b = 13750$ compared to the $Re_b = 5600$. The reason behind this observation is not clear at this moment. For wall-normal particle fluctuations, $\phi_l (T^{*3/2}_{xx} \times St \times 2\delta/d_p)$ is the dominant source, and it is lower for the $B_2$ as the $T^*_{xx}$ is lower compared to $B_1$ (except very close to the wall) as shown in Fig.~\ref{particle_stat_Re} (b). The above discussion explains the statistics of the particle phase shown in Fig.~\ref{particle_stat_Ldp} - \ref{particle_stat_Re}.

\section{Conclusion }
In the present work, the simulations are performed for vertical channel flow for different volume fractions, system sizes, and Reynolds numbers to critically assess those parameters' effects on turbulence modulation. The fluid phase is described in the Eulerian framework using the large eddy simulation. A dynamic one-equation model has been adopted to model the subgrid-scale stress. The individual particles are tracked using a Lagrangian technique with point-particle assumption. It is observed that the fluid fluctuations decrease with an increase in volume fraction up to a critical particle volume loading at which turbulence collapses. The extent of turbulence attenuation increases with an increase in system size ($2\delta/d_p$) for the same volume fraction while keeping the Reynolds number fixed. But, for the same volume fraction and fixed channel dimension, the extent of attenuation is low at a higher Reynolds number. 

Different terms from the momentum and energy balance equations are plotted, and it is concluded that the variation of the particle feedback term is the dominant term when the volume fraction is lower than the critical volume fraction. It is observed that the dissipation due to the particle significantly increases with an increase in system size for the same volume fraction and fixed Reynolds number. However, the dissipation due to particle decreases significantly when the Reynolds number is increased while keeping channel dimensions same, which reduces the extent of turbulence attenuation. From the analysis of the variation of scaled feedback force, it is found that the system size and bulk Reynolds number are the two important parameters that significantly affect the turbulence modulation compared to the others.

The spatial correlation of fluid fluctuations is plotted in the streamwise and spanwise directions, and it is observed that the correlation coefficient for the streamwise component decays slower with an increase in system size, which indicates that the streamwise turbulent structures become elongated along the channel length. The isocontours of the streamwise fluctuations also depict a similar observation. The isocontours show that the high-speed streaks become elongated and fewer with an increase in system size for a fixed Reynolds number. This may be because the presence of particles affects the breakdown of coherent structures with an increase in system size. The correlation coefficient decays faster for a higher Reynolds number than a lower one. The isocontours of the streamwise fluctuations show that the streaks are smaller, thinner, and closely packed for a higher Reynolds number. 

The particle fluctuations, scaled with fluid bulk velocities, are analyzed for the effect of system size and Reynolds number. For a constant system size and Reynolds number, the streamwise particle fluctuations decrease, while wall-normal and spanwise particle fluctuations increase with an increase in particle volume fraction due to the increase in particle collision frequency. It is observed that the particle fluctuations increase with an increase in system size for a constant Reynolds number and volume fraction. However, the particle fluctuations, except for the streamwise particle fluctuations in the near-wall, decrease with an increase in Reynolds number for fixed channel dimension and volume fraction. Qualitative analysis for the source of particle streamwise and wall-normal fluctuations is discussed. It is observed that the streamwise particle fluctuations generated due to wall-normal migration are dominant compared to the other two sources (force due to streamwise fluid fluctuations and collision-induced fluctuations). The source due to particle migration in the wall-normal direction increases with an increase in system size. For the wall-normal fluctuations, the source due to collision induced by streamwise particle fluctuations is dominant in the near-wall and channel center. The present analysis highlights the multiscale nature of particle-laden flows in the sense that an increase in Stokes number of particles universally demonstrates the turbulence modulation, but the system size is to be considered. Therefore, this study will be a guideline for designing industrial instruments while using them for gas-solid operations.

\section{Appendix}
\label{sec:appendix}

\begin{figure*}[htb]
	\begin{subfigure}[b]{1\textwidth}
	\minipage{0.4\textwidth}
	\includegraphics[width=\textwidth]{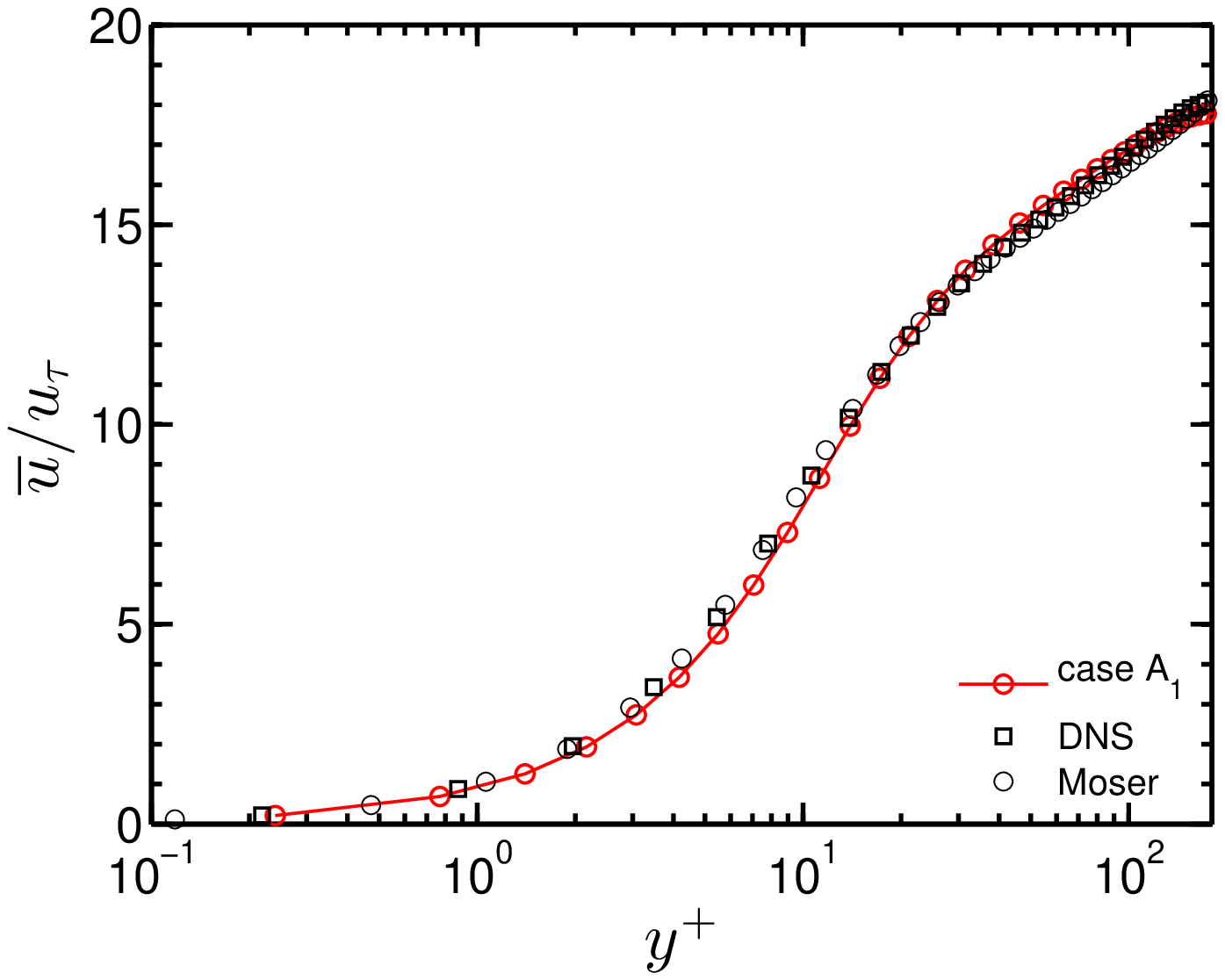}
	\caption{}
	\endminipage 
	\minipage{0.4\textwidth}
	\includegraphics[width=\textwidth]{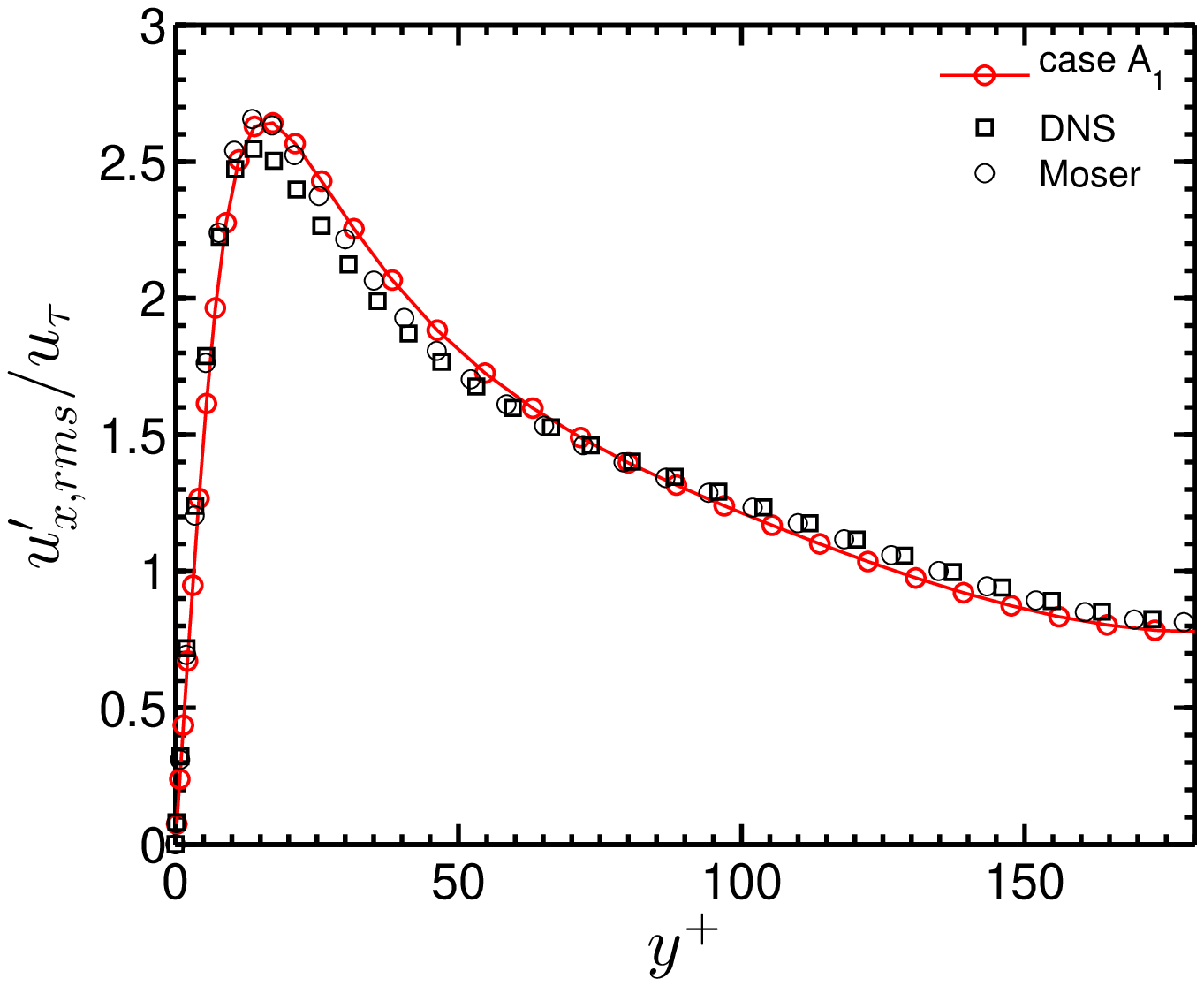}
	\caption{}
	\endminipage \\
	\minipage{0.4\textwidth}
	\includegraphics[width=\textwidth]{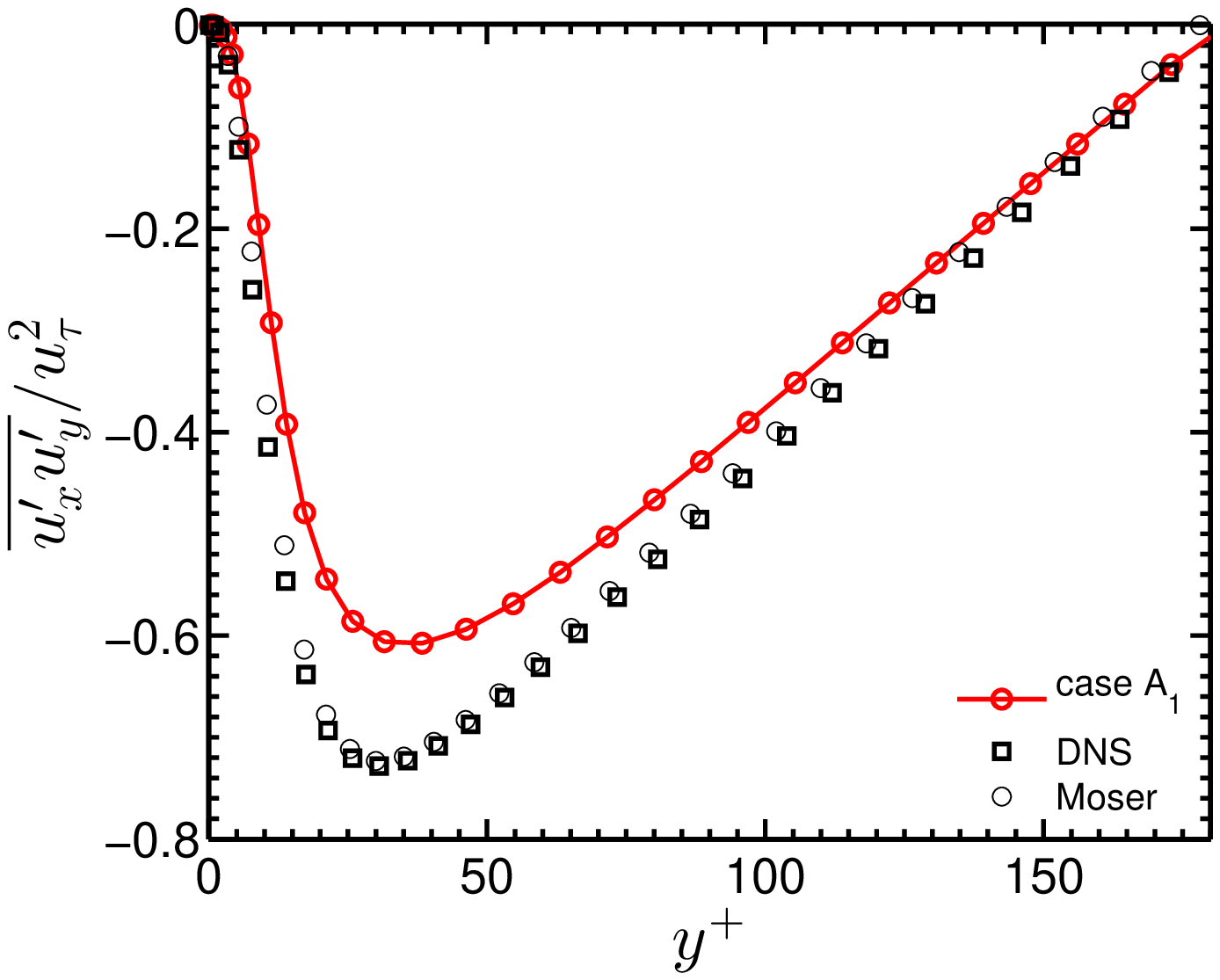}
	\caption{}
	\endminipage	
	\minipage{0.4\textwidth}
	\includegraphics[width=\textwidth]{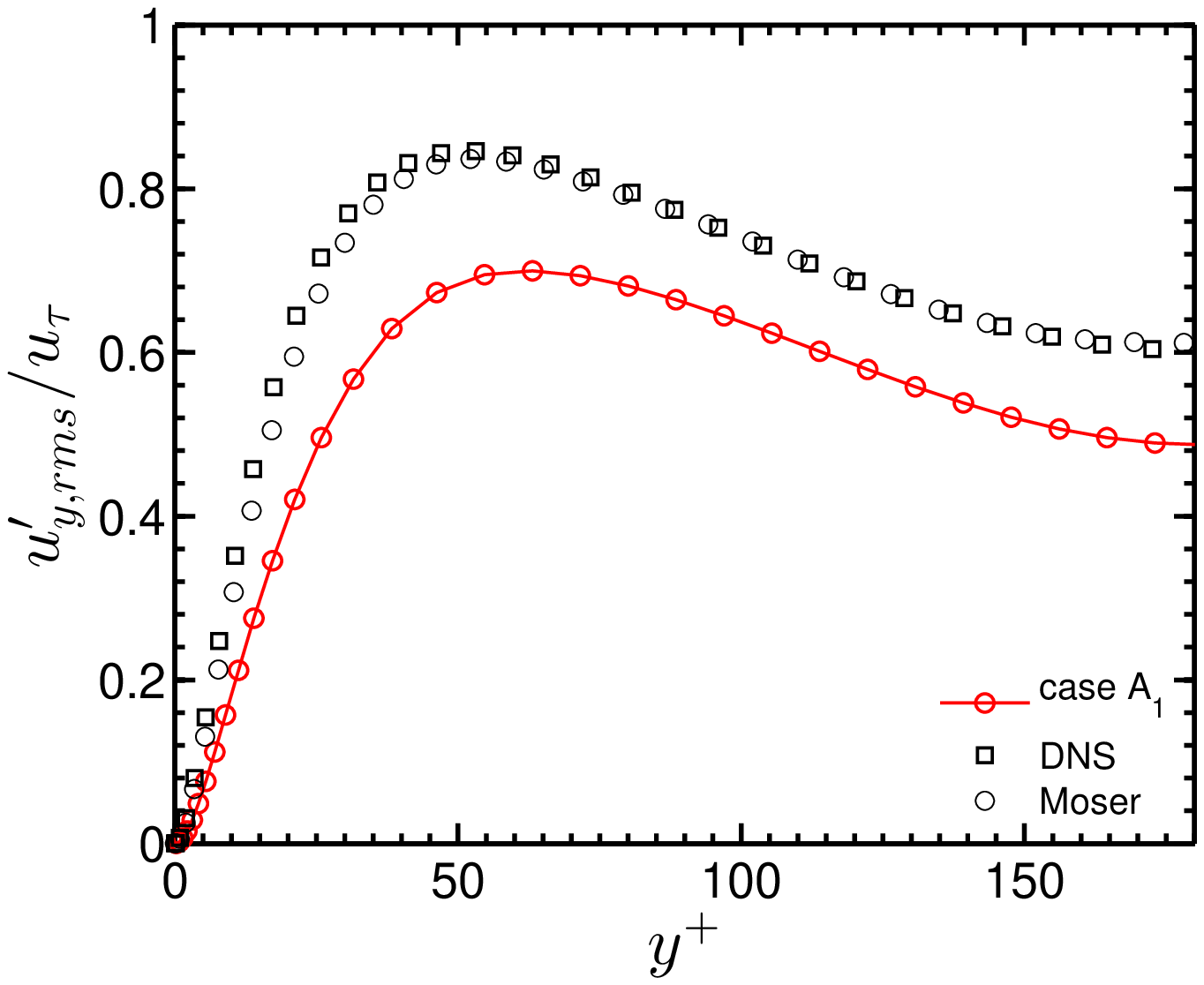}
	\caption{}
	\endminipage 
	\end{subfigure} 
	\caption{The unladen fluid (a) mean velocity, root mean square fluid fluctuations in (b) streamwise, (c) cross-stream stress, and (d)  wall-normal direction. The fluid velocities and wall-normal distance are normalized by viscous scales at $ Re_b = 5600 $. `DNS' is the data from \citet{muramulla2020disruption}, `Moser' is the DNS data from \citet{Moser1999}.}
	\label{unladen_Re180}
\end{figure*}
 
The unladen fluid statistics for $Re_b = 5600$ is presented in Fig. \ref{unladen_Re180}. The simulations are verfied against inHouse DNS \cite{muramulla2020disruption} and DNS data of \citet{Moser1999}. The fluid phase velocities are normalized with unladen friction velocity ($u_\tau$), and the wall-normal distance is normalized with $u_\tau$ and half-channel width ($\delta$). The streamwise mean velocity and fluctuations plotted in the wall-normal direction, are in good agreement with both the DNS datas as shown in Fig. \ref{unladen_Re180} (a and b). The cross-stream stress and  wall-normal fluctuations are underpredicted by LES within 15\% and 20\% deviations compared to DNS datas, respectively.

\begin{figure*}[t]
	\begin{subfigure}[b]{1\textwidth}
	\minipage{0.4\textwidth}
	\includegraphics[width=\textwidth]{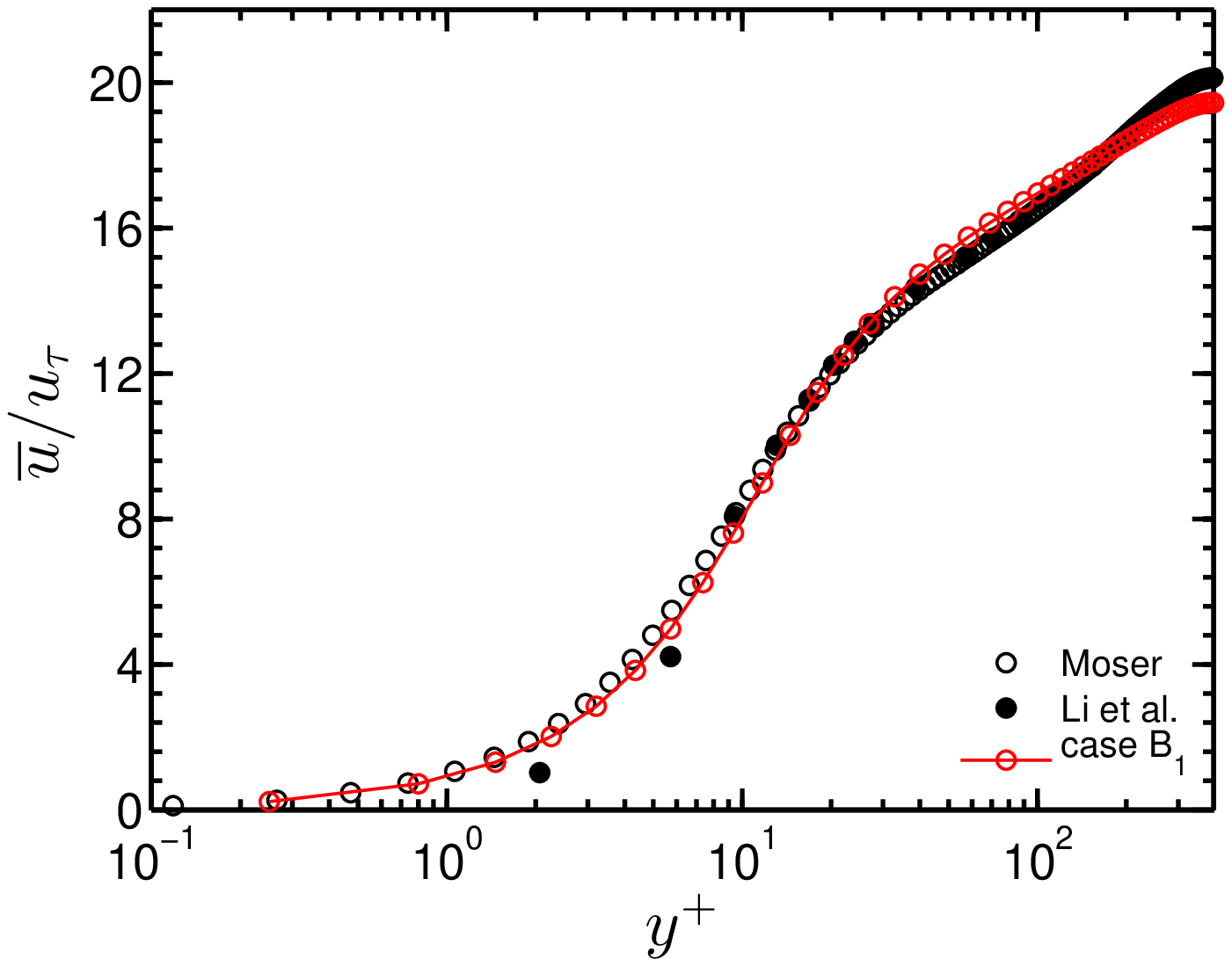}
	\caption{}
	\endminipage 
	\minipage{0.4\textwidth}
	\includegraphics[width=\textwidth]{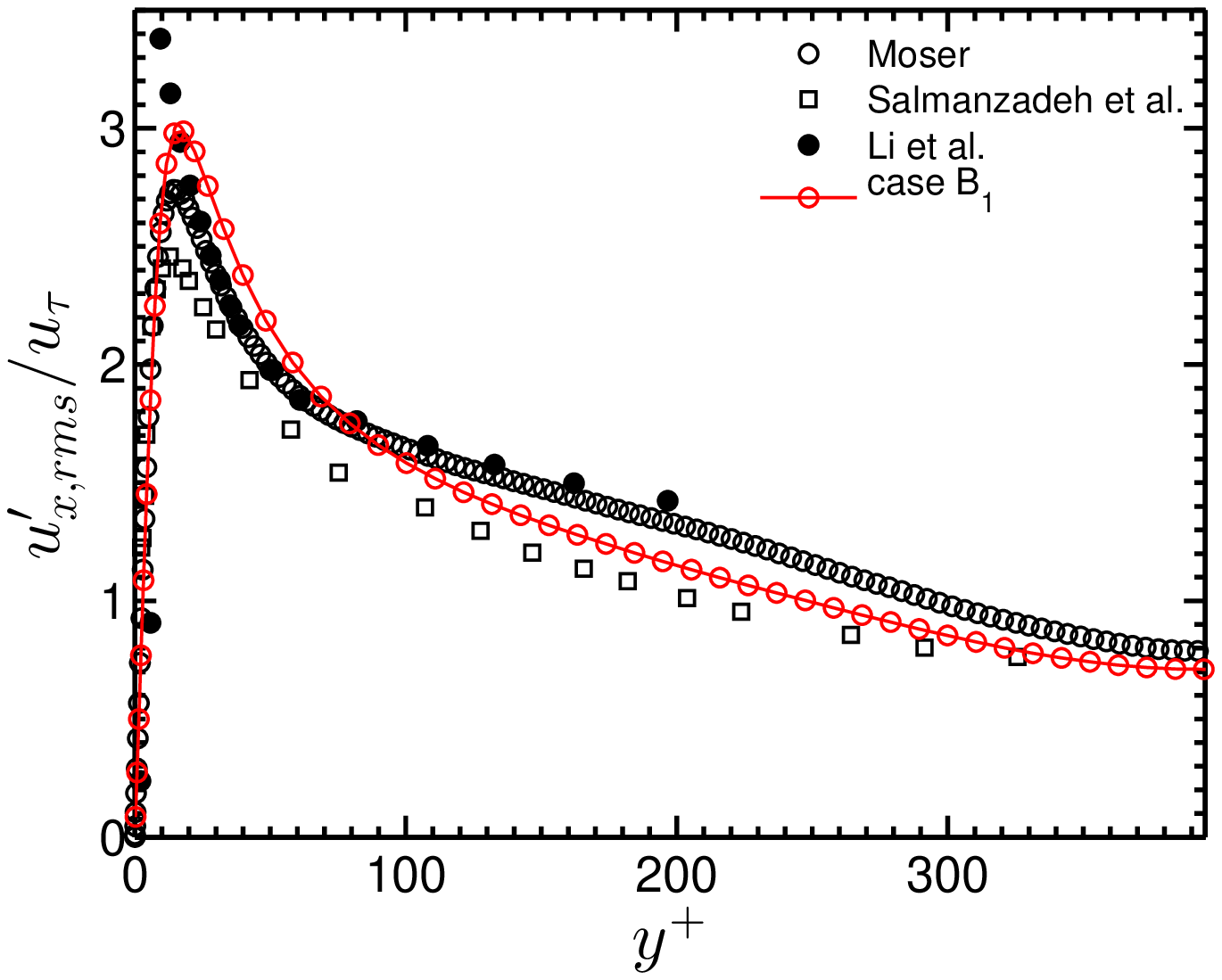}
	\caption{}
	\endminipage \\
	\minipage{0.4\textwidth}
	\includegraphics[width=\textwidth]{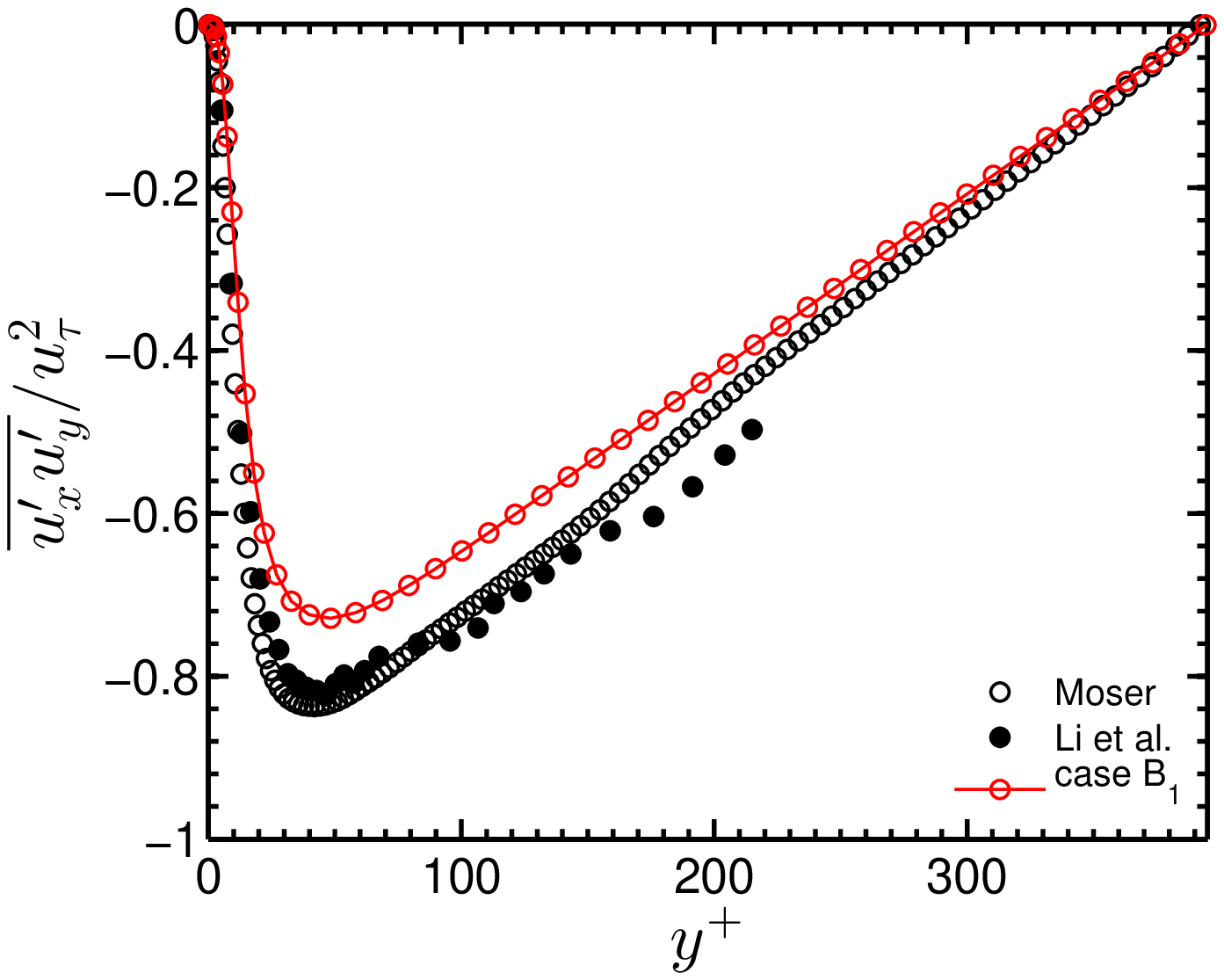}
	\caption{}
	\endminipage	
	\minipage{0.4\textwidth}
	\includegraphics[width=\textwidth]{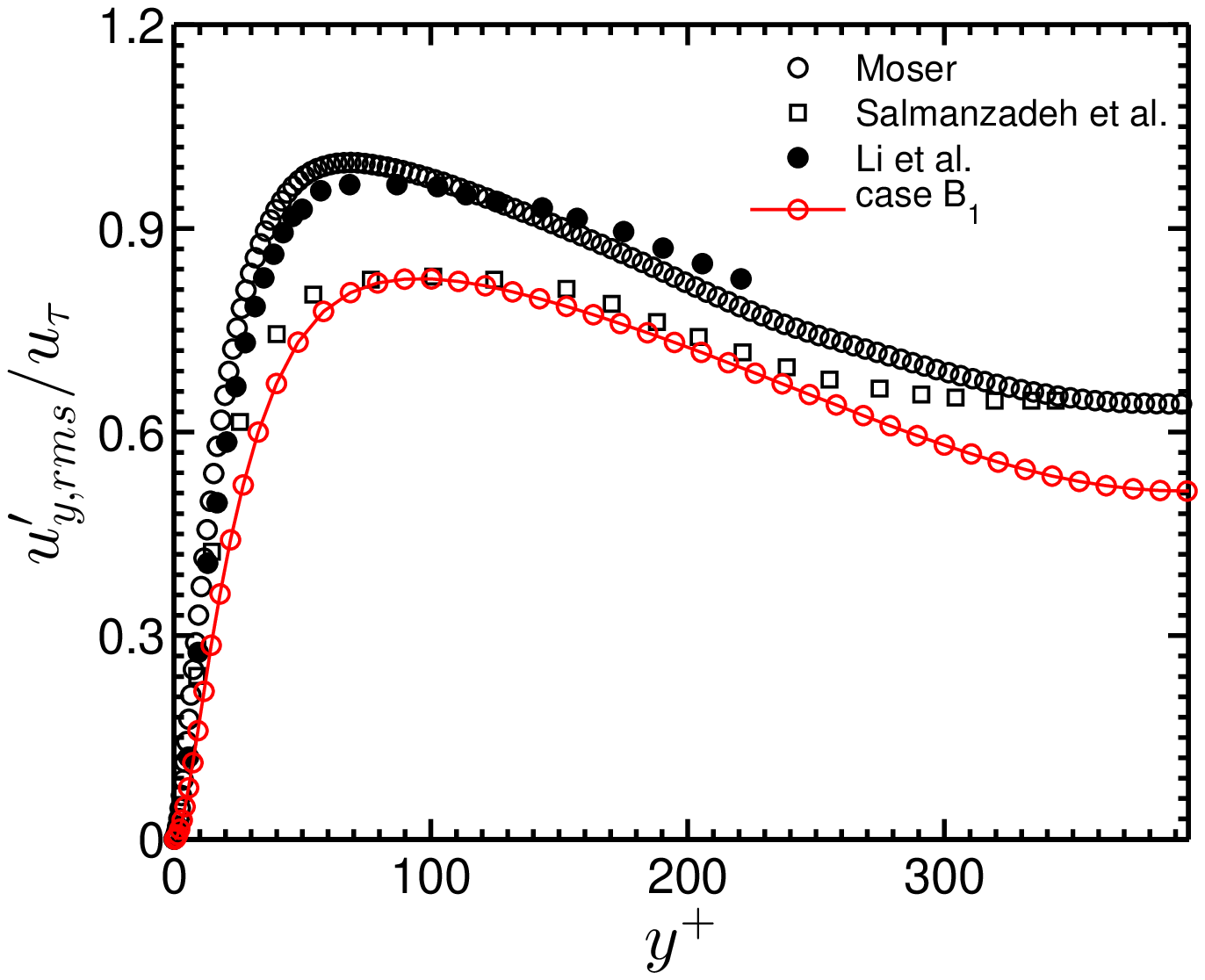}
	\caption{}
	\endminipage 
	\end{subfigure} 
	\caption{The unladen fluid (a) mean velocity, root mean square fluid fluctuations in (b) streamwise, (c) cross-stream stress, and (d)  wall-normal direction. The fluid velocities and wall-normal distance are normalized by viscous scales at $ Re_b = 13750 $. The results are compared with DNS data of \citet{Moser1999} at $Re_\tau= 395$, experimental data of \citet{li2012experimental} at $Re_\tau = 430$, and LES data of \citet{Salmanzadeh2010} at $Re_\tau = 350$.}
	\label{unladen_Re395}	
\end{figure*}

In Fig.~\ref{unladen_Re395}, the unladen fluid statistics for $Re_b = 13750$ is validated against the DNS data of \citet{Moser1999} at $Re_\tau= 395$, experimental data of \citet{li2012experimental} at $Re_\tau = 430$, and LES data of \citet{Salmanzadeh2010} at $Re_\tau = 350$. The mean velocity and fluid fluctuations predicted by LES model are in good agreement. The peak value of fluid fluctuations are within 15\% deviation compared to DNS data of \citet{Moser1999}.

\section*{Acknowledgement}
We would like to thank the Science and Engineering Research Board (SERB), Department of Science and Technology (DST), Government of India, for their financial support.
\section*{References}
\cite{Armenio1999}

\bibliography{mybib}

\end{document}